         \documentclass[11pt,draftcls,onecolumn,letterpaper]{IEEEtran}

        \usepackage[T1]{fontenc}                        

        \usepackage{theorem}
        \usepackage{latexsym}
        \usepackage{amssymb}
        \usepackage{subfigure}

        \usepackage{dsfont} 
        \usepackage{graphicx} 
        \graphicspath{{./Fig/}}     

        \input ./Macro/alphabet.tex     
        \input ./Macro/abrege.tex               
\def\cro#1{\left[#1\right]}

\def\sinc{{\mathrm{sinc}}\,}

\newsavebox{\fminibox}
\newlength{\fminilength}

 \def\T{^\tD} \def\+{^\dagger}

\def\nequiv{\not\kern-.05em\equiv}
\def\egal{\kern-.5em=\kern-.5em}        
\def\propt{\kern-.2em\propto\kern-.2em} 

\def\froc#1#2{{#1/#2}}                  

\def\intdouble{\int\kern-0.3em\int}
\def\inttriple{\int\kern-0.3em\int\kern-0.3em\int}

\def\rond#1{\overset{\kern-0.33em~_\circ}{#1}}
\def\rondit[#1]#2{\overset{\kern#1~_\circ}{#2}}

                \def\erf{\mbox{erf}}
                \newcommand{\field}[1]{\mathbb{#1}}
                \newcommand{\bR}{\field{R}}
                \newtheorem{remark}{Remark}

\begin{document}

%

\title{Data inversion for over-resolved \\ spectral imaging in astronomy}

\author{T. Rodet, F. Orieux, J.-F. Giovannelli and A. Abergel\thanks{T. Rodet, F. Orieux, J.-F. Giovannelli are with the Laboratoire des Signaux et Systèmes (\textsc{cnrs} -- Supélec -- \textsc{ups}), Supélec, Plateau de Moulon, 3 rue Joliot-Curie, 91192 Gif-sur-Yvette Cedex, France. E-mail: \texttt{{orieux,rodet,giova}@lss.supelec.fr}.}
\thanks{A. Abergel is with the Institut d'Astrophysique Spatiale, batiment 121,
Université Paris Sud 11, 91405 Orsay Cedex. E-mail: \texttt{abergel@ias.u-psud.fr.}}}

\markboth{Submitted to IEEE JOURNAL OF SELECTED TOPICS IN SIGNAL PROCESSING}{Signal Processing for Astronomical and Space Research Applications}

\maketitle
\begin{abstract}

We present 
an original
method for reconstructing a three-dimensional object having
two spatial dimensions and one spectral dimension from data provided by the
infrared slit spectrograph
on board the Spitzer Space Telescope. 
During acquisition, the light flux is deformed
by a complex process comprising four main elements (the telescope aperture,
the slit, the diffraction grating and optical distortion) before it reaches
the 
two-dimensional sensor. 

The originality of this work lies in the physical modelling, in integral
form, of this process of data formation in \textit{continuous variables}. The inversion is also approached with \textit{continuous variables} in a
semi-parametric format decomposing the object into a family of Gaussian functions. The estimate is built in a deterministic regularization framework as the
minimizer of a quadratic criterion. 

These specificities give our method the power to over-resolve. Its performance 
is illustrated using real and simulated data. We also present a study
of the resolution showing a 1.5-fold improvement relative 
to conventional methods.
\end{abstract}


 
%

\section{Introduction}\label{sec:introduction}

Since the end of the 1970's, infra-red to millimetric observations of the
sky from space have brought about a revolution in practically all fields
of astrophysics. It has become possible to observe distant galaxies,
perform 
detailed physicochemical studies of interstellar matter. Observations in
the far infra-red are now possible thanks to new types of sensors (Ge:Ga
Si:As semiconductors and bolometer arrays). The properties of these new sensors
encouraged the astrophysicists of the Institut d'Astrophysique Spatiale (IAS) 
to work with researchers at the Laboratoire des Signaux et Systèmes (L2S) in
order to develop suitable processing methods. The spectral imaging work
presented here was carried out in the framework of this cooperative effort.
The aim is to reconstruct an over-resolved object having two spatial 
dimensions
($\alpha,\beta$)\footnote{In this paper, the spatial dimensions are angles in radian} 
and one spectral dimension $\lambda$. 
 Data provided by 
the Infrared Spectrograph (IRS)~\cite{houck04} on board the
american Spitzer Space Telescope launched in 2003 
are used to illustrate our work. 
Several sets of two-dimensional data 
are delivered by the Spitzer Science Center (SSC), each set being
the result of an acquisition for a given satellite pointing direction. The
data were acquired using a slit spectrograph, the operation of which is described
in detail in part \ref{sec:modele-direct-cont}. This instrument is located
in the focal plane of the 
telescope. When the telescope is pointed
towards a region of the sky, the spectrograph slit selects a direction of
space 
$\alpha$. The photon flux is then dispersed perpendicularly
to the 
slit direction with a diffraction grating. The measurement is made using a 
two-dimensional sensor. A signal containing one spatial dimension $\alpha$ and the spectral dimension $\lambda$ is thus obtained. The second spatial dimension $\beta$ is obtained by scanning the sky (modifying telescope pointing). This scanning has two notable characteristics: 
\begin{itemize}
\item it is irregular, because the telescope control is not perfect;
\item it is, however, measured with sub-pixel accuracy (eighth of a pixel).
\end{itemize}
In addition, for a given pointing direction, 
: the telescope optics, the slit width and the sensor 
integration limit the spatial resolution while the grating, the slit and 
the sensor integration limit
the spectral resolution. The specificity of systems of this type is that
the width of impulse response
 depends on the wavelength.
A phenomenon of spectral also 
aliasing appears for the shortest wavelengths.
Finally, the scanning results in irregular sampling along the spatial
direction $\beta$.
The problem to be solved is thus one of inverting the spectral aliasing
(i.e. the over-resolution) using a finite number of discrete data provided
by a complex system. The solution proposed here is based on precise modelling
of the instrument and, in particular, the integral equations containing the
continuous variables ($\alpha$, $\beta$ and $\lambda$) of the optics and
sensing system. The model input is naturally a function of these continuous
variables $\phi(\alpha,\beta,\lambda)$ and the output is a finite set $\yb$
of discrete data items. 

The approach used for solving the inverse problem, i.e. reconstructing an
object having three continuous variables from the discrete data

\begin{itemize}

\item comes within the framework of regularization by penalization; 

\item uses a semi-parametric format where the object is decomposed into a
family of functions.

\end{itemize}

There is a multitude of families of functions available (possibly forming
a basis of the chosen functional space). The most noteworthy are Shannon,
Fourier, wavelet and pixel-indicator families or those of spline, Gaussian
Kaiser-Bessel, etc.  
Work on 3D tomographic reconstruction has used a family of Kaiser-Bessel
functions having spherical symmetry in order to calculate the projections
more efficiently \cite{Lewitt90,Lewitt92,Matej96,Andreyev06}.
In a different domain, the signal processing community has been working on
the reconstruction of over-resolved images from a series of low resolution
images \cite{Park03}.
A generic direct model can be described \cite{Park03},
starting 
with a continuous scene, to which are applied $k$ shift or deformation
operators including at least one translation. This step gives $k$ deformed,
high-resolution images. A convolution operator modelling the optics and sensor
cells is then applied to each of the images. After subsampling, the $k$ low-resolution
images that constitute the data are obtained. Recent work on the direct model
has mainly concerned modelling the shift by introducing a rotation
of the image \cite{Elad97,Elad99} and a magnifying factor \cite{Rochefort06}.
Other works have modelled the shift during sensor integration by modifying
the convolution operator \cite{Patti97}.
To the best of our knowledge, in most works, the initial discretization step
is performed on pixel indicators \cite{Hardie97,Elad97,Patti97,Elad99,Park03,Woods06}. 
On this point, a noteworthy contribution has been made by P. Vandewalle \textit{et al.}
 who discretize the scene on a truncated discrete Fourier basis \cite{Vandewalle07}.
However, their decomposition tends to make the images periodic leading
 to create artefacts on 
the image side. Thus we have decided not to use 
this approach. Recently, the problem of X-ray imaging spectroscopy has been solved 
in the Fourier space \cite{Piana07}, but each spectral component has been 
estimated independantly.

The two major contributions of 
our paper are (1) the modelling of the
measurement system as a whole with continuous variables and (2) the continuous
variable decomposition of the three dimensional object over a family of Gaussian
functions. Modelling with continuous variables enables a faithful description
to be made 
of the physical phenomena involved in the acquisition and avoids
to carry out any prior data interpolation. 
In our case, 
computing the model output requires six integrals (two for the response
of the optics, two for the grating response, and two for the sensor integration)
and the choice of a Gaussian family allows five of these six integrals to
be explicitly stated.
Our paper is organised as follows: 

Part \ref{sec:modele-direct-cont}
describes the continuous model of the instrument comprising:
the diffraction at the aperture, the truncation by the slit, the response
of the grating, the distortion of the light flux, the sensor integration and the scanning of the sky. 
In part \ref{sec:decomp-sur-une}, the object
with continuous variables is decomposed over a family of Gaussian functions.
The aperture and grating responses are approximated by Gaussian functions. 
This part concludes with the obtention of a precise, efficient model of the
measuring system. 
The inverse problem is solved in a regularized framework
in part \ref{sec:inversion}. Finally, part \ref{sec:resultats} gives an evaluation of the resolving power of the method and a comparison with a standard data co-addition method
using both simulated and real data.

\section{Continuous direct model}\label{sec:modele-direct-cont}

\begin{figure}[htbp]
  \centering
\includegraphics[width=0.49\textwidth]{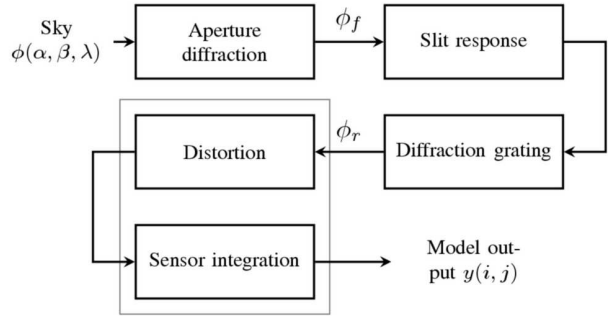}
   \caption{Block diagram of the direct model for one acquisition:
 from a continuously defined sky $\phi$ to a discrete output $\yb$ describing the data.
The flux $\phi_f$ is a convolution of the flux $\phi$ and the PSF of the primary miror.
$\phi_f$ is truncated by a rectangular slit and is dispersed by the gratting.
Finally, the sensor provide a discrete output $\yb$.
\label{fig:schem-bloc}}
  
\end{figure}

The aim of the instrument model is to reproduce the data, $\yb$, acquired by
the spectral imager from a flux $\phi(\alpha,\beta,\lambda)$ of incoherent 
light. 
 Fig.~\ref{fig:schem-bloc} illustrates the instrument model for 
one acquisition (the telescope remains stationary). To simplify, we present 
the scanning procedure in section \ref{sec:scann-proc-sky}.
First, we have the response of the primary mirror (aperture), 
which corresponds to a convolution. 
Second, there is a truncation due to a rectugular slit.
Third, a grating disperses the light.
Finnally,  the sensor integration provides the discrete data $\yb$. Distortion
of the luminous flux is modelled in the sensor integration. 

\subsection{Aperture diffraction }\label{sec:diffr-de-louv}
\begin{figure}[htbp]
  \centering
\includegraphics[width=0.35\textwidth]{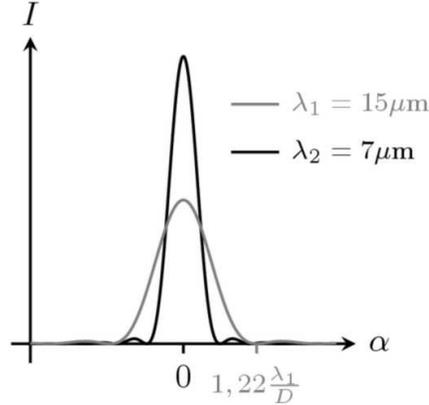}
   \caption{ 
   Profile of an \textit{Airy} disk (PSF for a circular
   aperture) for two wavelengths. 
 \label{fig:airy1D}}
\end{figure}

Under 
some hypotheses, the propagation of a light wave 
which passes through an aperture 
is determined by
\textsc{Fresnel} diffraction \cite{Goodman72} and the result in the focal
plane is a convolution of the 
input flux $\phi$ with 
the Point Spread Function (PSF) $h_a$
illustrated in Fig. \ref{fig:airy1D} 
for a circular aperture. 
This PSF, which is 
a low pass filter, has a width proportional to the wavelength of the incident flux. 
For a circular aperture, it can be written: 
\begin{equation}\label{eq:2}
  h_a(\alpha,\beta,\lambda) = 
  A \left[2 \frac{J_1(\pi D \sqrt{\alpha^2 + \beta^2} / \lambda )}{ \pi D
      \sqrt{\alpha^2 + \beta^2} /\lambda } \right]^2 
\end{equation}
where $J_{1}$ is 
the first order Bessel function of the first kind, $A$ is
an amplitude factor and $D$ is the diameter of the mirror.

The flux in the focal plane, $\phi_f$, is written in integral form:
\begin{equation}\label{eq:3}
 \phi_f(\alpha',\beta',\lambda) = \iint_{\alpha,\beta}
 \phi(\alpha,\beta,\lambda)\,  h_a(\alpha - \alpha', \beta -
 \beta',\lambda)\, \dD\alpha\, \dD\beta
\end{equation}

\begin{figure*}[!t]
  \centering
  \includegraphics[width=0.8\textwidth]{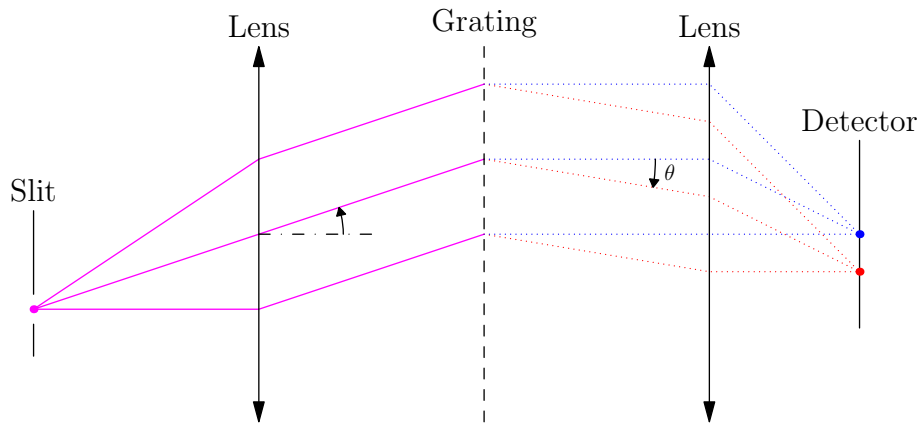} 
  \caption{Optical scheme of IRS instrument: the slit is on the focal plan of the telescope \textsc{Spitzer}.
 The gratting disperses the light and the detector collectes the dispersed flux }
  \label{fig:LightPath}
\end{figure*}

\subsection{Slit and diffraction grating}\label{sec:le-rseau-de}

\paragraph{Slit}\label{sec:la-fente}
Ideally, the slit and grating enable the dispersion of the wavelengths in
spatial dimension $\beta$ previously ``suppressed'' by the slit (see Fig. \ref{fig:LightPath}).
In practice, the slit cannot be infinitely 
narrow because the flux would be zero. 
The slit thus has a width $\gamma$ of
about two pixels.

\begin{figure}[htbp]
  \centering
\includegraphics[width=0.35\textwidth]{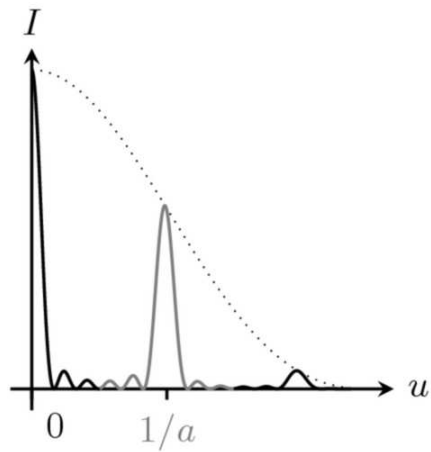}
  \caption{Diffraction grating response. The grey curve corresponds to the
  first mode. In reality, the response width is smaller.
 \label{fig:psfReseau}}
\end{figure}

\paragraph{Diffraction grating}\label{sec:le-reseau-de}
Ideally, the grating gives a diffracted wave 
with an output angle $\theta$
linearly dependent on the wavelength $\lambda$
(see Fig. \ref{fig:LightPath}). 
In a more accurate model,
the dependencies become more complex. Let us introduce a variable $u$ in
order to define an invariant response $ h_{r}$ of the system \cite{perez04}.
\begin{equation}
\label{eq:10}
  u= \frac{\sin\theta - \sin\beta'}{\lambda}\approx 
\frac{\sin\theta - \beta'}{\lambda}
\end{equation}
where $\beta'$ is the angle of incidence of the wave on the grating
, and $|\beta'| \leq \gamma/2$ where $\gamma$ is the angular slit width (5.6 arcseconds).
 The response of the grating centred on mode $m$ ($m=0,1,\dots$)
 can, with some approximations, be written as the square of a cardinal sine
 centred on $m/a$ \cite{perez04}.
\begin{equation}\label{eq:4}
   h_{r}(\theta, \beta', \lambda) = B\,\sinc^{2} \left( \pi L (u -m/a)\right)
 \end{equation}
 where $L$ is the width of the grating and $a$ the grid step (distance
between two grooves). This response centred on the first mode ($m=1$) is
plotted in Fig. \ref{fig:psfReseau}.

As the flux is an incoherent light source, the expression for the signal
at the output of the grating is written in the form of an integral over
$\beta'$ and $\lambda$: 

\begin{equation}\label{eq:5}
  \phi_r(\alpha',\theta) = \int_{\lambda} \int_{|\beta'| \leq \gamma/2}
  \phi_{f}(\alpha',\beta',\lambda) h_{r}(\beta',\lambda,\theta)
  \dD \beta' \dD \lambda
\end{equation}
where $l$ is the 
slit width. 

\subsection{Sensor integration}\label{sec:lintgration-capteur}

Once the flux has passed through the grating and the wavelengths have been
dispersed according to $\theta$, the light flux 
is focused on 
the sensor composed of square detectors. 
The sensor is simply modelled by integrating the flux $\phi_r$ on square areas of side $d$.
The flux is integrated along the direction $\alpha$, which is not modified
by the diffraction grating, and the dimension $\theta$, a combination of $\beta$
and $\lambda$, to obtain the discrete values
\begin{equation}\label{eq:7}
  y(i,j) = \int_{id}^{(i+1)d}\int_{jd + e_{ij}^{1}}^{(j+1)d + e_{ij}^{2}}
  \phi_{r}(\alpha',\theta) \dD \alpha' \dD \theta.
\end{equation}

\begin{figure}[htbp]
  \centering
\includegraphics[width=0.49\textwidth]{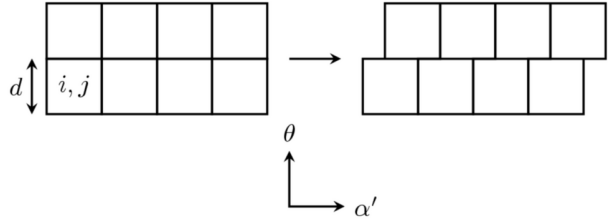}
  \caption{Modelling the distortion: sensor integration limits are shifted
  according to the dimensions $\alpha'$.\label{fig:distor}}
  
\end{figure}

The integration limits are modified by the terms $e_{ij}^{n}$ 
in order to take into account the data distortion as illustrated in Fig. \ref{fig:distor}.

\subsection{Scanning procedure of the sky}
\label{sec:scann-proc-sky}

In a direction parallel to the slit width a scanning procedure 
(illustrated Fig.~\ref{fig:balai}) is applied. 
This scanning procedure is composed of $Q$ acquisitions.
Between the first and the $q^{th}$ acquisitions, the instrument
is moved by $\Delta_\alpha (q)$ (resp. $\Delta_\beta (q)$) in the direction 
$\alpha$ (resp. $\beta$).
To taking into account the motion of the instrument, we substitue 
$\phi(\alpha,\beta,\lambda)$ 
for $\phi(\alpha-\Delta_\alpha (q),\beta-\Delta_\beta (q),\lambda)$ in the 
previous equations.
In practice, we fix the $\alpha$ axis in
the direction of the slit and the $\beta$ axis
perpendicular to the slit (see Fig.~\ref{fig:balai}). 
In consequence, $\Delta_\alpha (q)$ is equal to zero.


\subsection{Complete model}\label{sec:modle-complet}
%
By combining expression (\ref{eq:2}), (\ref{eq:3}), (\ref{eq:5}),(\ref{eq:4})
and (\ref{eq:7}), we obtain
a continuous direct model in the form
%

\begin{multline}\label{eq:8}
  y(i,j,q) = \mathcal{A} \int_{id}^{(i+1)d}\int_{jd +
    e_{ij}^{1}}^{(j+1)d +e_{ij}^{2}} \int_{\lambda} \int_{|\beta'| \leq
    \gamma/2}\\
  \int_{\alpha}\int_{\beta} \phi(\alpha-\Delta_\alpha (q),\beta-\Delta_\beta (q),\lambda)
  h_{a}(\alpha - \alpha',\beta - \beta',\lambda) \dD \alpha \dD \beta\\
  h_{r}(\theta,\beta',\lambda) \dD \beta' \dD\lambda \dD \alpha' \dD
  \theta
\end{multline}
%
where $\mathcal{A}$ is a scale factor.

The equation (\ref{eq:8}) can rewritten: 
\begin{equation}
  \label{eq:17}
   y(i,j,q) = \int_{\alpha}\int_{\beta} \int_{\lambda} 
\phi(\alpha,\beta,\lambda)h_{tot}^{i,j,q}(\alpha,\beta,\lambda)
\dD \alpha \dD \beta \dD\lambda 
\end{equation}
with 
\begin{multline}
  \label{eq:18}
  h_{tot}^{i,j,q}(\alpha,\beta,\lambda)= \mathcal{A} \int_{id}^{(i+1)d}\int_{jd +
    e_{ij}^{1}}^{(j+1)d +e_{ij}^{2}} \int_{-\gamma/2}^{\gamma/2}\\ 
  h_{a}(\alpha - \alpha'-\Delta_\alpha (q),\beta - \beta'-\Delta_\beta (q),\lambda)
  h_{r}(\theta,\beta',\lambda)
 \dD \alpha' \dD \theta\dD \beta'
\end{multline}
We have been developed a model relying the continuous 
sky $\phi(\alpha,\beta,\lambda)$ and discrete data $\yb$. 
Our model is linear not-shift-invariant, because the 
aperture response and the grating response depend on the 
wavelength. 

\section{Decomposition over a family and Gaussian approximation}\label{sec:decomp-sur-une}

In the previous part, we 
have seen that obtaining the output from the model requires 
the six integrals of equation~(\ref{eq:8}) to be calculated. 
The estimation of $\hat\phi$ in $L^2(\bR)$ by 
inversion of this model is quite tricky, so we prefer 
to decompose the object over a family of functions. 
As we can see in the introduction, a lot of
such decomposition functions 
can be used. The
most traditional are Fourier bases, wavelets, cardinal sines, splines and
pixel indicators. 
The choice does not have any great influence on the final result if the continuous
object is decomposed over a sufficiently large number of functions. We therefore
chose our decomposition functions in such a way as to reduce the computing
time for the instrument model. 
First, we chose the  $\alpha$ axis in
the direction of the slit and the $\beta$ axis
perpendicular to the slit (see Fig.~\ref{fig:balai}).
Second, we have two spatial variables $(\alpha,\beta)$
and one spectral variable $\lambda$, so to simplify the calculus, 
we chose decomposition functions
that are separable into $(\alpha,\beta)$ and $\lambda$.
Thrid,  the object is convolved by the response of the optics, which has
circular symmetry. So we choose functions 
possessing the same circular symmetry in order to make this calculation 
explicit.
Finally, 
the slit and the grating have an impact in the $\beta$ direction only (\ref{eq:5}),
which motivates us to choose functions that are separable into $\alpha$
and $\beta$.
These considerations led us to choose Gaussian functions along the spatial
directions. Finally the complexity of the $\lambda$ dependence encouraged
us to choose Dirac impulses for the spectral direction.

\subsection{Decomposition over a family of Gaussian functions}\label{sec:decomp-sur-une-1}

The flux $\phi$ is a continous function decomposed over a family of
separable functions:
%
\begin{eqnarray}\label{eq:1}
          \phi(\alpha, \beta,\lambda) &=& \sum_k \sum_l \sum_p x(k,l,p)
 \nonumber\\
  &&\Pi(\alpha - kT_\alpha)\,\Phi(\beta - lT_\beta)\,\Gamma(\lambda -
  pT_\lambda)\nonumber\\
&=&\sum_k \sum_l \sum_p x(k,l,p)\Psi_{k,l,p}(\alpha,\beta,\lambda)
\end{eqnarray}
where $x(k,l,p)$ are the decomposition coefficients, $T_\alpha$,
$T_\beta$ and $T_\lambda$ are the sampling steps, and with:

\begin{eqnarray}\label{eq:9}
 \Pi(\alpha)\Phi(\beta) & = &\frac 1{2\pi\sigma^2}\exp
  \left(-\frac{1}{2}\frac{\alpha^2+\beta^2}{\sigma^2}\right)\\ 
\Gamma(\lambda)  & = & \delta(\lambda)
\end{eqnarray}

With such decomposition, the inverse problem becomes one of estimating a finite
number of coefficients $x(k,l,p)$ from discrete data 
$y(i,j,q)$.
By combining equations (\ref{eq:17}) and (\ref{eq:1}), we obtain: 
\begin{multline}
\label{eq:20}
   y(i,j,q) =\sum_k \sum_l \sum_p x(k,l,p)
 \int_{\alpha}\int_{\beta} \int_{\lambda}\Psi_{k,l,p}(\alpha,\beta,\lambda)\\
  h_{tot}^{i,j,q}(\alpha,\beta,\lambda) \dD \alpha \dD \beta \dD\lambda 
\end{multline}
If the $y(i,j,q)$ and $x(k,l,p)$ are gathered in vectors $\yb$ and $\xb$\footnote{In this paper, we use 
the following convention: bold, lower-case variables represent vectors and bold, 
upper-case variables represent matrices.} 
respectively, the equation (\ref{eq:20}) can be formalized as a vector matrix product. 
%
\begin{equation}\label{eq:DirectModel}
  \yb=\Hb\xb
\end{equation}
%
with each component of the matrix $\Hb$ is calculated using the integral part of the equation~(\ref{eq:20}). 
The $n(k,l,p)$
-th column of matrix $\Hb$ constitutes the output when the model
is used with the $n$-th decomposition function ($\Psi_{k,l,p}$).
The model output for $\Psi_{k,l,p}$ 
 is
calculated in the next two sections.

\subsection{Impulse responses approximated by Gaussian functions}\label{sec:repons-impuls-}

\subsubsection{Approximation of the PSF 
 }\label{sec:approximation-de-la-1}
Equation (\ref{eq:3}) comes down to convolutions of a 
squared
Bessel function
and Gaussians. This integral is not explicit and, in order to carry out
the calculations, the PSF is approximated by a Gaussian
$$
\tilde{h}_a (\alpha, \beta,\lambda) = \frac{1}{2\pi \sigma_{\lambda}^2}
\exp\left(-\frac{1}{2}\frac{\alpha^2 + \beta^2}{\sigma_{\lambda}^2}\right)
$$
with a standard deviation $\sigma_{\lambda}$ 
depending on the wavelength. 
 Indeed, the Bessel functions cross zero at the first time 
in $1.22 \froc{\lambda}{D}$.  $\sigma_{\lambda}$ is determined numerically
 by minimizing the quadratic error 
between the Gaussian kernel and the squared Bessel function, which gives 
for our instrument $\sigma_{\lambda}\approx \lambda/2$.
The relative quadratic error $err_{L2}=\froc{\|\text{Bessel}-\text{Gaussian}\|^2_2\;}{\;\|\text{Bessel}\|^2_2}$ is equal to $0.15 \%$ for our instrument.
If we caculate the relative absolute error $err_{L1}=\froc{\|\text{Bessel}-\text{Gaussian}\|_1\;}{\;\|\text{Bessel}\|_1}$, we obtain 5 \%. 
We can conclude that most of the energy of the squared bessel function is 
localized in the primary lobe. 
 Another advantage of using
the Gaussian approximation is that the convolution kernel is separable
into $\alpha$ and $\beta$.
Finally, the result of the convolution of two Gaussian functions is a standard one and is also a Gaussian:

\begin{equation}\label{eq:11}
\tilde{h}_a (\alpha, \beta,\lambda)\star \Pi(\alpha)\Phi(\beta) 
=\frac{1}{2\pi(\sigma_{\lambda}^2+\sigma^2)}
\exp \left(- \frac{\alpha'^2+\beta'^2}{2(\sigma_{\lambda}^2+\sigma^2)}\right)
\end{equation}

\subsubsection{Approximation of the grating response}\label{sec:approximation-de-la}

The presence of the slit means that integral (\ref{eq:5}) is bounded 
over $\beta'$ and is not easily calculable. Since the preceding expressions
use Gaussian functions, we approximate the squared cardinal sine by
a Gaussian to make the calculations easier:
\begin{multline}
\text{sinc} ^2 \left(\pi L \left(\frac{\sin \theta -
      \beta'}{\lambda} - \frac{m}{a} \right) \right) \approx \\
\frac{1}{\sqrt{2\pi}\lambda\sigma_{s}}\exp \left( -\frac{1}{2} \frac{(\frac{\sin\theta-\beta'}{\lambda}-\frac{ m}{a})^2}{\sigma_{s}^2} \right)  
\end{multline}
$\sigma_{s}$ is determined numerically by minimizing the quadratic error 
between the Gaussian kernel and the squared cardinal sine, which gives 
for our instrument $\sigma_{s} \approx 25.5\;\mathrm{m}^{-1}$.
The relative errors made are larger than the bessel case ($err_{L2}=0.43 \%, err_{L1}=10.7 \%$), but this
Gaussian approximation of the grating response allows the flux $\phi_r$ coming out of the grating to be known explicitly.

The error introduced here is larger than for the Gaussian approximation of the PSF described in the previous section.
However,
our goal is to have a good model 
of
 the spatial
dimension of the array. Furthermore, with respect to the current method,
the fact of taking the response of the grating into consideration, even as an approximation, is already a strong improvement.

\begin{eqnarray}
  \label{eq:6}
  \phi_r(\alpha',\theta) &=& \int_{\lambda} \int_{-\gamma/2}^{\gamma/2}
  \int_{\alpha}\int_{\beta}\Pi(\alpha-\alpha_k)\Phi(\beta-\beta_l-\Delta_\beta(q)) \nonumber\\
&&\Gamma(\lambda-\lambda_p)
 \tilde{h}_{a}(\alpha - \alpha',\beta - \beta',\lambda) 
 \tilde{h}_{r}(\theta,\beta',\lambda)\nonumber\\
&& \dD\alpha \dD\beta \dD\beta'\dD\lambda \nonumber\\
&&\nonumber\\
&=&\mathcal{A}
\exp \left(- \frac{(\alpha'-\alpha_k)^2}{2(\sigma_{\lambda}^2+\sigma^2)}\right)
\exp \left(- \frac{(\sin \theta-\nu)^2}{2\Sigma^2}\right)\nonumber\\
& & \times\left[\erf\left(\frac{\gamma/2-\mu}{\Sigma'\sqrt 2}\right)- 
\erf\left(\frac{-\gamma/2-\mu}{\Sigma'\sqrt 2}\right)\right]
\end{eqnarray}
with
\begin{equation*}
  \left\{
    \begin{array}{lcc}
\mathcal{A}&=&  \displaystyle\frac{1}{4\pi}\sqrt{\frac{1}{(\sigma_{\lambda}^2+\sigma^2)(\sigma_{\lambda}^2+\sigma^2
+\lambda_p^2\sigma_s^2)}}\\
      \Sigma^2&=& (\sigma_{\lambda}^2+\sigma^2)+\lambda_p^2\sigma_s^2\\
           \nu&=& \displaystyle\frac{m\lambda_p}{a}+\beta_l+\Delta_\beta(q)\\
       \Sigma'&=& \displaystyle\frac{\sqrt{ \sigma_{\lambda}^2+\sigma^2}\lambda_p\sigma_s}
       {\sqrt{ \sigma_{\lambda}^2+\sigma^2+\lambda_p^2\sigma_s^2}}\\
          \mu&=& \displaystyle\frac{
\left(\sin\theta-\frac{m\lambda_p}{a}\right)( \sigma_{\lambda}^2+\sigma^2)}
{ (\sigma_{\lambda}^2+\sigma^2)+\lambda_p^2\sigma_s^2} \\
\erf(a) &=&\displaystyle\frac{2}{\sqrt{\pi}} \int_0^a e^{-t^2} \dD t
    \end{array}
\right.
\end{equation*}
In equation (\ref{eq:6}), it can be seen that $\phi_r$ is separable into $\alpha'$ and $\theta$. 
Let us introduce the functions $f$ and $g$ such that:
\begin{equation}
  \label{eq:13}
  \phi_r(\alpha',\theta)=\mathcal{A}f(\alpha')g(\theta)
\end{equation}

\subsection{Sensor integration}\label{sec:modele-discretise}

First, we calculate the sensor integration in the $\alpha'$ direction.
\begin{multline}
  \label{eq:12}
  \int_{jd + e_{ij}^{1}}^{(j+1)d + e_{ij}^{2}}f(\alpha')\dD \alpha'=\\ 
\mathcal{K}\left[\erf\left(\frac{(j+1)d+e^2_{ij}-\alpha_k}{\sqrt {2(\sigma_{\lambda}^2+\sigma^2)}}\right)- 
\erf\left(\frac{jd+e^1_{ij}-\alpha_k}{\sqrt {2(\sigma_{\lambda}^2+\sigma^2)}}\right)\right]
\end{multline}
with $\mathcal{K}=\sqrt{\pi (\sigma_{\lambda}^2+\sigma^2) /2}$

The integral of $g$ is calculated numerically as the presence of $\erf$ functions
in equation (\ref{eq:6}) does not allow  analytical calculations.

We obtain the expression for the $n$-th column of matrix $\Hb$, which now
contains only a single integral: 
\begin{multline}
  \label{eq:14}
y(i,j,q)=\mathcal{AK}\int_{id}^{(i+1)d}g(\theta)\dD \theta\\
\times\left[\erf\left(\frac{(j+1)d+e^2_{ij}-\alpha_k}{\sqrt {2(\sigma_{\lambda}^2+\sigma^2)}}\right)- 
\erf\left(\frac{jd+e^1_{ij}-\alpha_k}{\sqrt {2(\sigma_{\lambda}^2+\sigma^2)}}\right)\right]
\end{multline}

Using expression (\ref{eq:14}), the elements of matrix $\Hb$ are  pre-computed
relatively rapidly.
Thanks to the sparsity of the matrix $\Hb$ to calculate  the
model output of Eq.~(\ref{eq:DirectModel}).


\section{Inversion}\label{sec:inversion}

The previous sections build the relationship~(\ref{eq:DirectModel}) between the object 
coefficients and the data: it describes a complex instrumental model but remains linear.
 The problem of input (sky) reconstruction is a typical inverse problem and the 
literarure on the suject is abundant.

The proposed inversion method resorts to linear processing. It is based on conventional 
approaches described in books such~\cite{Tikhonov77,Andrews77} or more
 recently~\cite{Idier08}. In this framework, the reader may also 
consider~\cite{Bertero85,Bertero88} for inversion based on specific decomposition. These 
methods rely on a quadratic criterion
\begin{equation}
  \label{eq:15}
  J(\xb) = || \yb - \Hb\xb ||^{2} + \mu_{\alpha\beta} || \Db_{\alpha\beta}\xb||^{2} + \mu_{\lambda}||\Db_{\lambda} \xb||^{2} \,.
\end{equation}
It involves a least squares term and two penality terms concerning the differences 
between neighbouring coefficients: one for the two spatial dimensions and one for the 
spectral dimension. They are weighted by $\mu_{\alpha\beta}$ and $\mu_{\lambda}$, 
respectively. 
%
The estimate $\hat \xb$ is chosen as the minimizer of this criterion. It is thus explicit
 and linear with respect to the data:
\begin{equation}
  \label{eq:16}
  \hat \xb = \left(\Hb\T\Hb +\mu_{\alpha \beta}\Db_{\alpha\beta}\T\Db_{\alpha\beta} +\mu_{\lambda}\Db_{\lambda}\T\Db_{\lambda}\right)^{-1}\Hb\T\yb
\end{equation}
and depends on the two regularization parameters $\mu_{\alpha \beta}$ and $\mu_\lambda$.

\begin{remark}~---~
This estimator can be interpreted in a Bayesian framework \cite{Demoment89} based on 
Gaussian models for the errors and the object. As far as the errors are concerned, the 
model is a white noise. As far as the object is concerned, the model is correlated and 
the inverse of the correlation matrix is proportional to 
$\mu_{\alpha \beta}\Db_{\alpha\beta}\T\Db_{\alpha\beta} + \mu_{\lambda}\Db_{\lambda}\T\Db_{\lambda}$, \ie it is a Gauss markov field. In this framework, the estimate maximizes the \apost 
law.
\end{remark}

\begin{remark}~---~
Many works in the field of over-resolved reconstruction concern edge preserving 
priors~\cite{Schultz96,Hardie97,Elad97,Nguyen01,Humblot06,Woods06}. In our application 
here, smooth interstellar dust clouds are under study, so preservation of edges is not 
appropriate. For the sake of simplicity of implementation, we chose a Gaussian object 
prior.
\end{remark}

The minimizer $\hat \xb$ given by relation (\ref{eq:16}) is explicit but, in practice, it 
cannot be calculated on standard computers, because the matrix to be inverted is too 
large. The solution $\hat \xb$ is therefore computed by a numerical optimization
 algorithm. Practically, the optimization relies on a standard gradient descent 
algorithm~\cite{Bertsekas99,Nocedal00}. More precisely, the direction descent is a 
approximate conjugate gradient direction \cite{Polak71} and the optimal step of descent 
is used. Finally, we initialise the method with zero ($\xb=0$).
\section{Results }\label{sec:resultats}

\begin{figure}[htb]
  \centering
\includegraphics[width=0.49\textwidth]{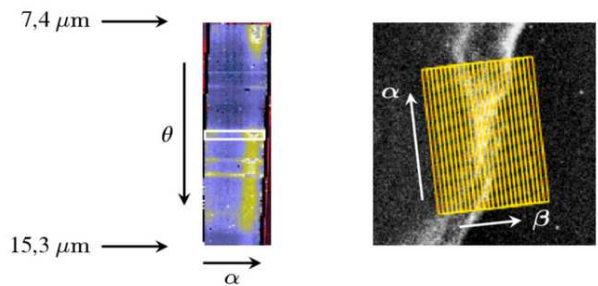}  
  \caption{Acquisition. The left-hand image represents the data acquired
  for one pointing position: the vertical axis shows the spectral dimension  $\theta$ and the horizontal axis is the spatial dimension $\alpha'$. The
 slit is represented schematically by a rectangle in the middle. The right-hand image illustrates the scanning strategy in the $\beta$ direction.}
  \label{fig:balai}
\end{figure}

As we have presented in part \ref{sec:decomp-sur-une} 
the $\alpha$  and $\beta$ axis are fix (see Fig.~\ref{fig:balai}, right).
%
%
The real data is composed of 23 acquisitions having a spatial dimension 
$\alpha'$ and a spectral dimension  $\theta$ of wavelength between 7.4 and 
15.3 $\mu \mathrm{m}$ 
(each acquisition is an image composed of $38\times128$ detector cells, see Fig.~\ref{fig:balai}, left).
 Between two 
acquisitions, the instrument is moved by half a slit width in the $\beta$ 
direction. Fig.~\ref{fig:balai}, right, shows the scanning procedure applied 
to the Horsehead nebula \cite{Compiegne07}.

Our results (Fig. \ref{fig:SimL2S}, solid line on Fig. \ref{fig:SimSpectre} and
Fig. \ref{fig:imL2S})
 can be compared with those obtained with
the conventional processing (Fig. \ref{fig:SimAlain}, dotted line on 
Fig.~\ref{fig:SimSpectre} and Fig. \ref{fig:imAlain}). For the conventional 
processing (described in Compi\`egne et al. 2007 \cite{Compiegne07}) 
an image of the slit is simply extracted for each wavelength from the data 
taken after each acquisition (e.g. left panel of Fig. \ref{fig:balai}) 
and projected and co-added on the output sky image, without any description 
of the instrument properties.


\subsection{Simulated data}
\label{sec:donnees-simulees-1}

\begin{figure}[htbp]
  \centering
  \subfigure[]{\includegraphics[width=0.22\textwidth]{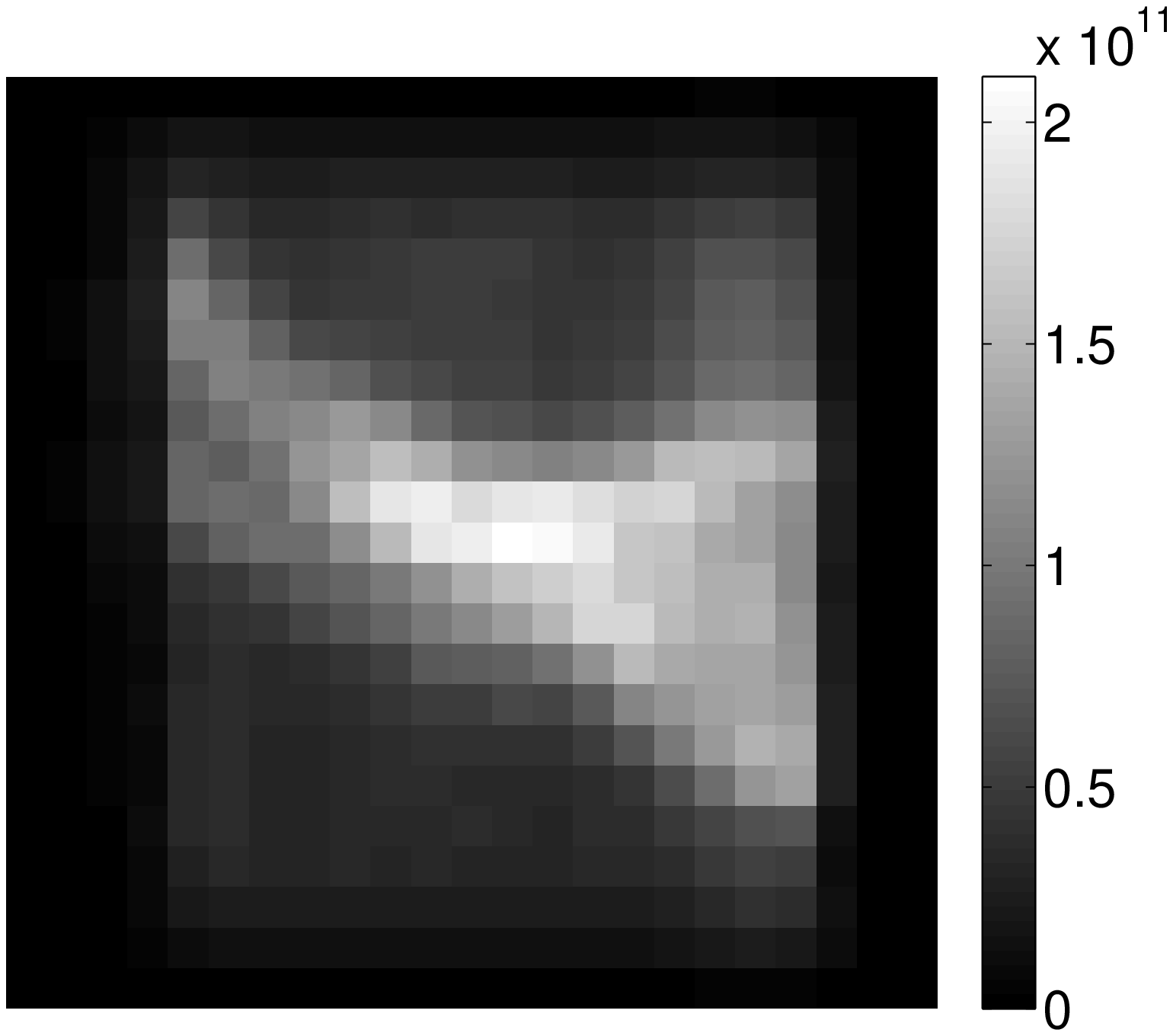} \label{fig:SimVrai}}
 \hfill
  \subfigure[]{\includegraphics[width=0.2\textwidth]{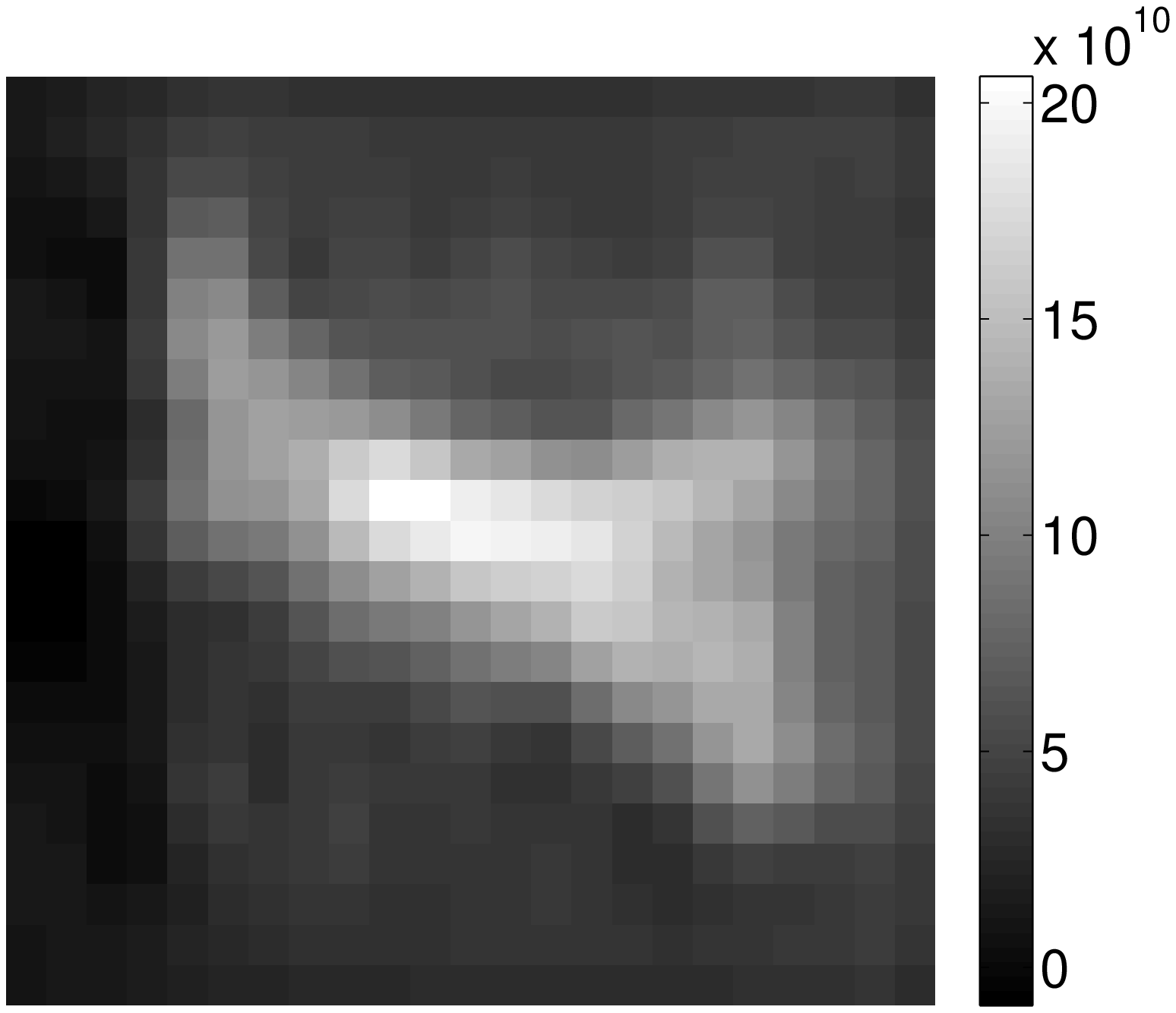} \label{fig:SimL2S}}
  \hfill
  \subfigure[]{\includegraphics[width=0.2\textwidth]{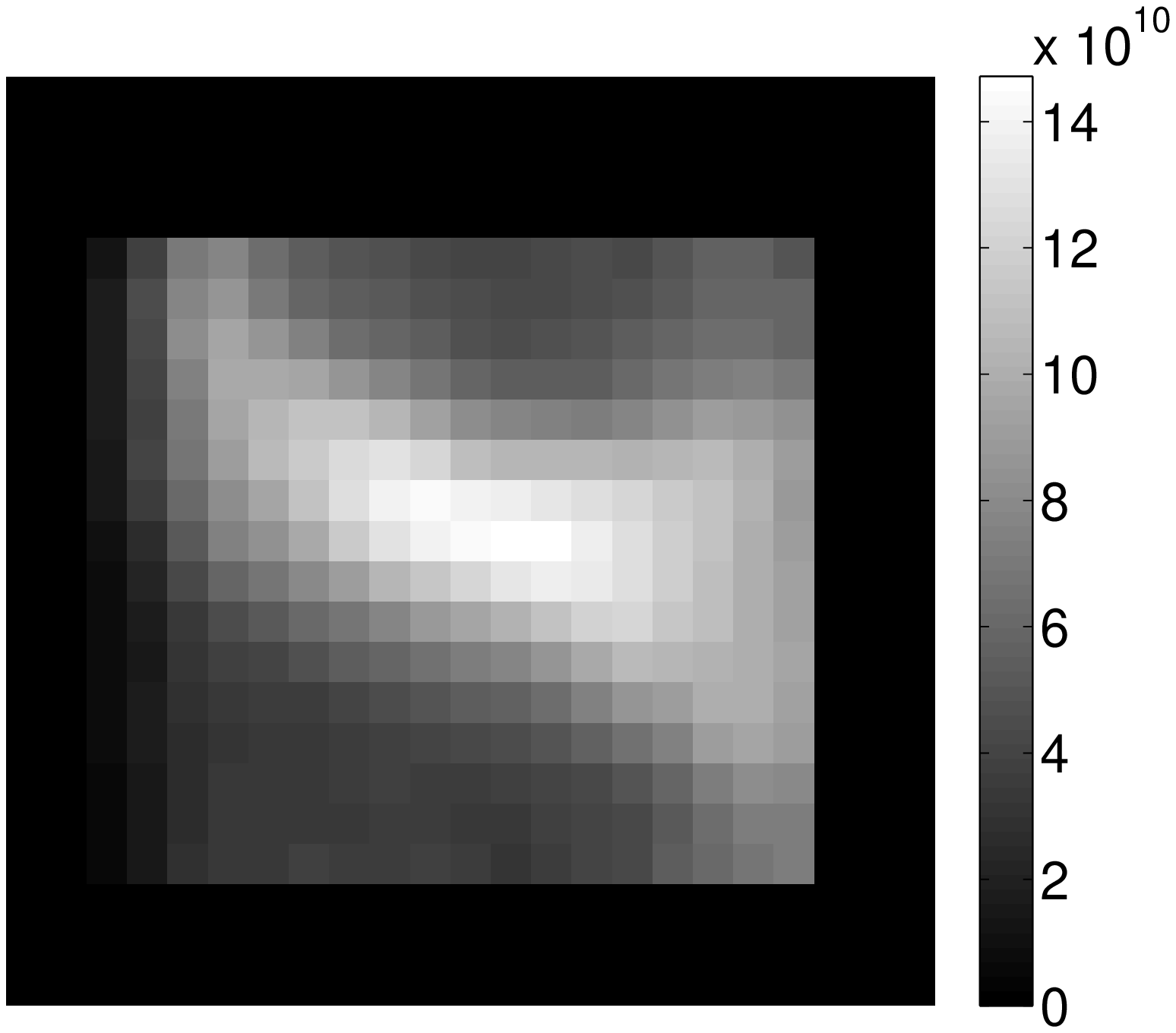} \label{fig:SimAlain}}
  \hfill
\caption{Image at $\lambda = 8.27\, \mu \mathrm{m}$:,   (a) simulated sky, (b) 
image estimated by our method, (c) image estimated by  a conventional 
 method.}
  \label{fig:Sim}
\end{figure}

In our first experiment, we reconstruct data simulated using our direct 
model. We choose an object with the same spatial morphology 
and the same spectral content as the 
 Horsehead nebula (see Fig. \ref{fig:SimVrai}).
However, in order to tune the regularization coefficient, we perform a large
number of reconstructions. Thus, we need to simulate a problem smaller 
than in our real case.
The data are composed of 14 acquisitions, and the virtual detector 
contained $18\times40$ pixels. We choose to reconstruct a 
volume with 15870 gaussians distributed on a cartesian grid 
$23\times23\times30$. 
Finally, we add to the output of the model
 a white Gaussian noise with the 
same variance as the real data.

\begin{figure}
  \centering
  \begin{tabular}{ccccc}
    \includegraphics[width=0.075\textwidth]{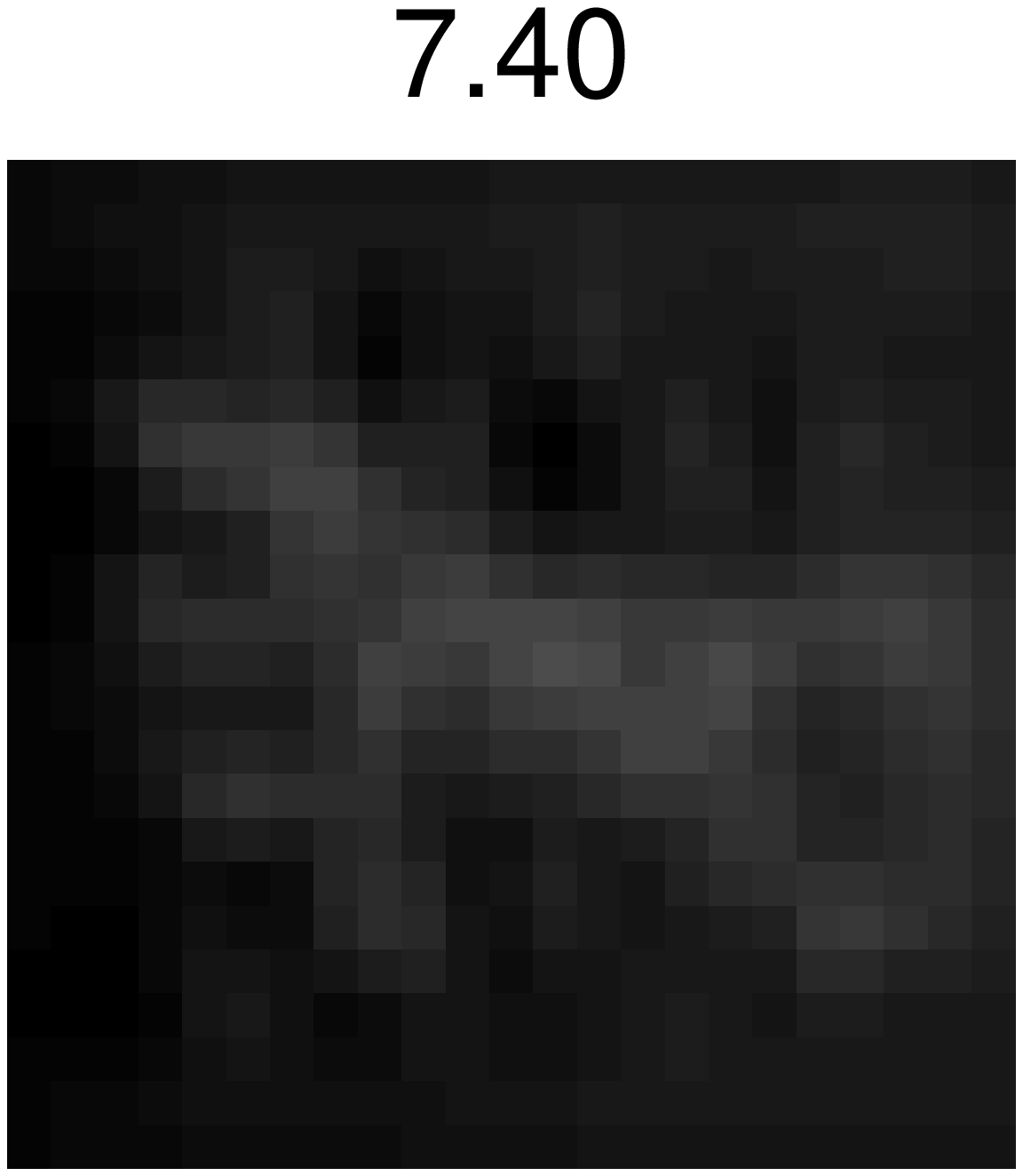}&
    \includegraphics[width=0.075\textwidth]{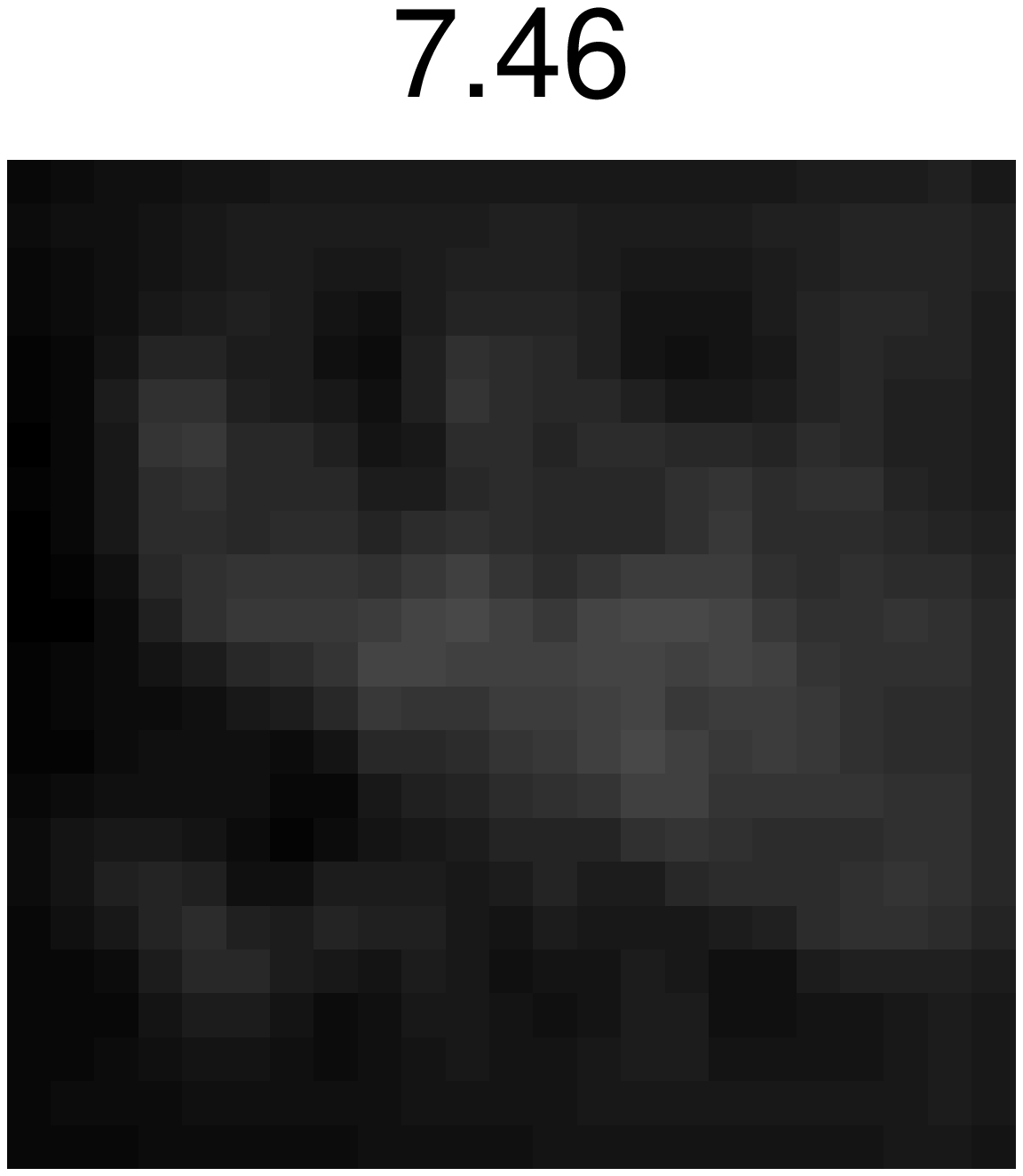}&
    \includegraphics[width=0.075\textwidth]{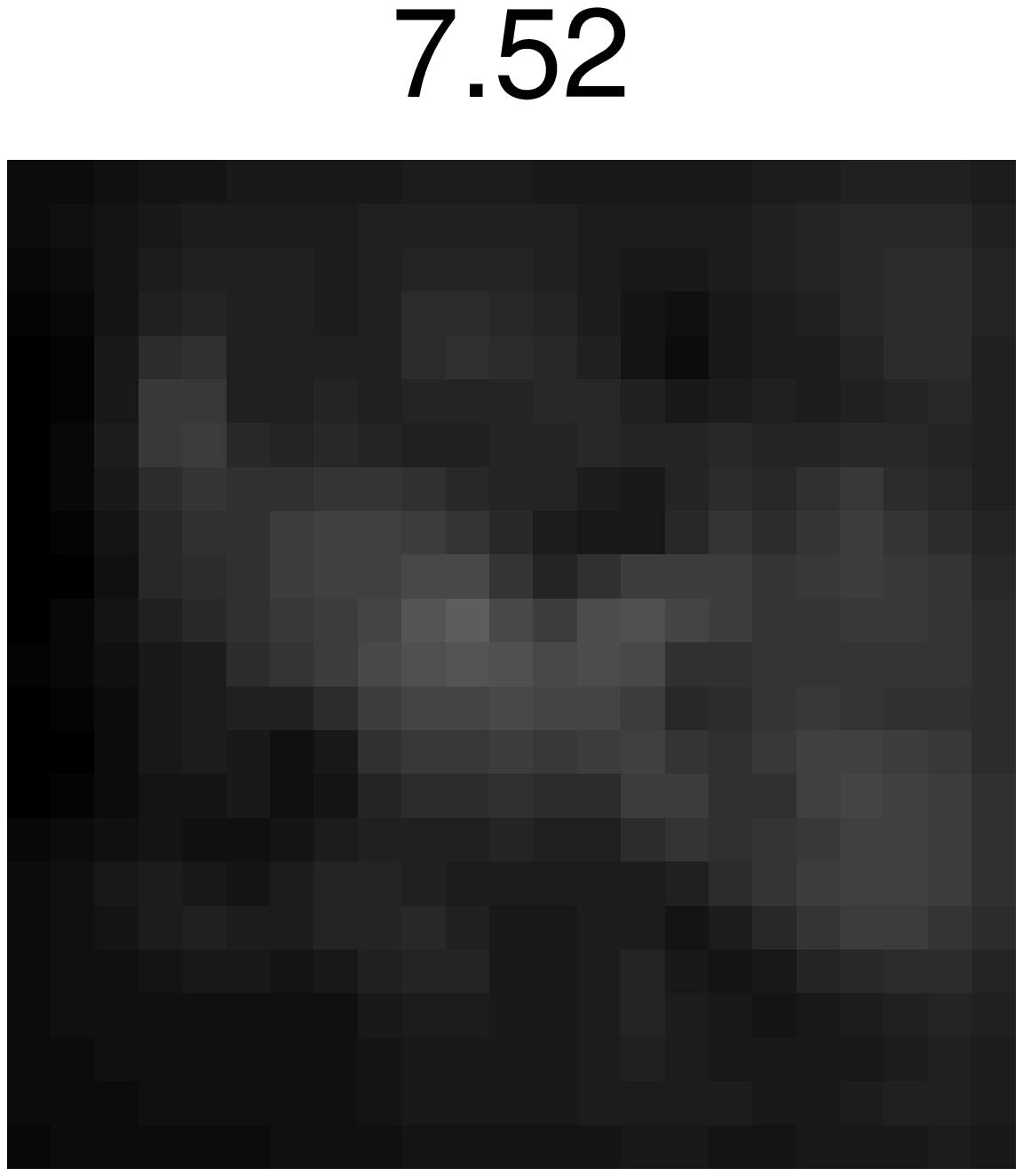}&
    \includegraphics[width=0.075\textwidth]{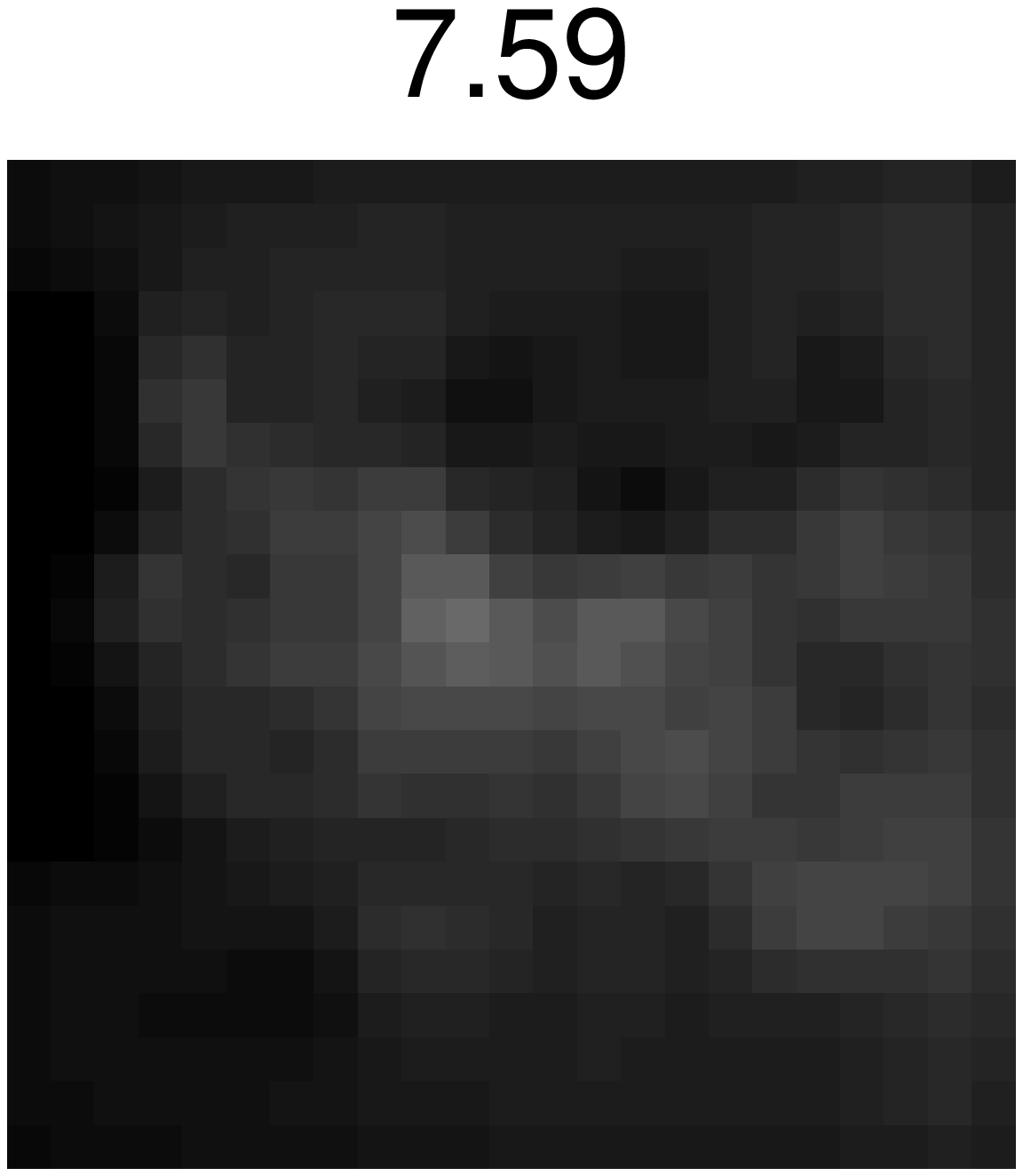}&
    \includegraphics[width=0.075\textwidth]{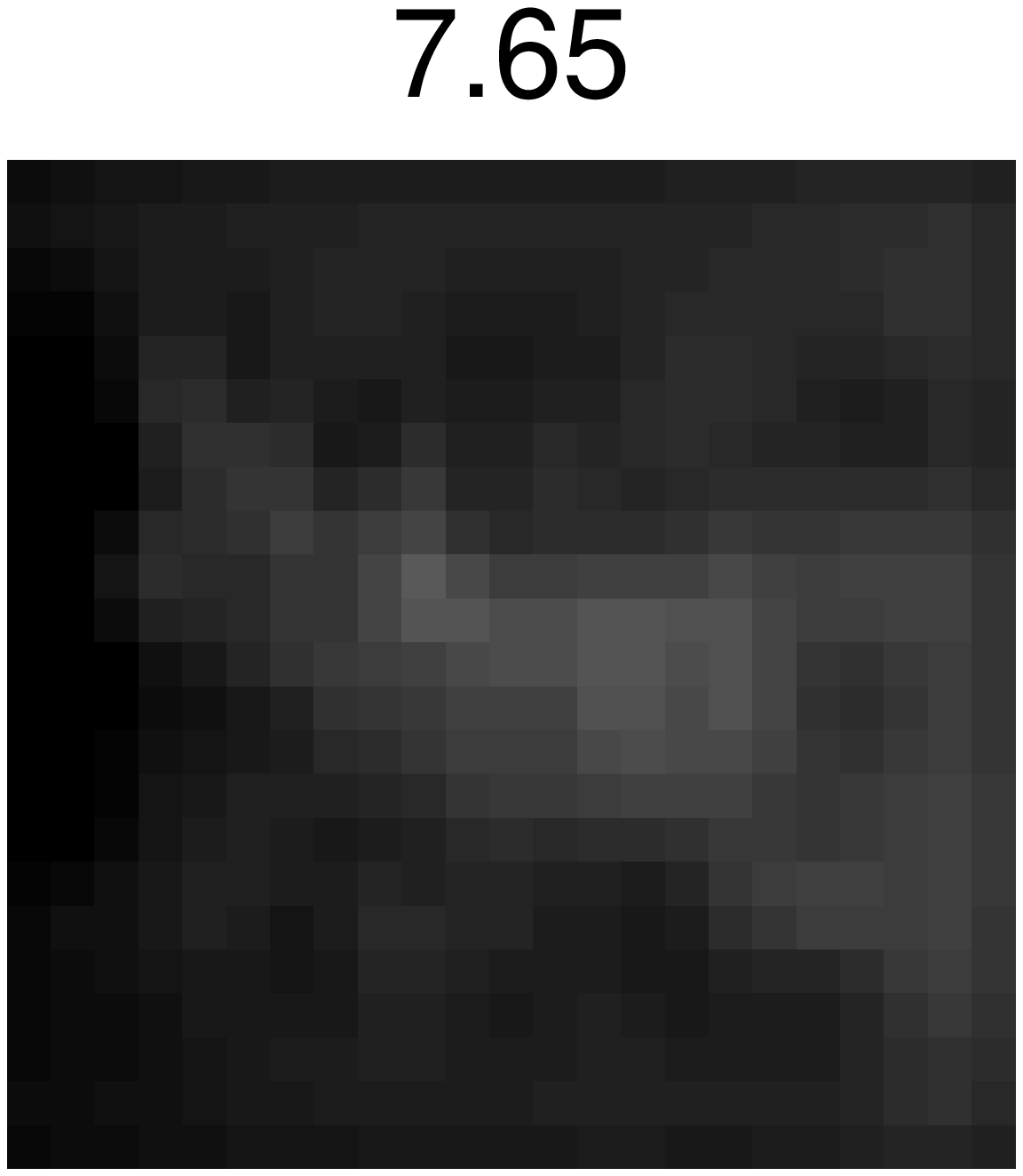}\\
    \includegraphics[width=0.075\textwidth]{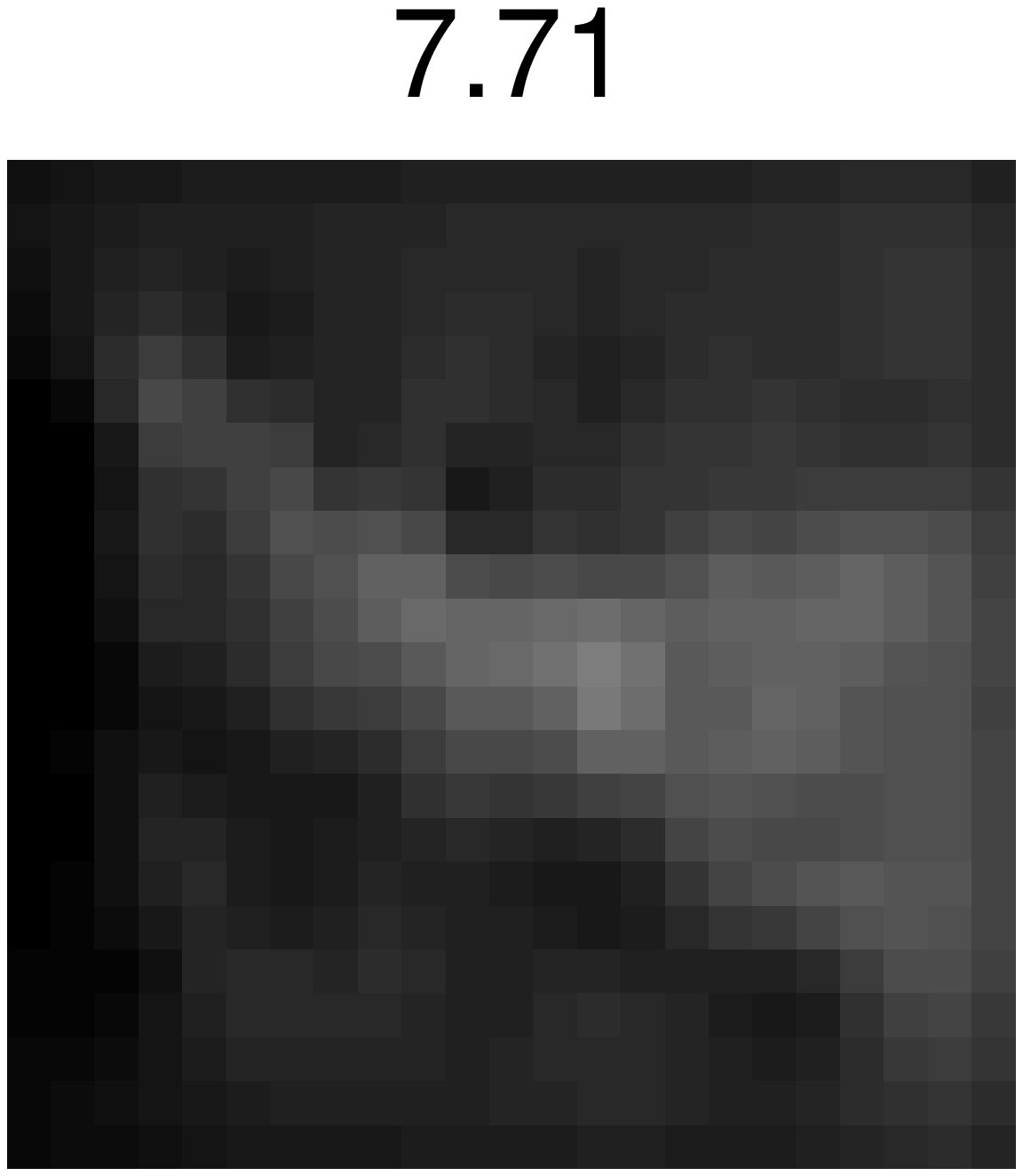}&
    \includegraphics[width=0.075\textwidth]{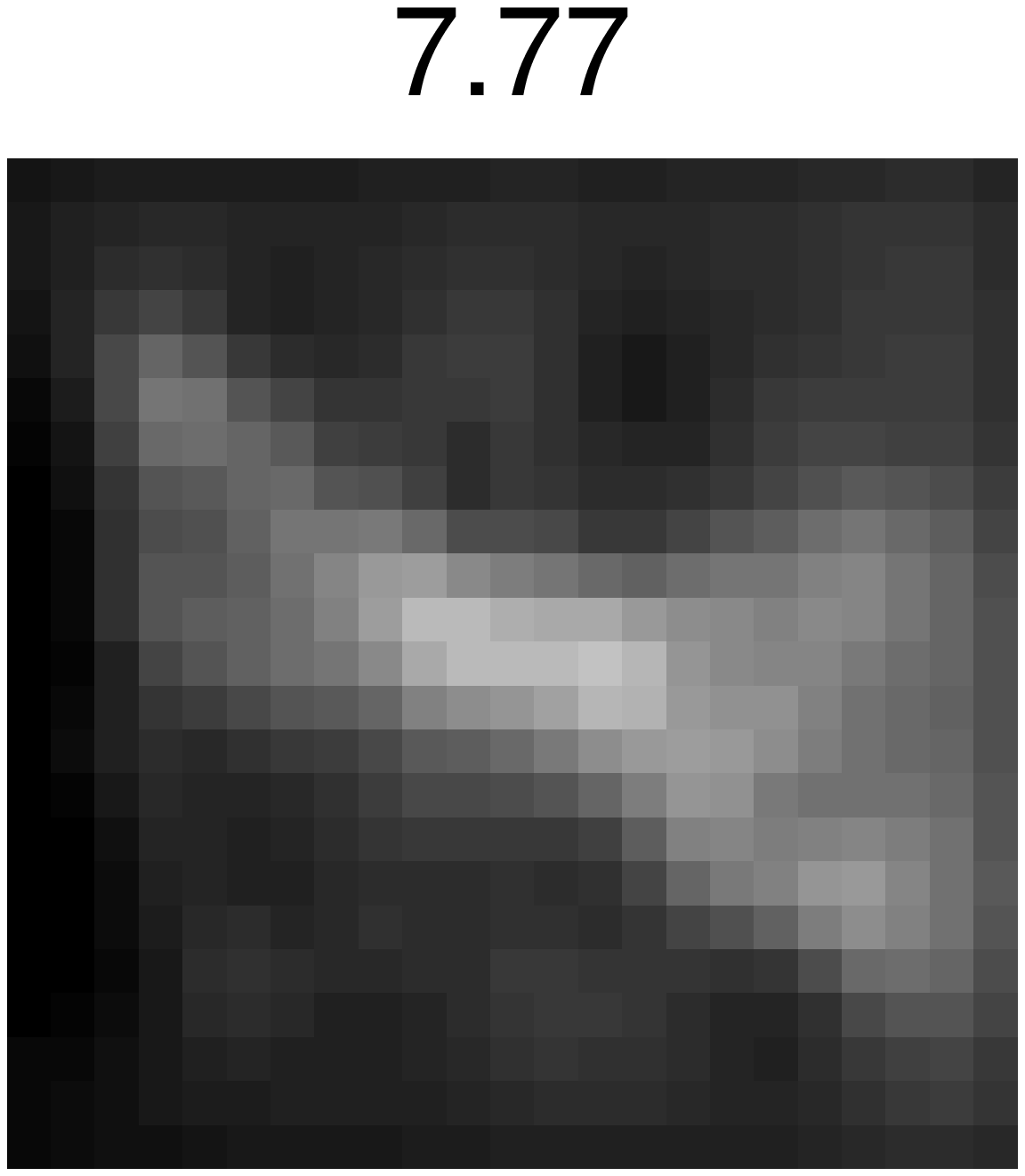}&
    \includegraphics[width=0.075\textwidth]{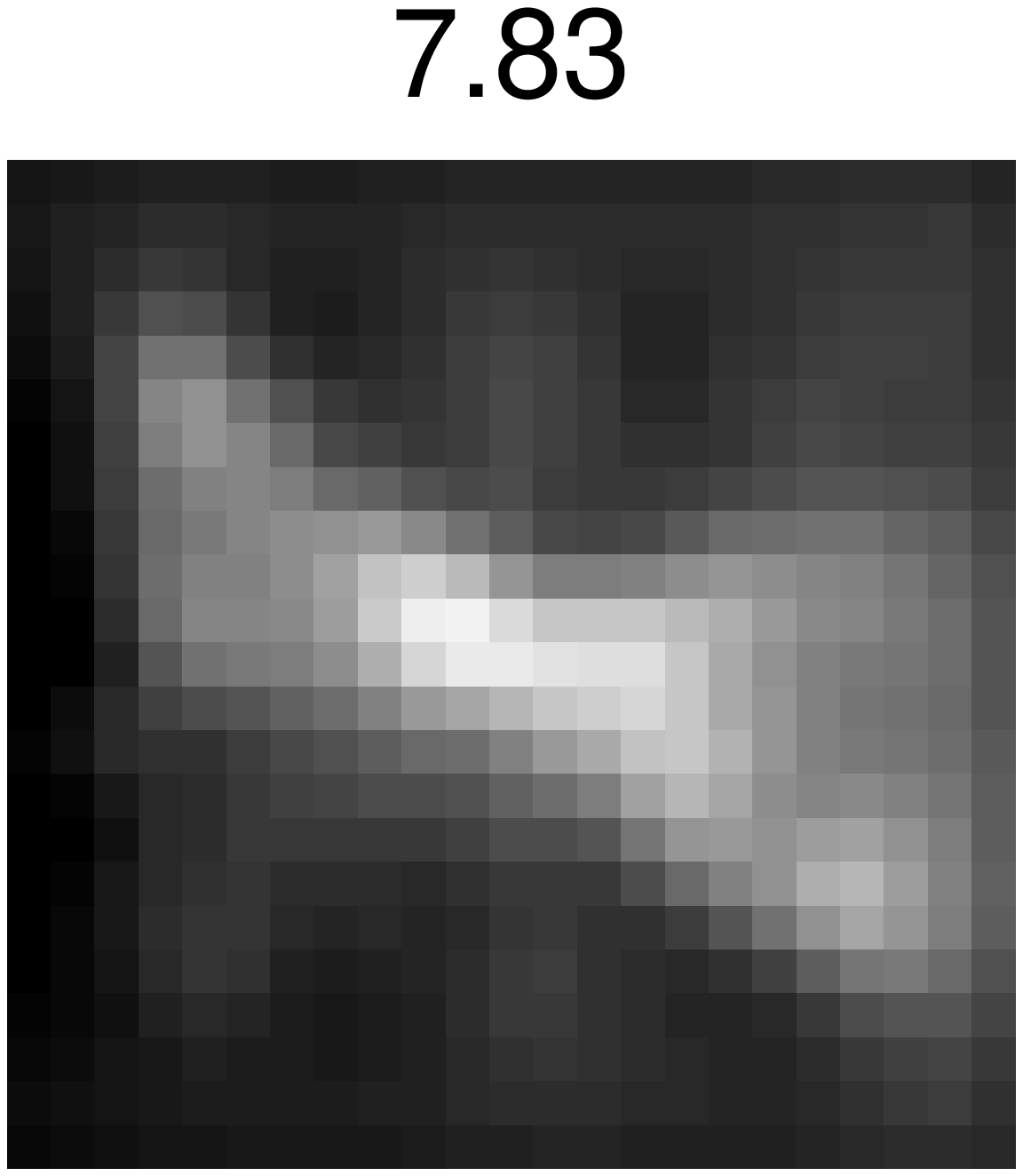}&
    \includegraphics[width=0.075\textwidth]{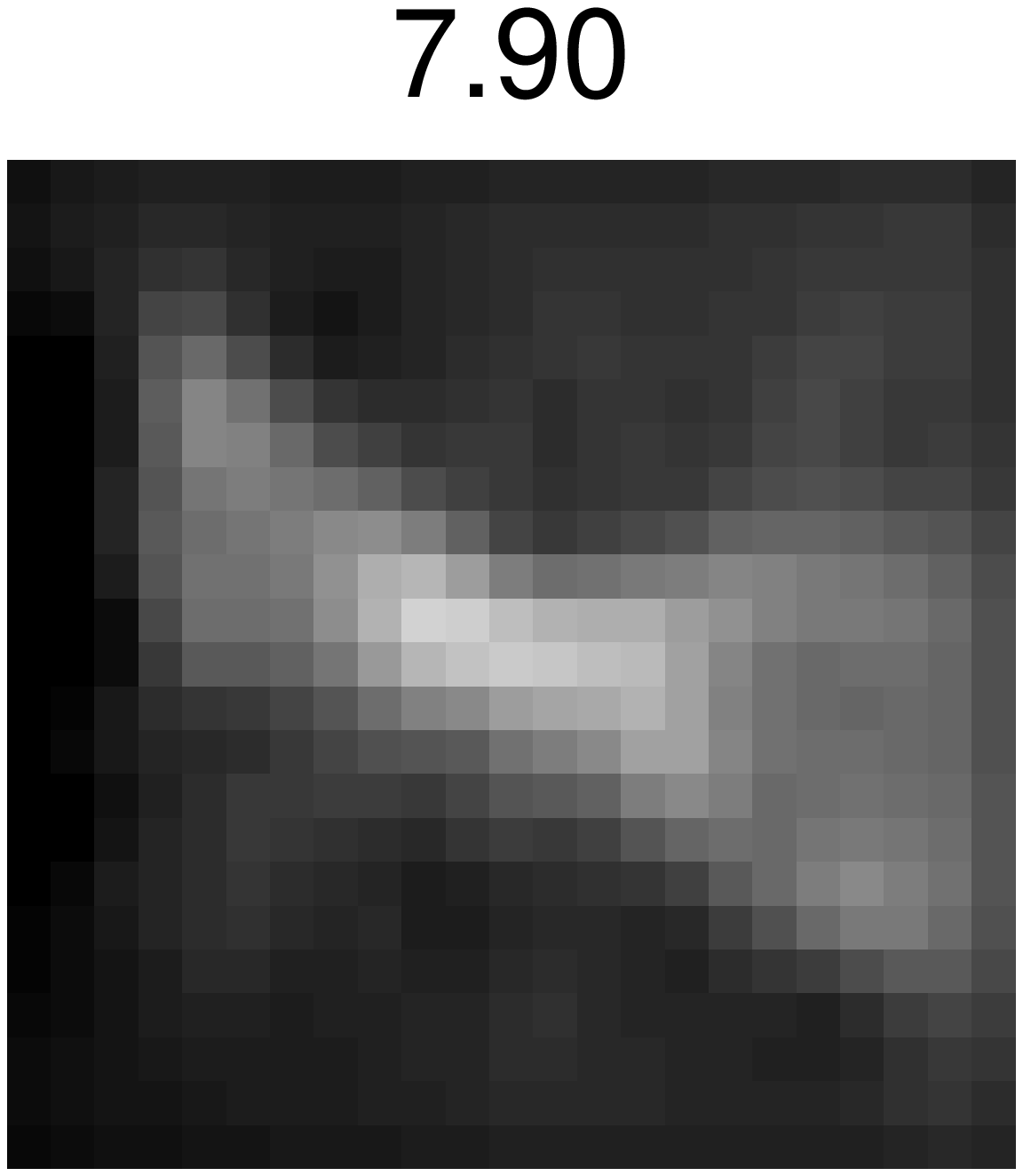}&
    \includegraphics[width=0.075\textwidth]{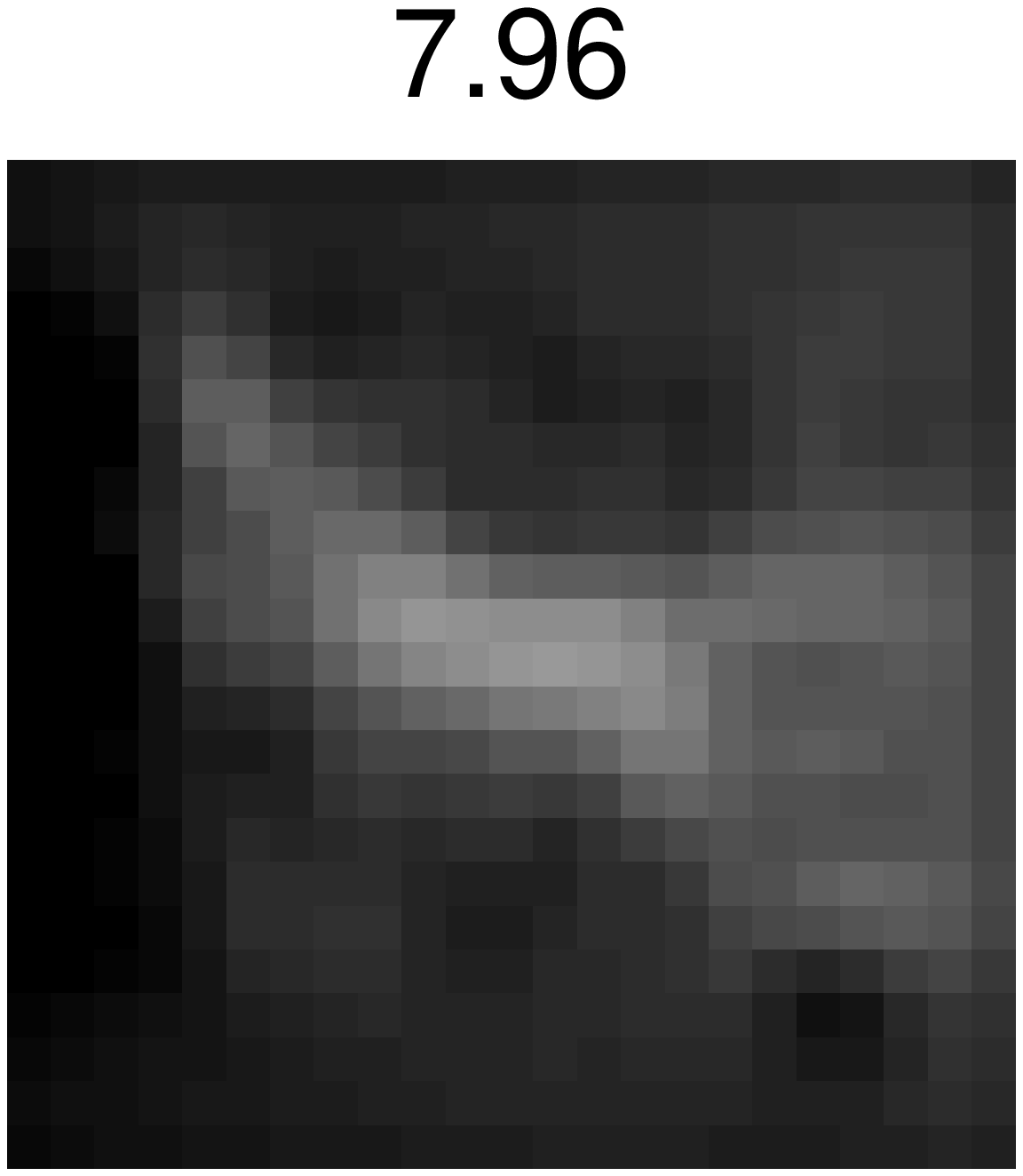}\\
    \includegraphics[width=0.075\textwidth]{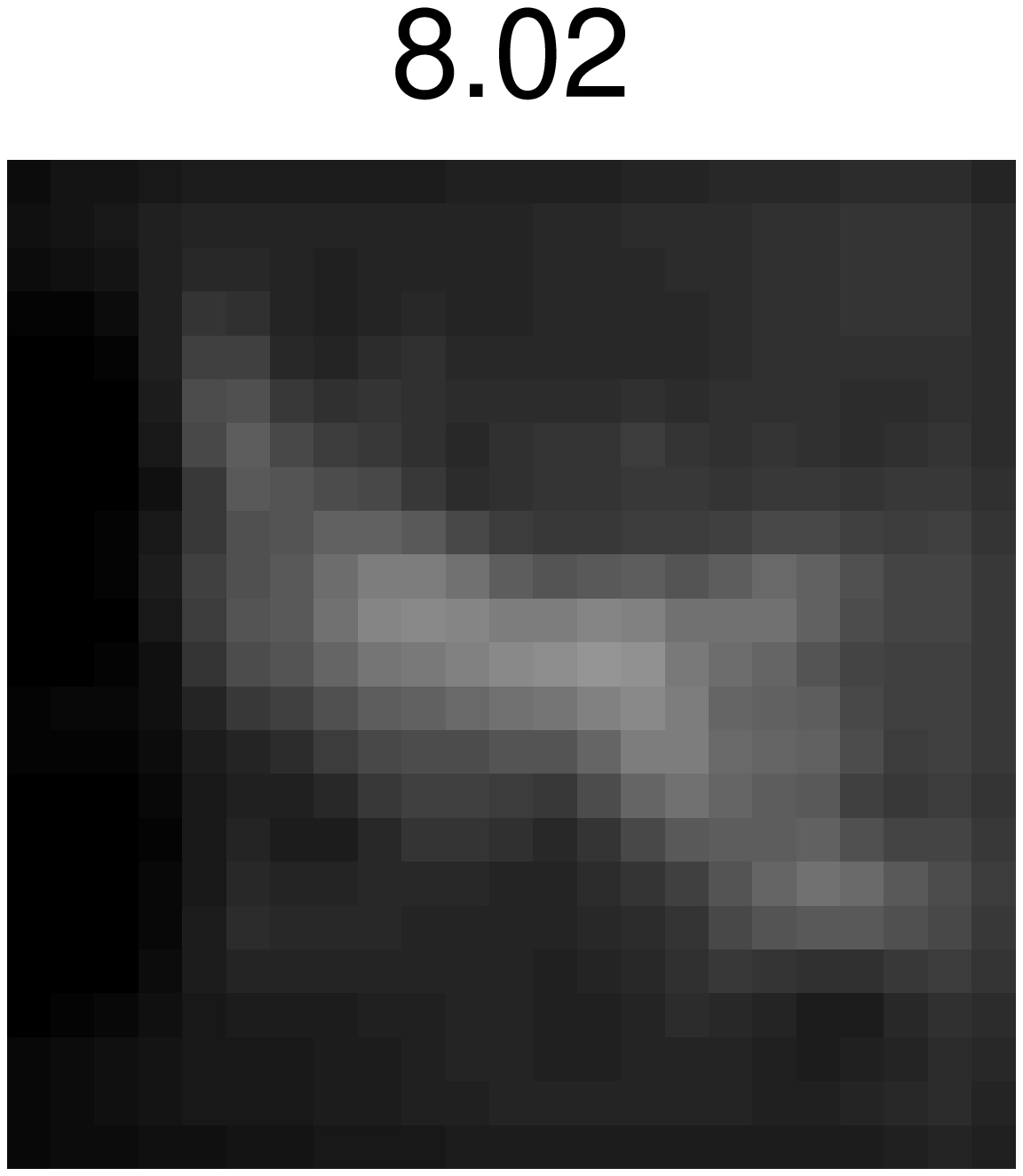}&
    \includegraphics[width=0.075\textwidth]{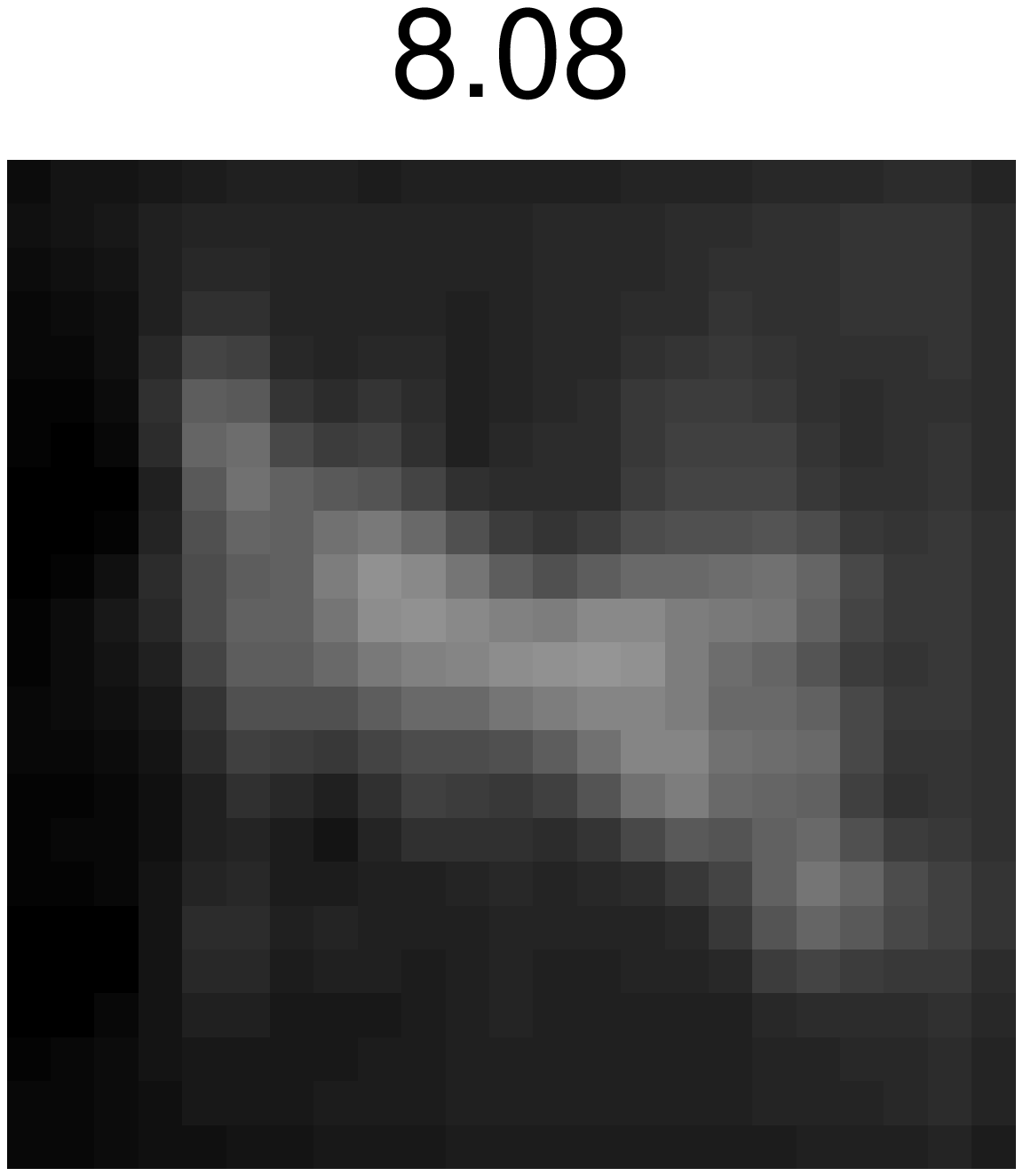}&
    \includegraphics[width=0.075\textwidth]{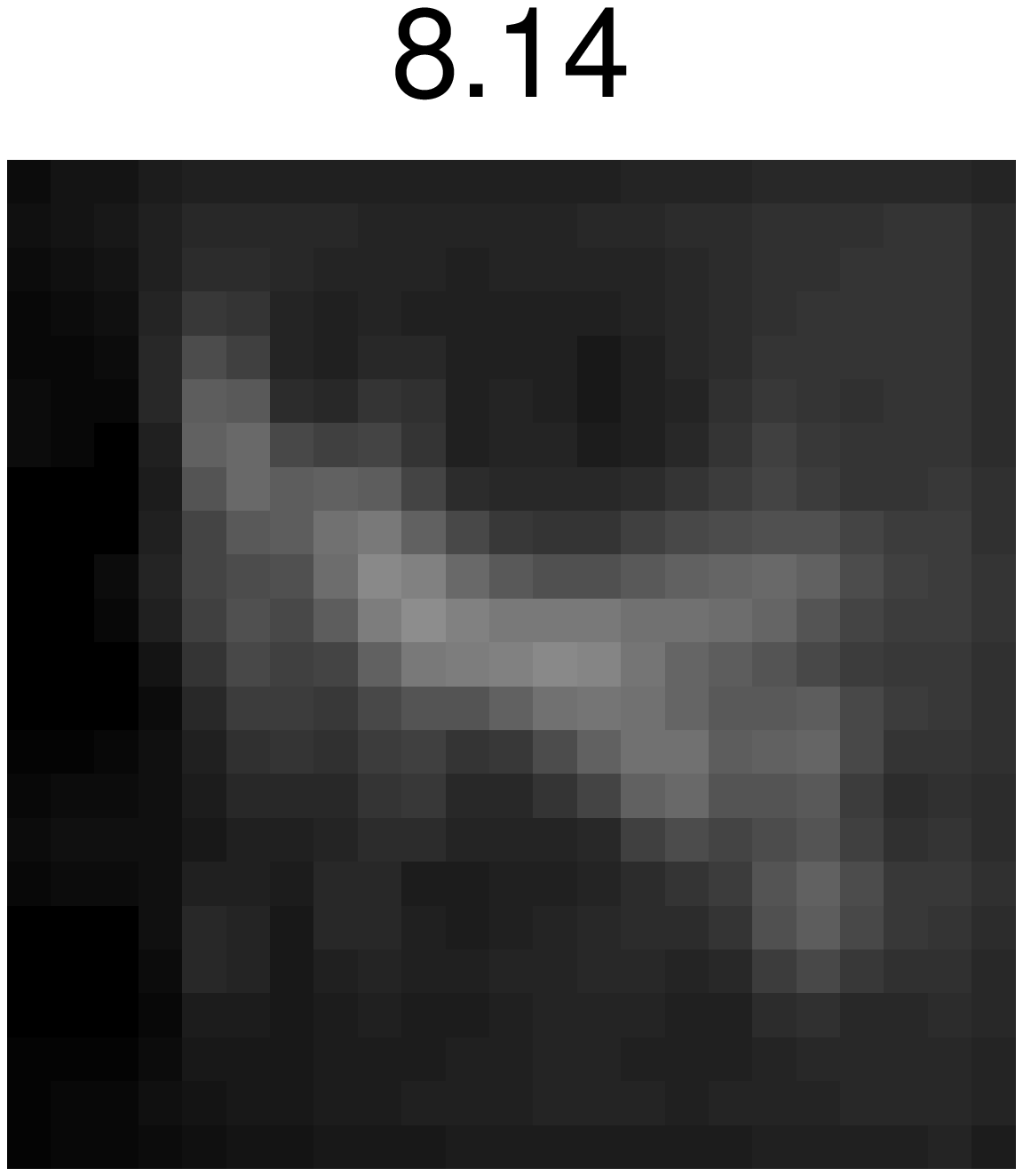}&
    \includegraphics[width=0.075\textwidth]{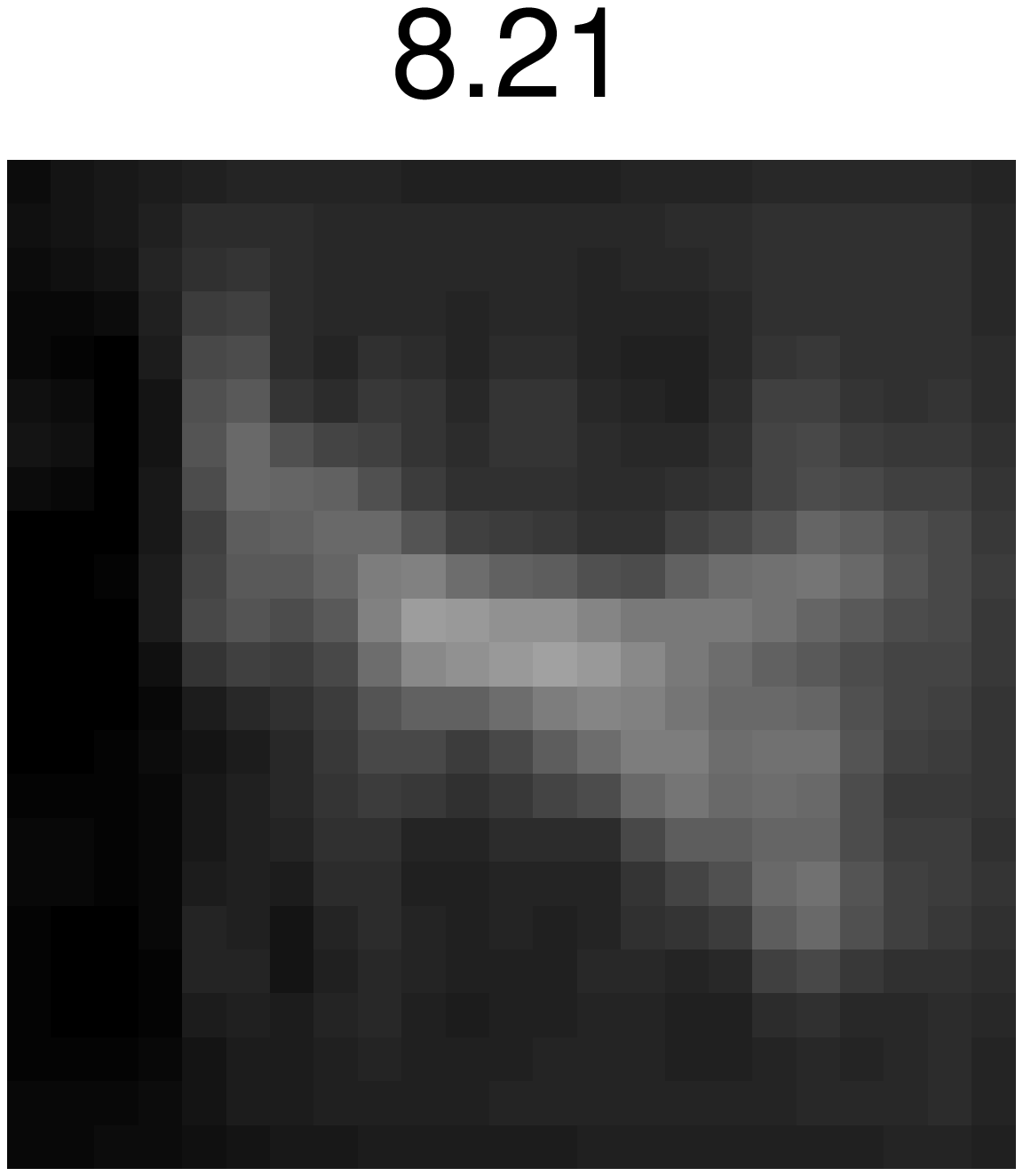}&
    \includegraphics[width=0.075\textwidth]{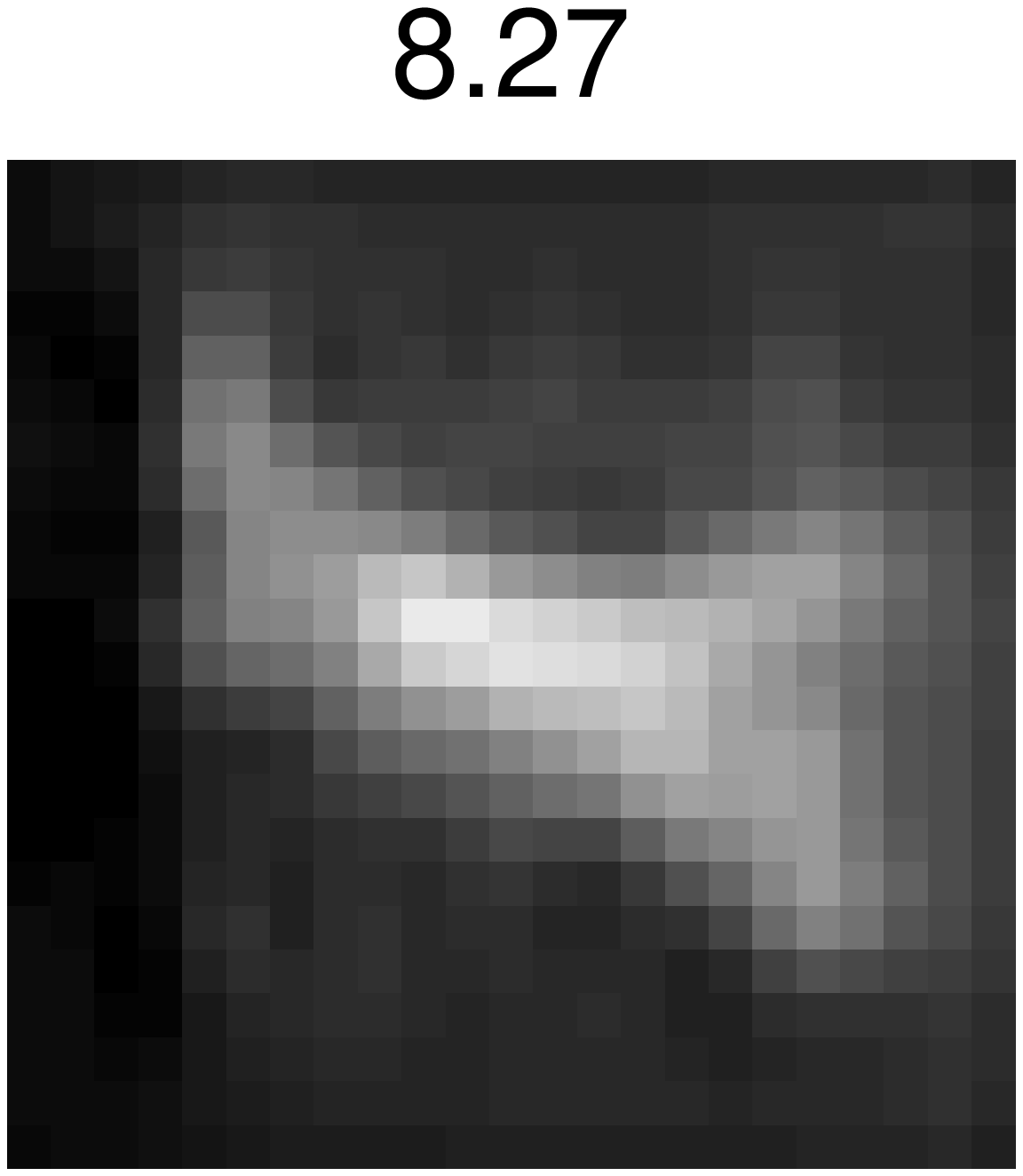}\\
    \includegraphics[width=0.075\textwidth]{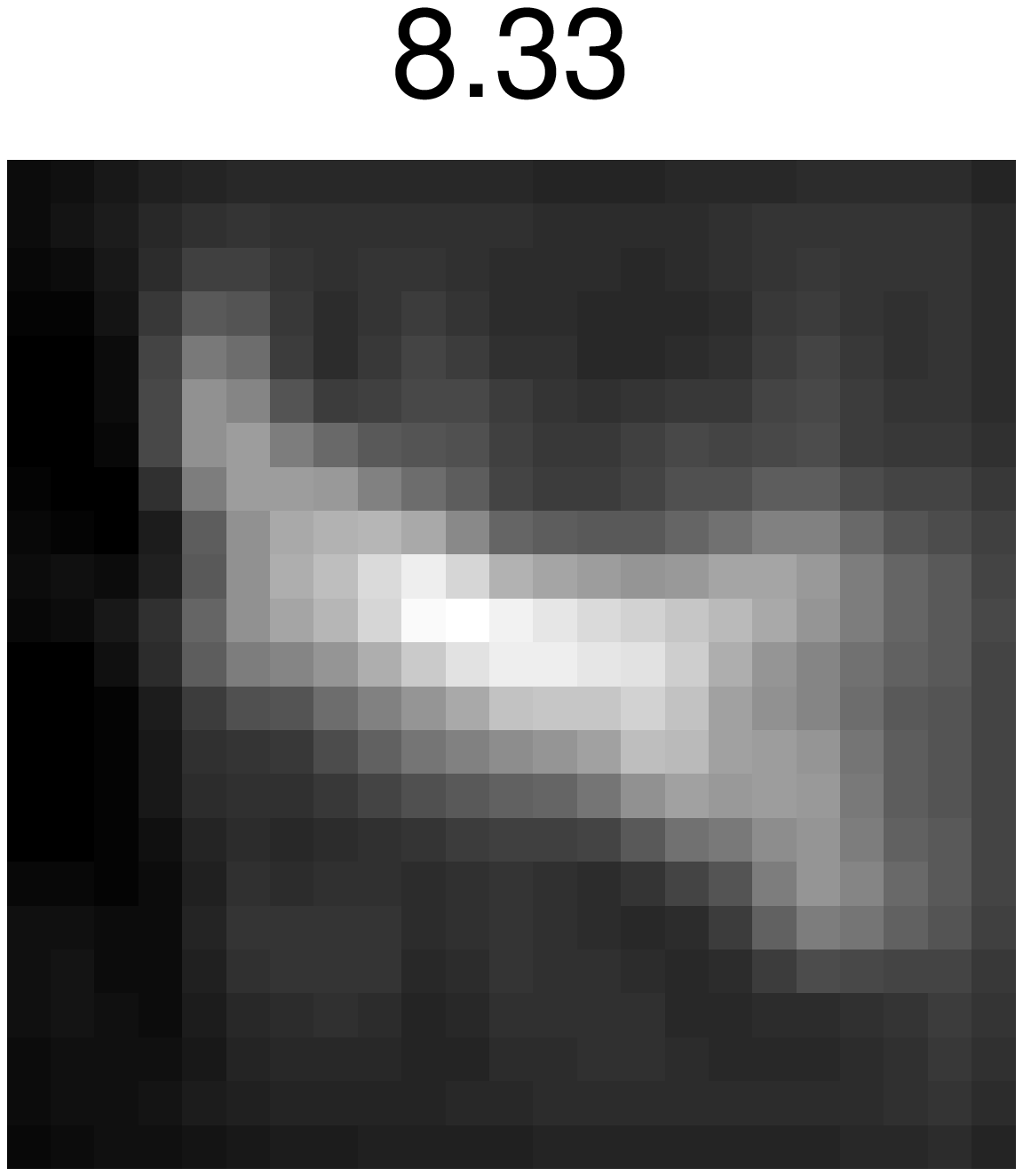}&
    \includegraphics[width=0.075\textwidth]{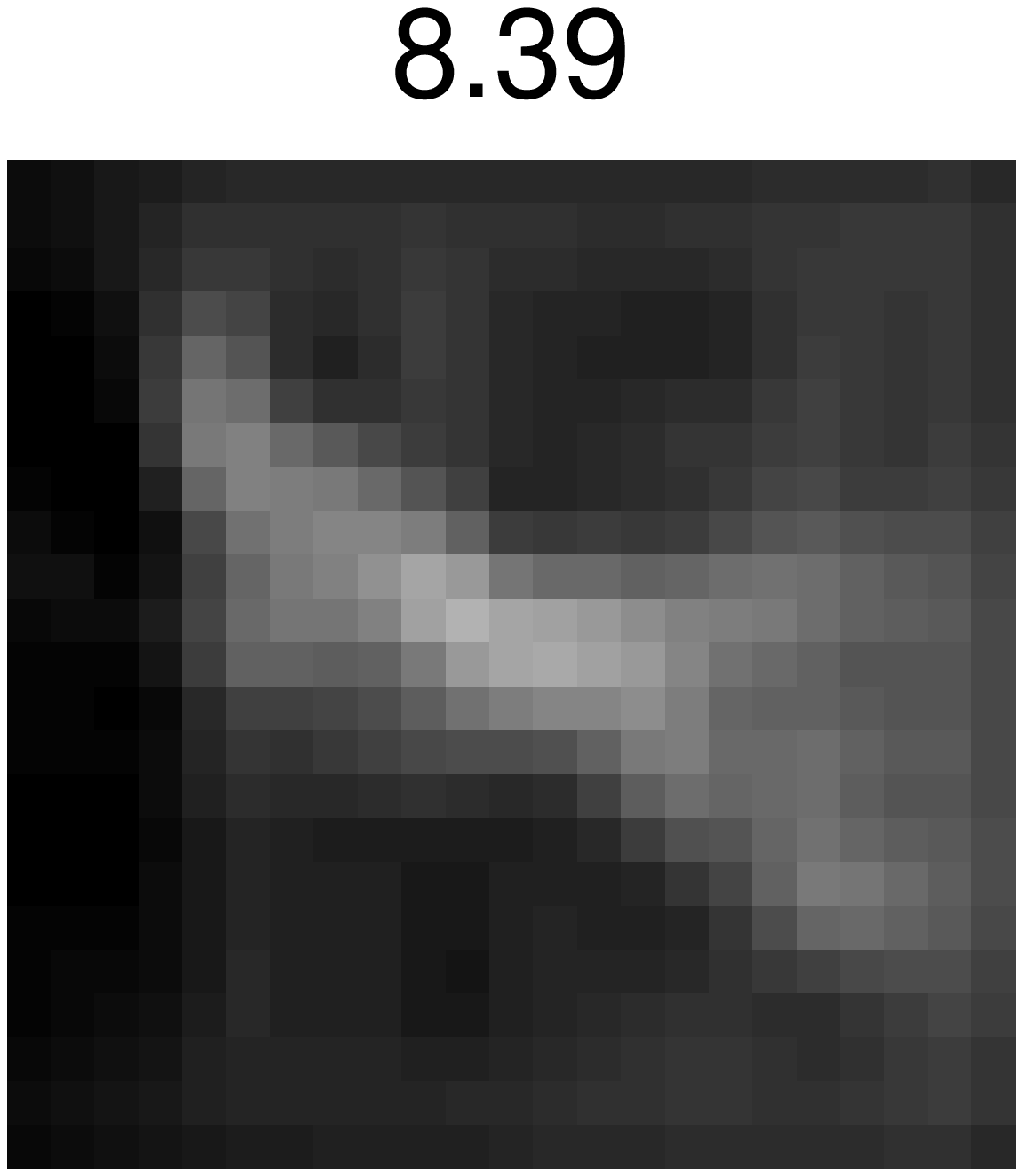}&
    \includegraphics[width=0.075\textwidth]{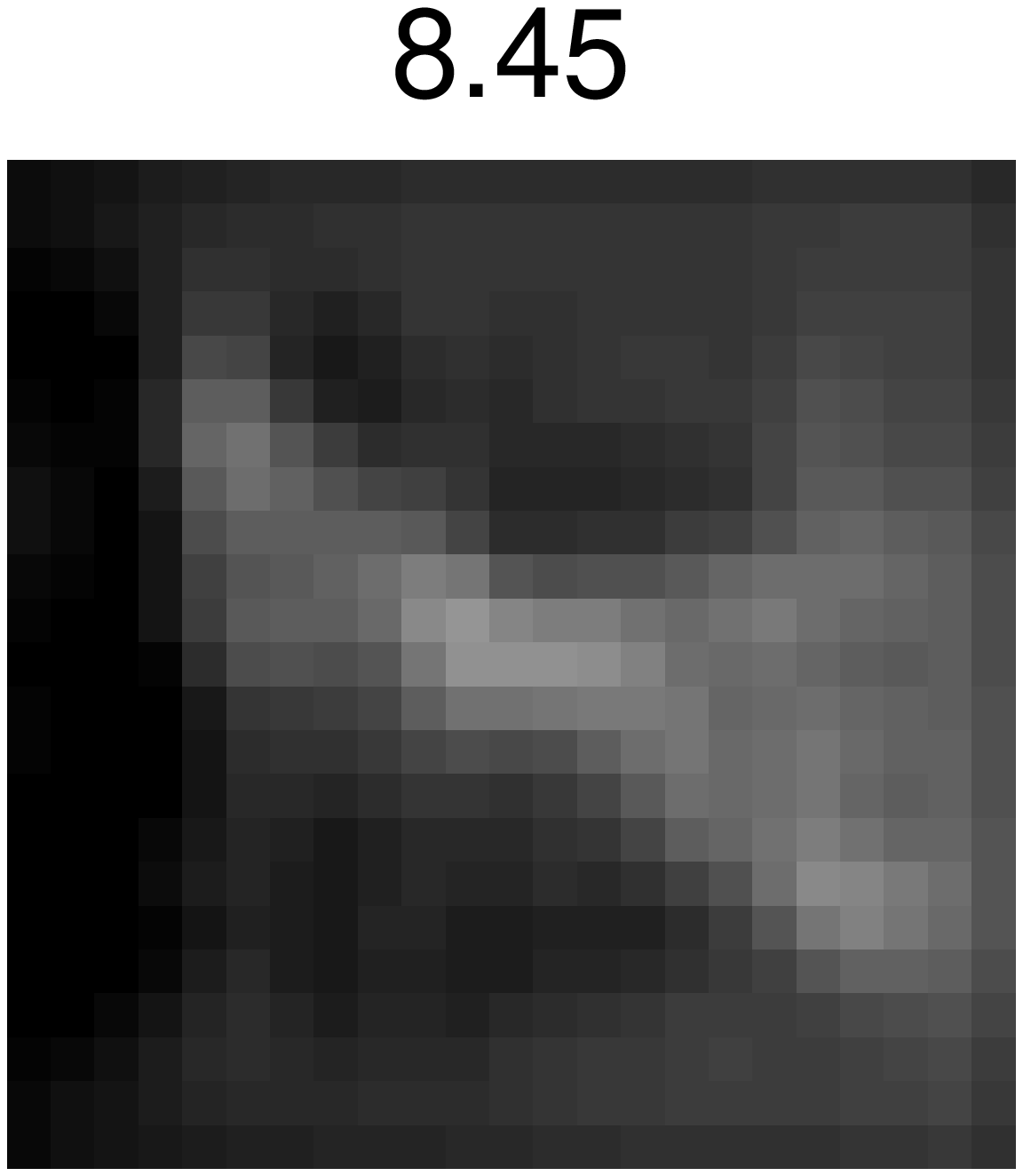}&
    \includegraphics[width=0.075\textwidth]{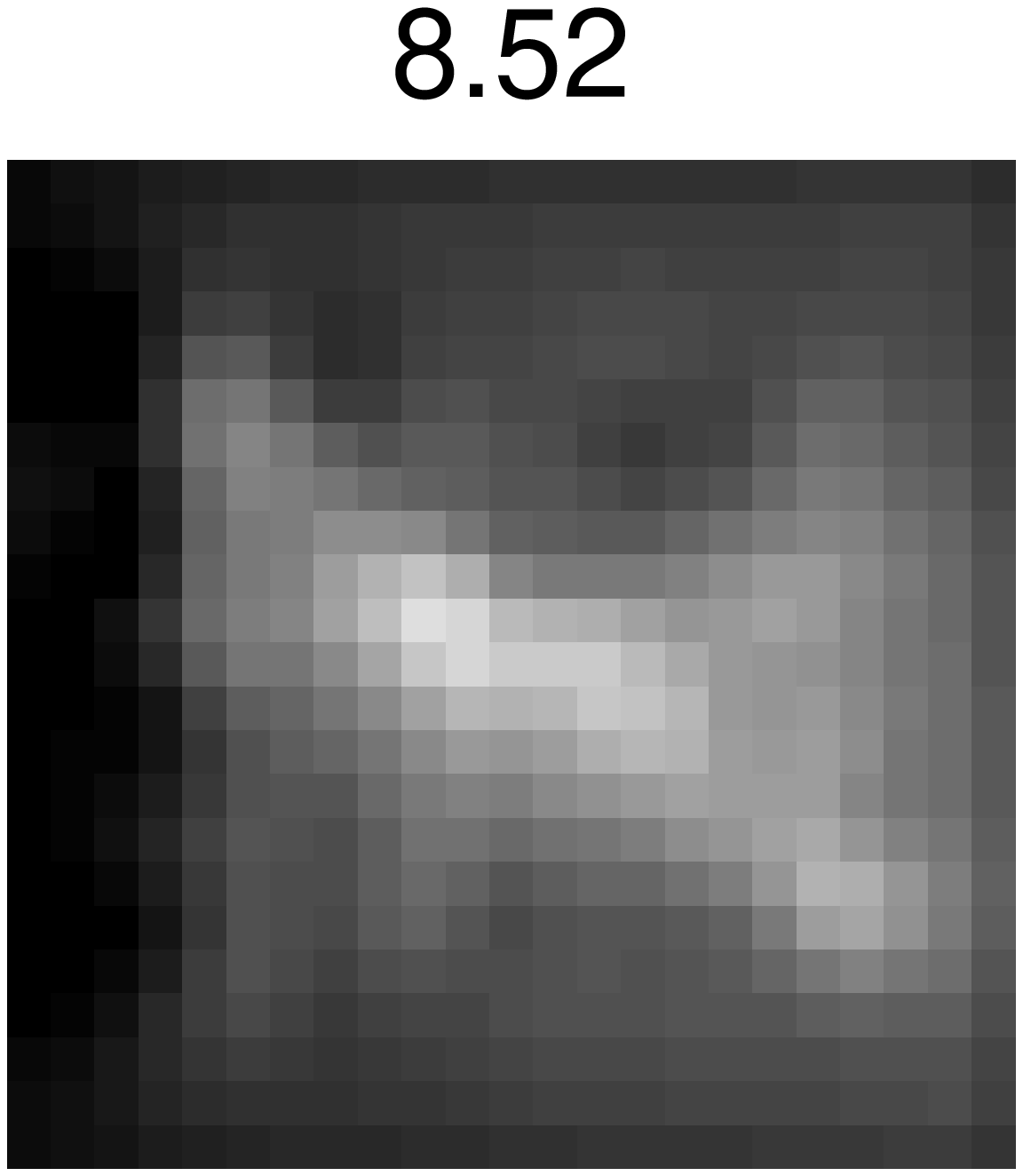}&
    \includegraphics[width=0.075\textwidth]{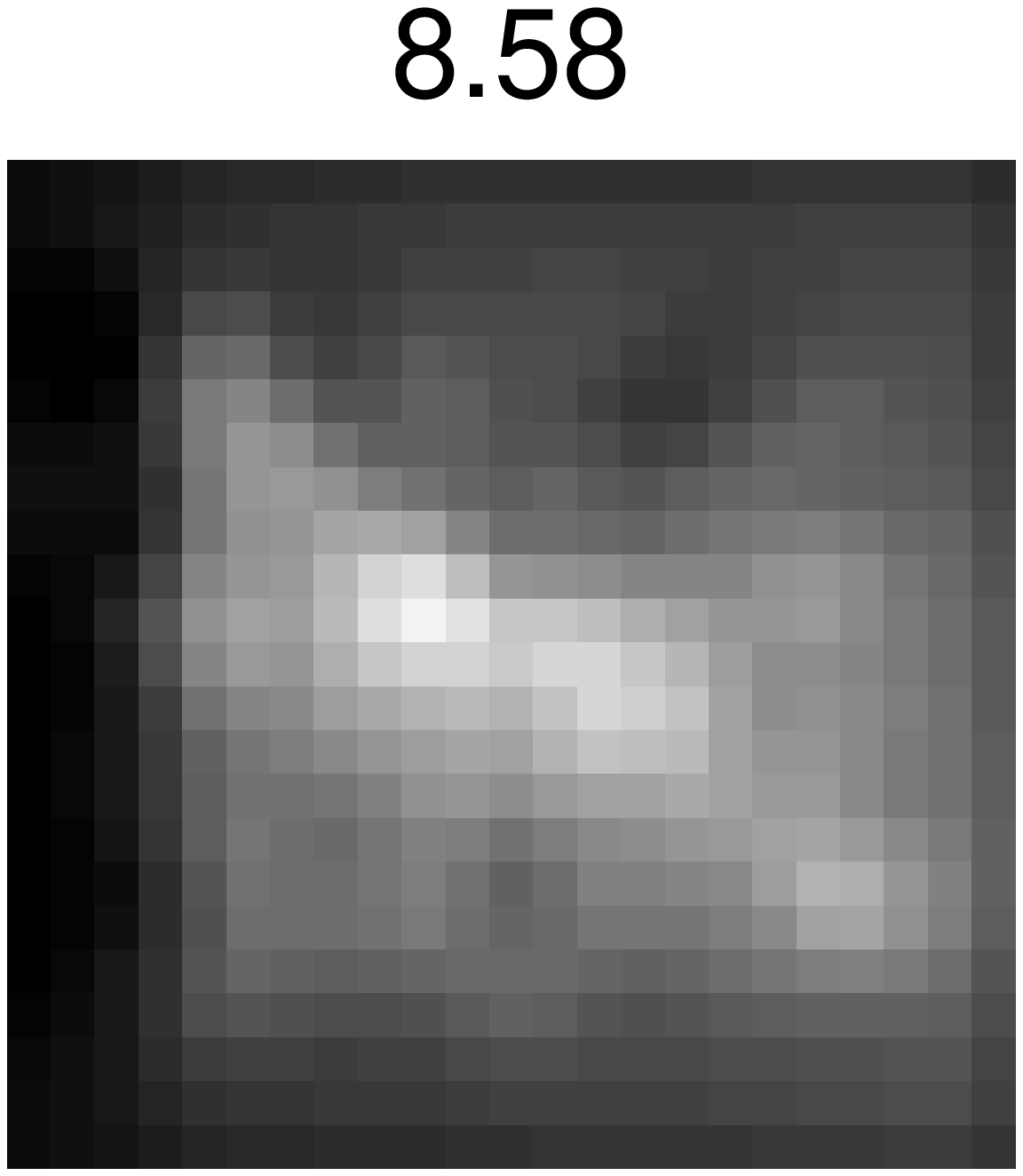}\\
    \includegraphics[width=0.075\textwidth]{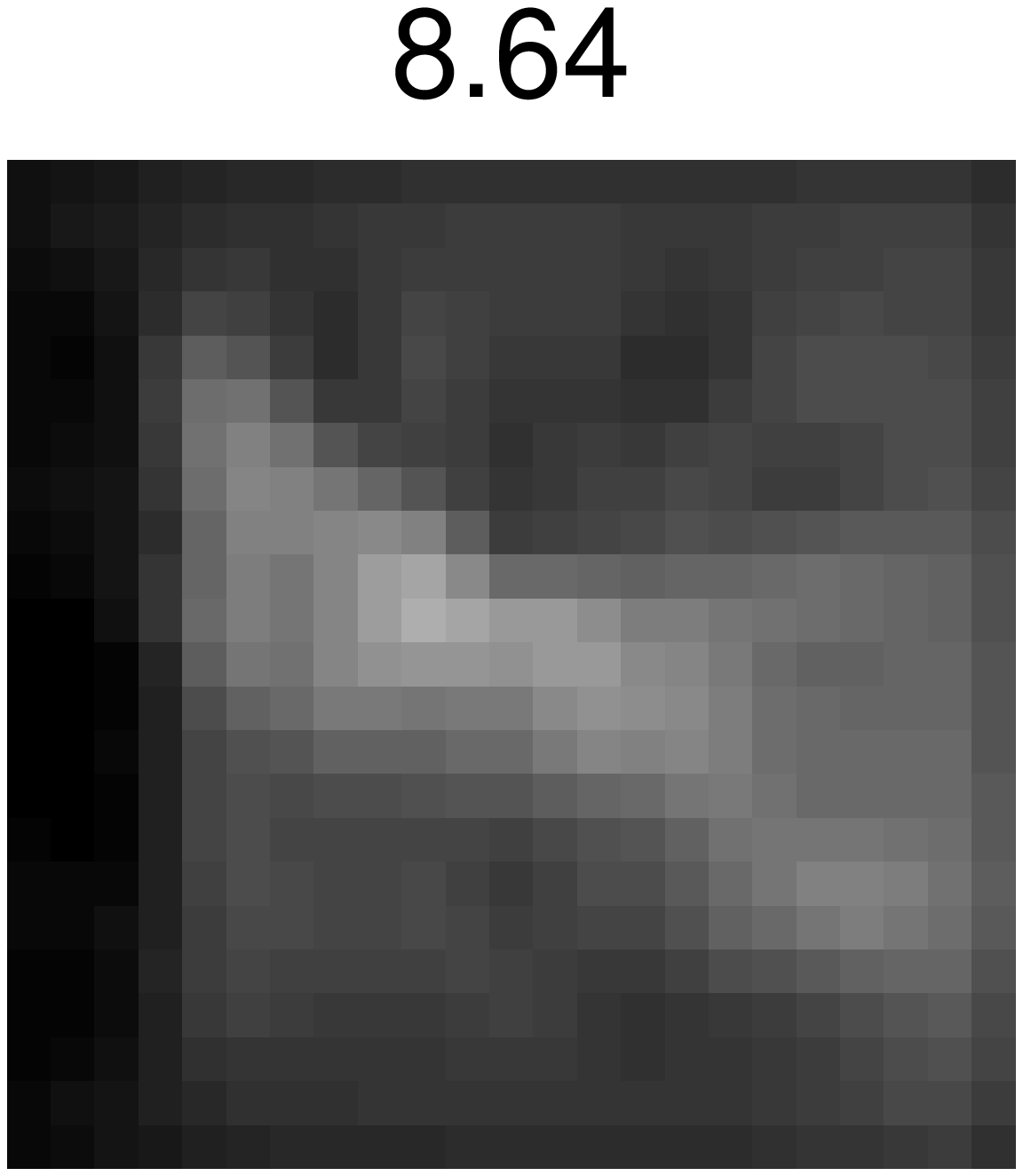}&
    \includegraphics[width=0.075\textwidth]{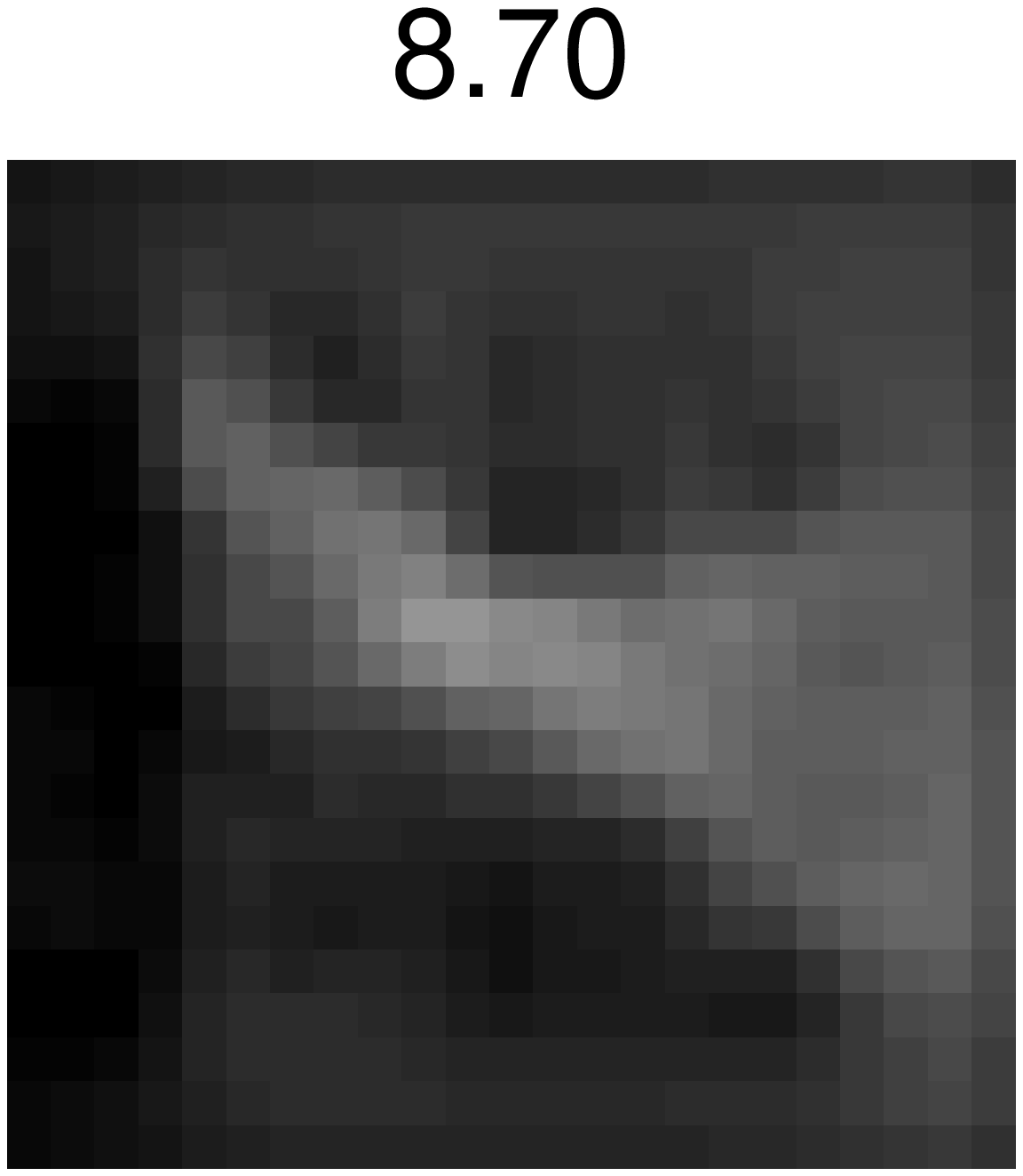}&
    \includegraphics[width=0.075\textwidth]{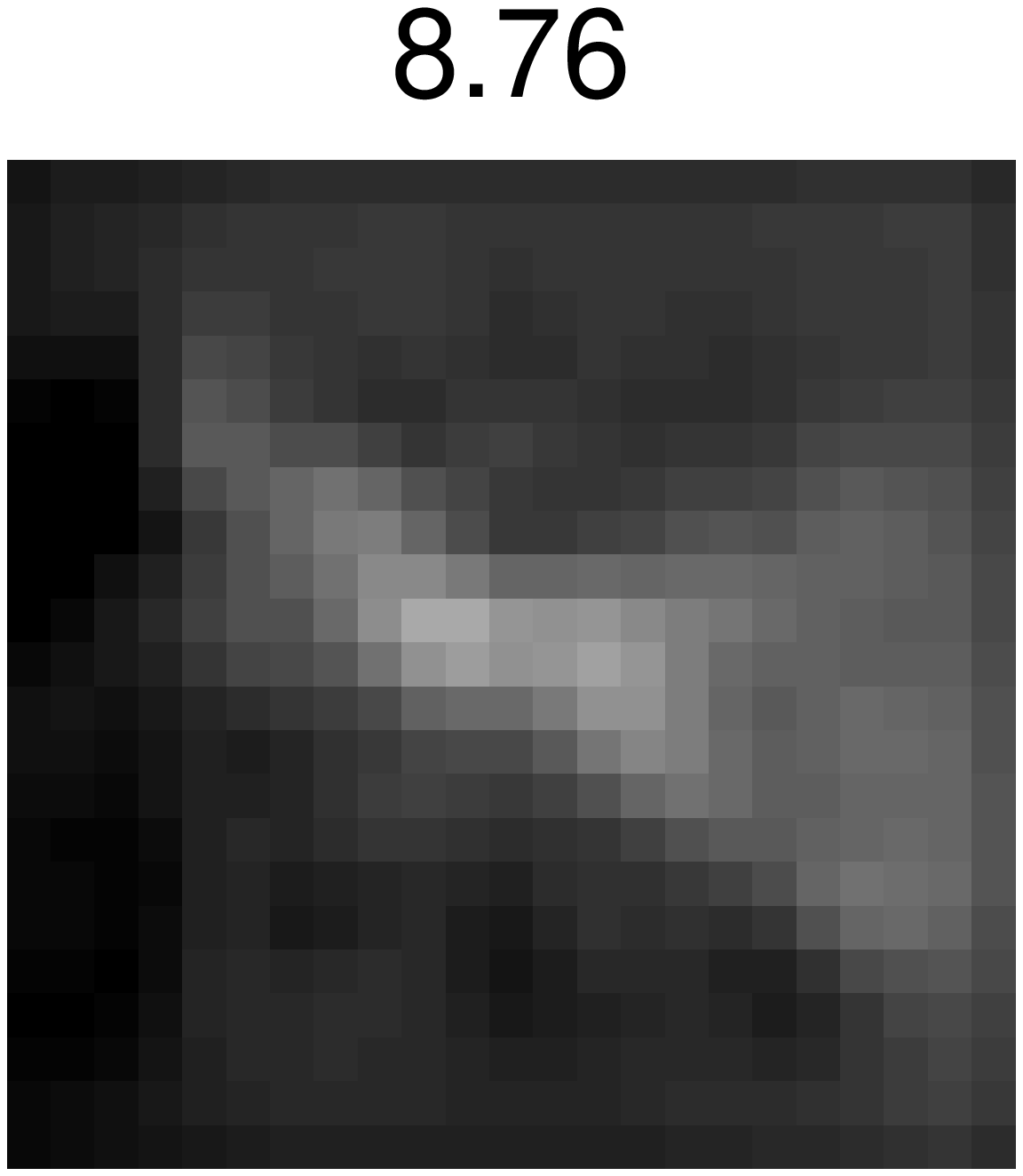}&
    \includegraphics[width=0.075\textwidth]{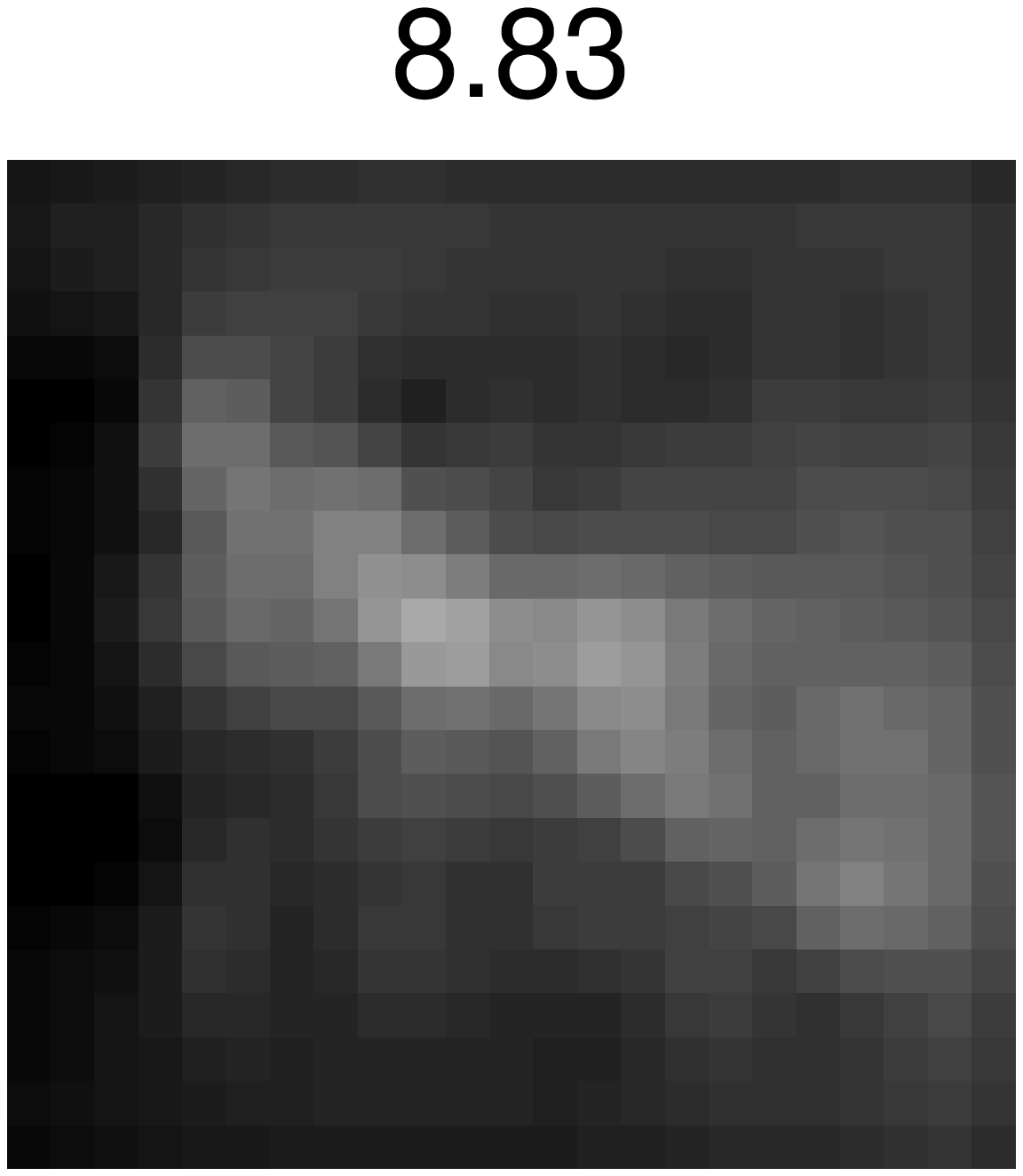}&
    \includegraphics[width=0.075\textwidth]{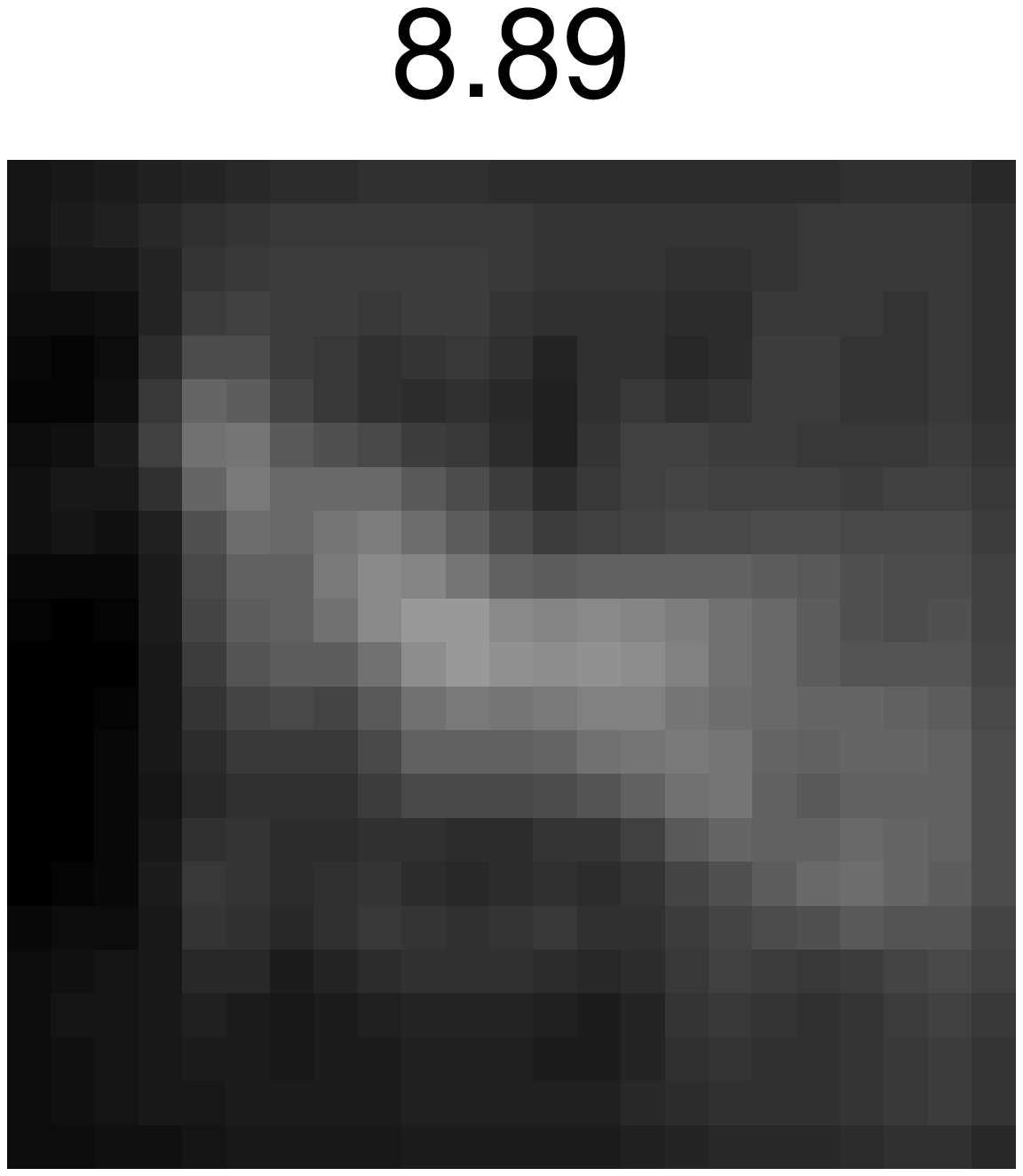}\\
    \includegraphics[width=0.075\textwidth]{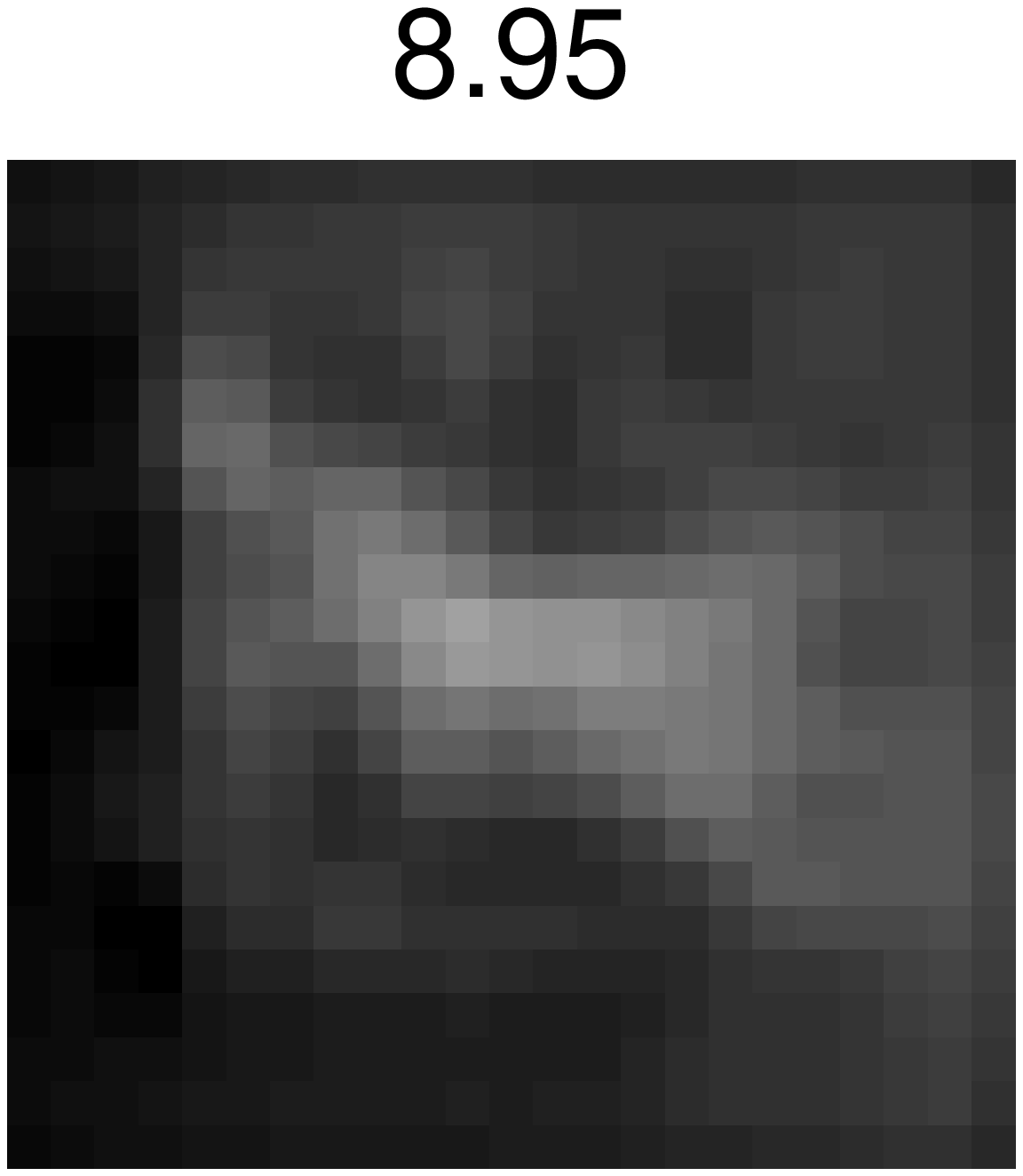}&
    \includegraphics[width=0.075\textwidth]{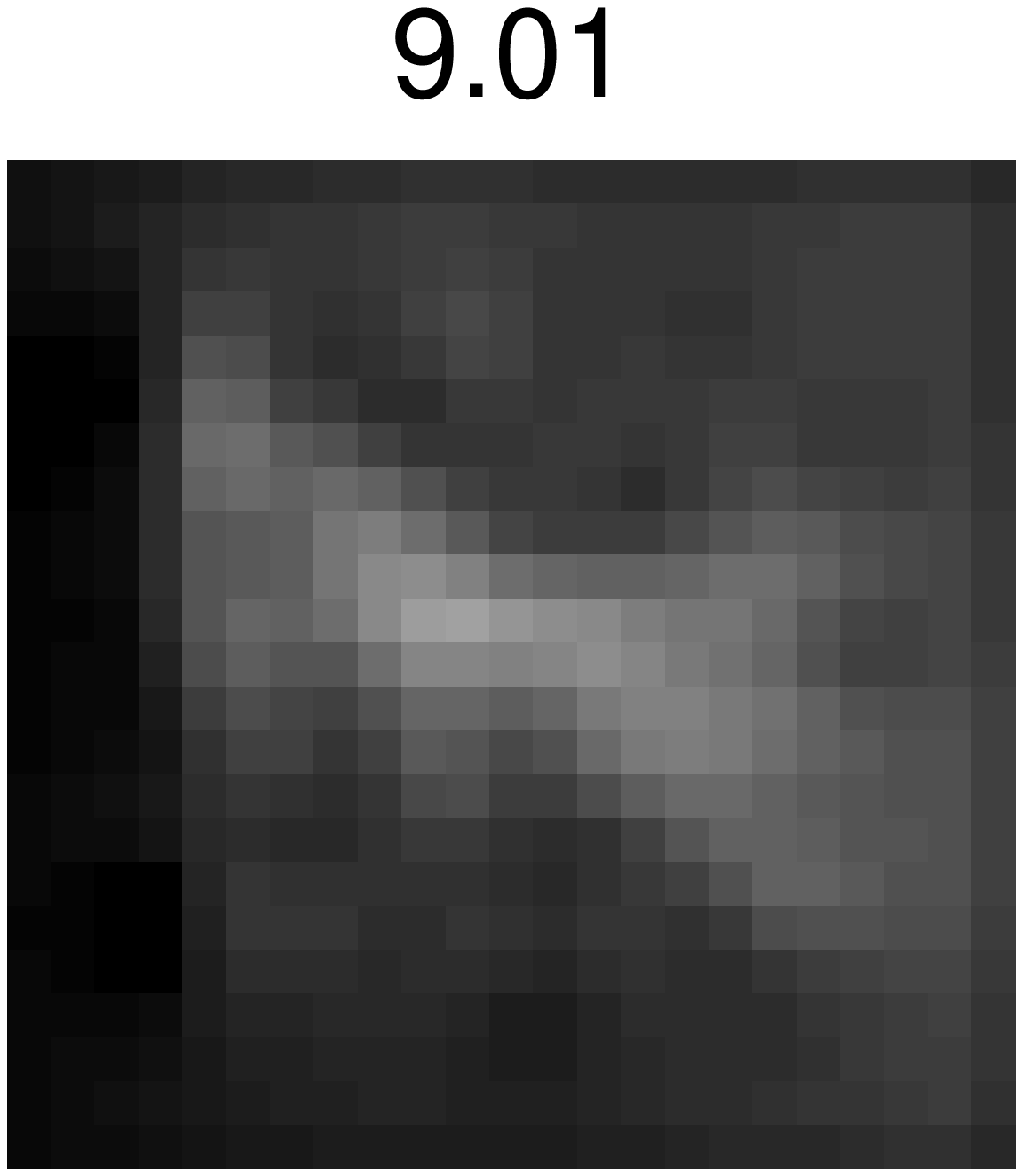}&
    \includegraphics[width=0.075\textwidth]{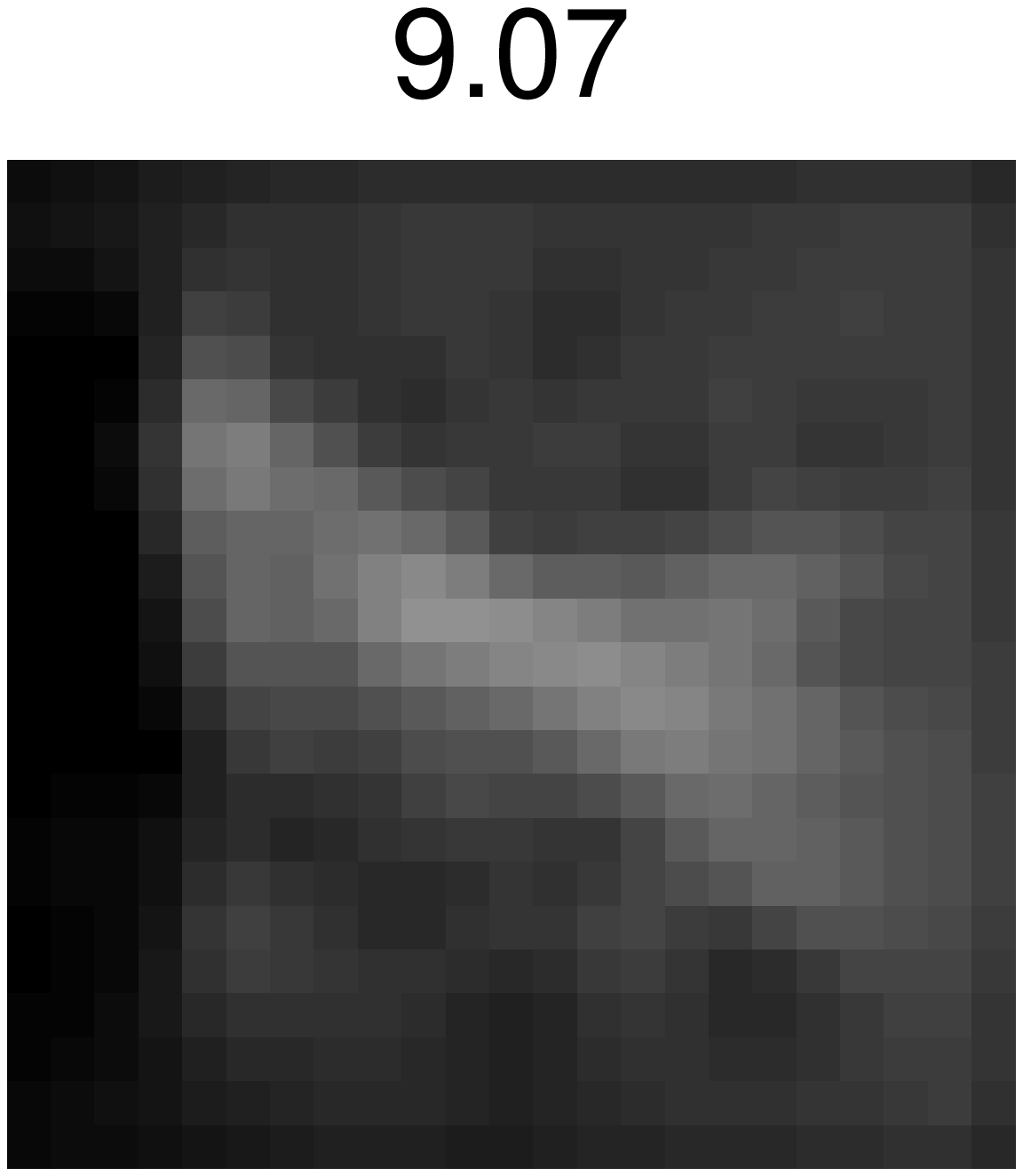}&
    \includegraphics[width=0.075\textwidth]{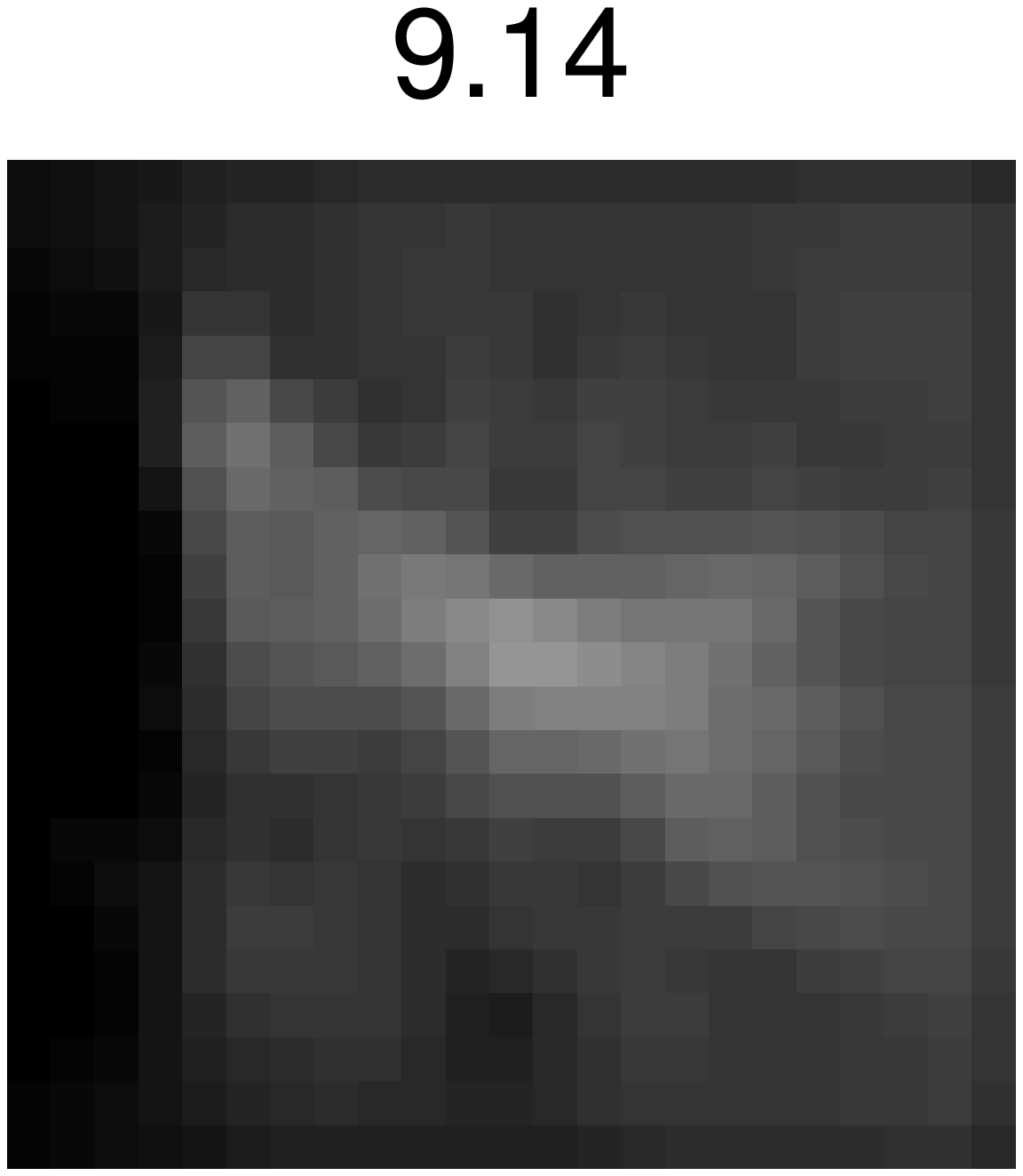}&
    \includegraphics[width=0.075\textwidth]{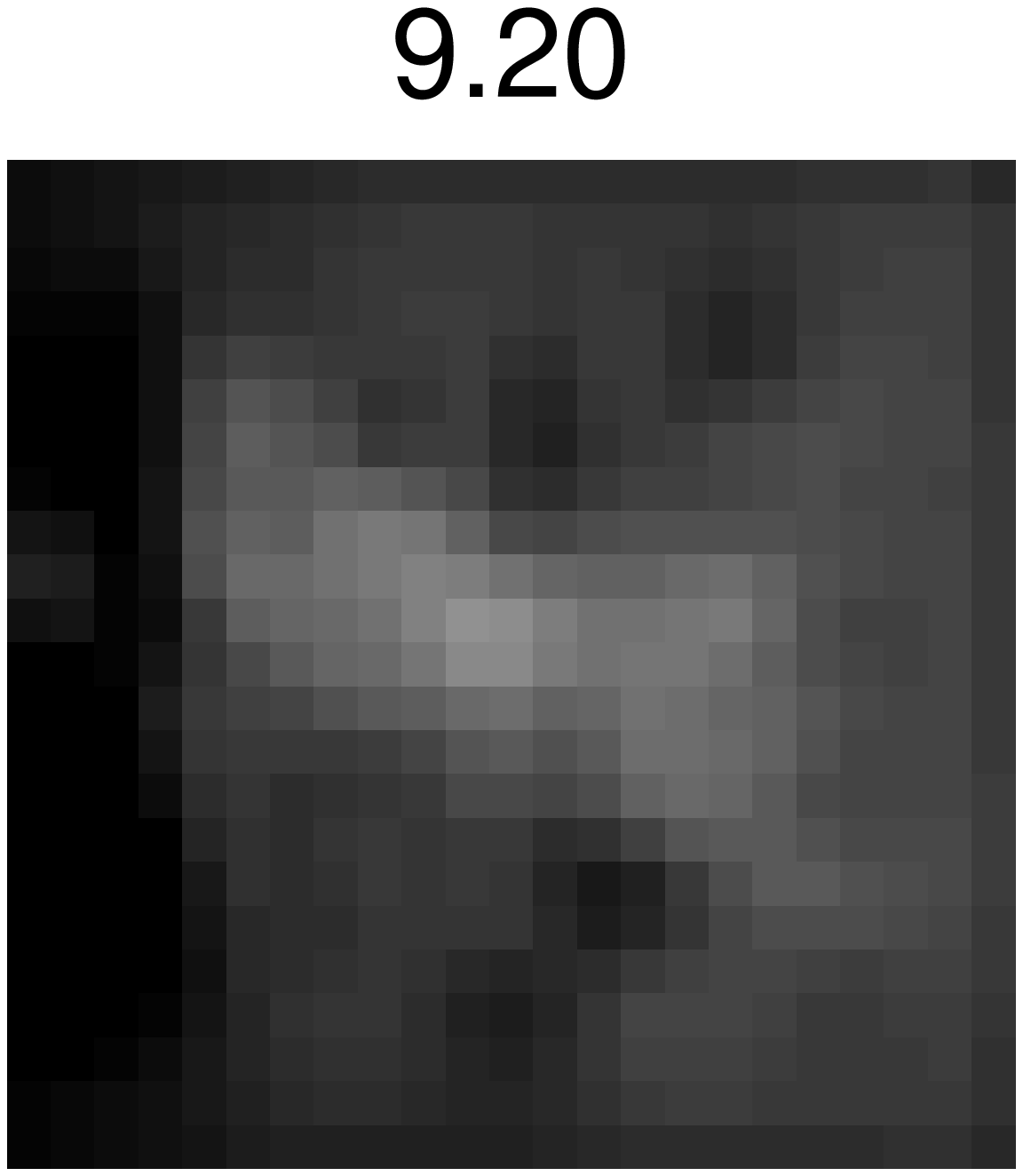}
  \end{tabular}
  \caption{Set of 30 images of our reconstruction from simulated data. Each image corresponds to 
one wavelength for 7.4 to 9.2\;$\mu m$ with a step of 0.062\;$\mu m$.}
  \label{fig:SimImages}
\end{figure}

\begin{figure}[htbp]
  \centering
  \includegraphics[width=0.35\textwidth]{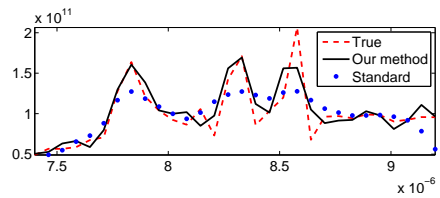}
  \caption{Spectrum of a pixel. The curves abscissa is the wavelenght in meters:
solid line: simulated sky, dashed line: our method, dotted line: 
conventional method.}
  \label{fig:SimSpectre}
\end{figure}

The results contain a set of 30 images (see Fig. \ref{fig:SimImages}).
Fig. \ref{fig:SimL2S} and solid line on Fig.\ref{fig:SimSpectre} illustrate our result 
for one wavelength (8.27\;$\mu m$) and one pixel, respectively.
The image computed with our method (Fig. \ref{fig:SimL2S}) appears comparable to the true image 
(Fig. \ref{fig:SimVrai}), 
while the image computed with the conventional processing (Fig. \ref{fig:SimAlain}) is smoother. 
 A comparison of solid line and dotted line in Fig. \ref{fig:SimSpectre} clearly 
also shows that
our method provides a spectrum comparable to the true spectrum, while the peaks obtained with the conventional processing are too broad.  

Our sky estimation depends on the regularization coefficients
$\mu_{\alpha\beta}$ and $\mu_{\lambda}$. 
We tune this parameters by minimizing numerically the quadratic error between the estimated 
object and the real object were selected. In this experiment we obtain 
$\mu_{\alpha\beta}=0.01,\;\mu_{\lambda}=0.005$.

\subsection{Real data}\label{sec:donnees-reelles}

Eeal data contain 23 acquisitions composed of $38\times128$ values. 
To obtain a over-resolved reconstruction, we describe our volume 
with 587264 gaussians destributes on a cartesian grid $74\times62\times128$. 
The spatial $(\alpha,\beta)$ sampling step is equal to a quater slit width,
and the spectral dimension is uniformly sampled between the wavelength 7.4 and 
15.3\;$\mu \mathrm{m}$.  
The reconstruction is computed after setting the regularization coefficients
$\mu_{\alpha\beta}$ and $\mu_{\lambda}$ empirically. Too low a value for
these coefficients produces an unstable method and a quasi explosive reconstruction. Too high a value produces images that are visibly too smooth. A compromise found by trial and error led us to $\mu_{\alpha\beta}=0.3$ and $\mu_{\lambda}=0.7$. The ratio between $\mu_{\alpha\beta}$ and  $\mu_{\lambda}$  is also based on our simulation. However, we cannot compare the regularization coefficients 
between the simulated and the real case, since 
the size of the problem modifies the weigth of the norm in the 
Eq. (\ref{eq:15}). 
Pratically, we take large value for the regularization coefficients, 
and we gradually reduce the value up that we are seeing noise.

\begin{figure}[htbp]
  \centering
  \hfill
  \subfigure[]{\includegraphics[width=0.22\textwidth]{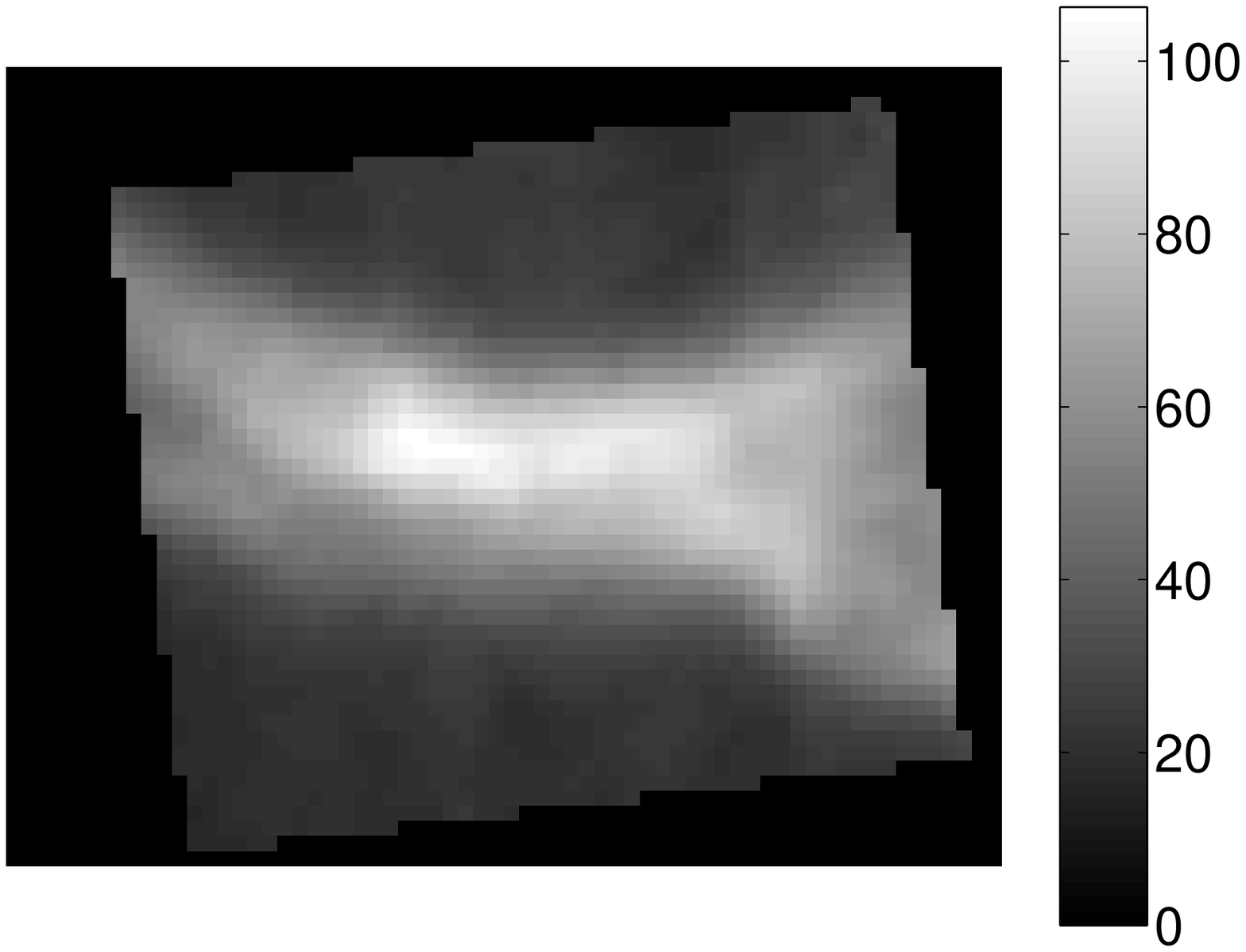} \label{fig:imAlain}}
  \hfill
  \subfigure[]{\includegraphics[width=0.22\textwidth]{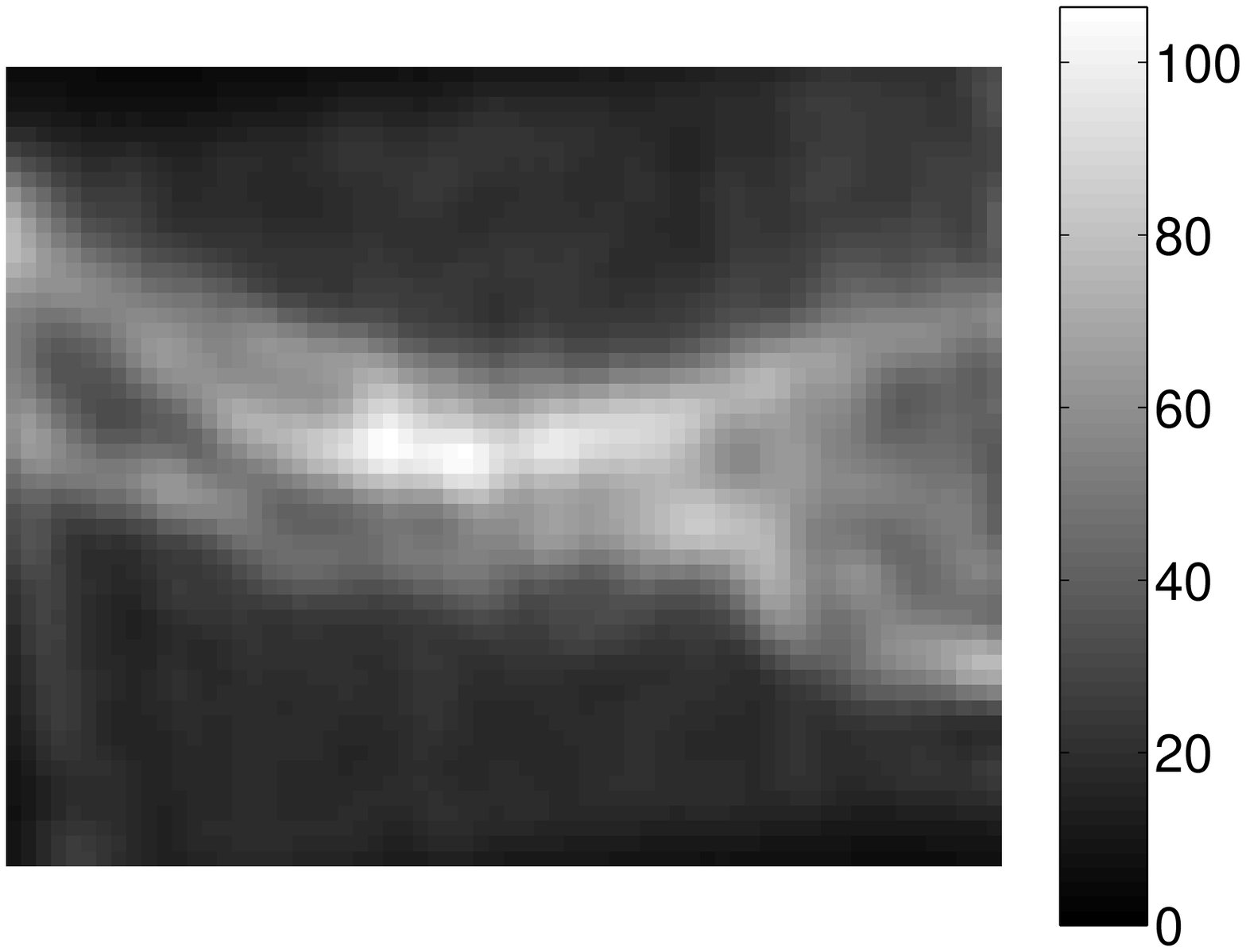} \label{fig:imL2S}}
  \hfill
  \subfigure[]{\includegraphics[width=0.2\textwidth]{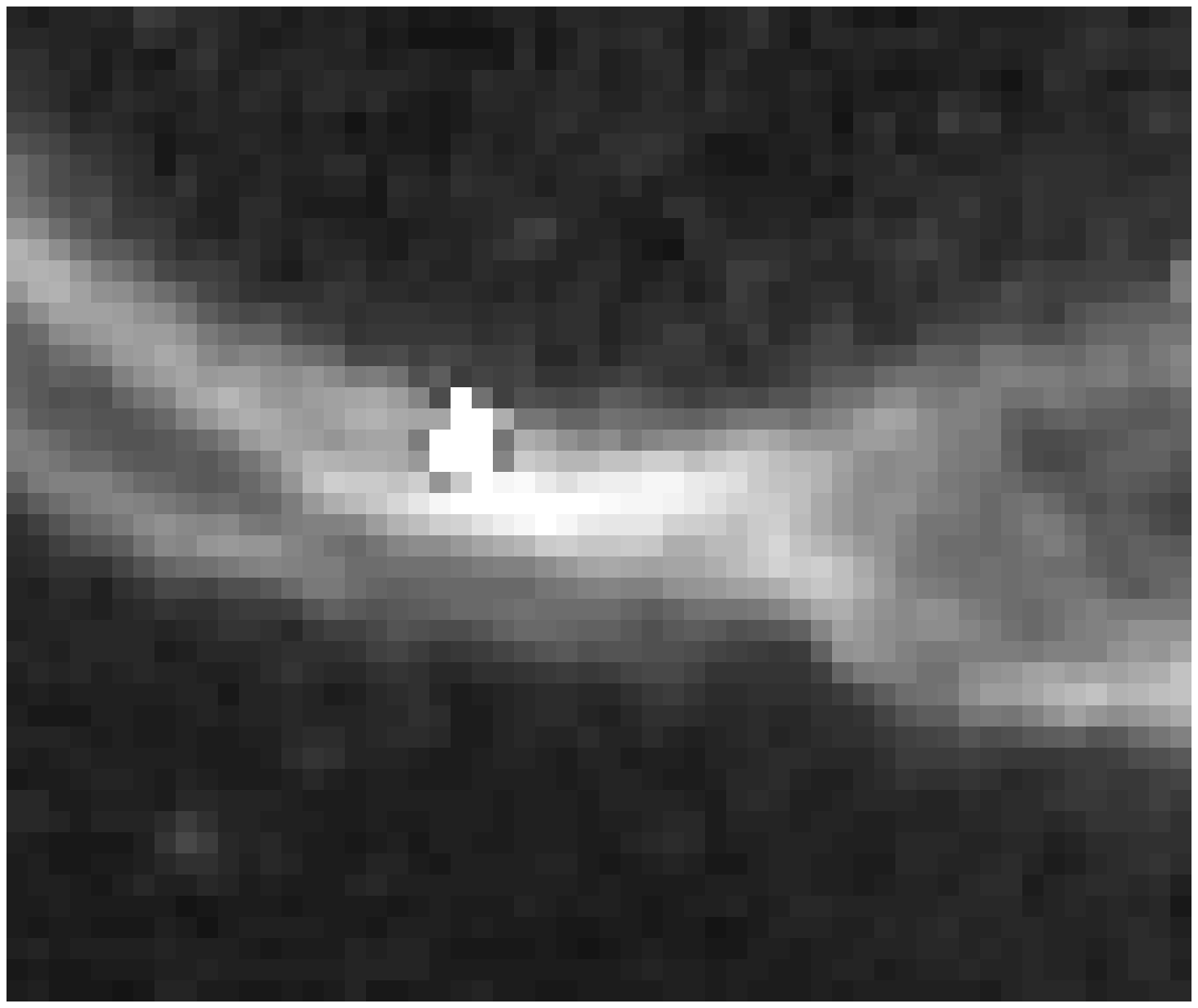} \label{fig:irac}}
  \caption{Reconstruction of a sky $\phi$ representing the Horsehead nebula: (a) image estimated at 11.37 $\mu \mathrm{m}$ by the conventional method (b) image estimated at 11.37 $\mu \mathrm{m}$ by 
  our method, (c) image obtained with the Infrared Array Camera IRAC on board the Spitzer Space Telescope having better resolution at 4.5 $\mu\mathrm{m}$.}
  \label{fig:Horse}
\end{figure}

\begin{figure}[htbp]
  \centering
  \subfigure[]{\includegraphics[width=0.45\textwidth]{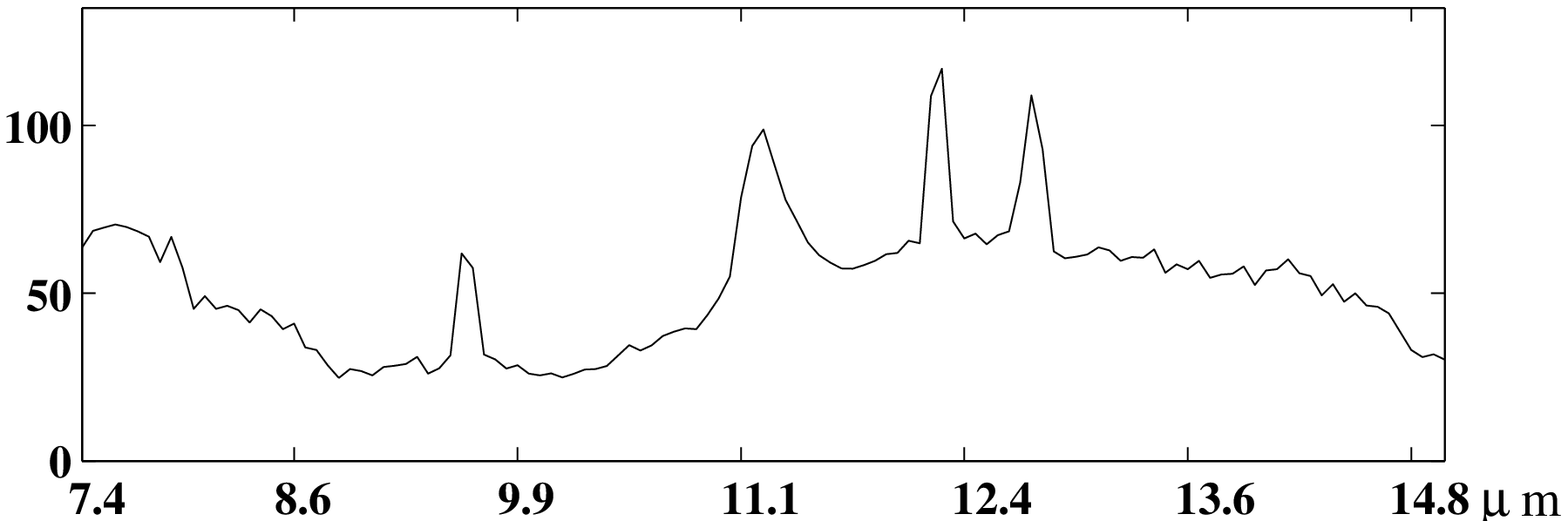}\label{fig:spec1}}
  \subfigure[]{\includegraphics[width=0.45\textwidth]{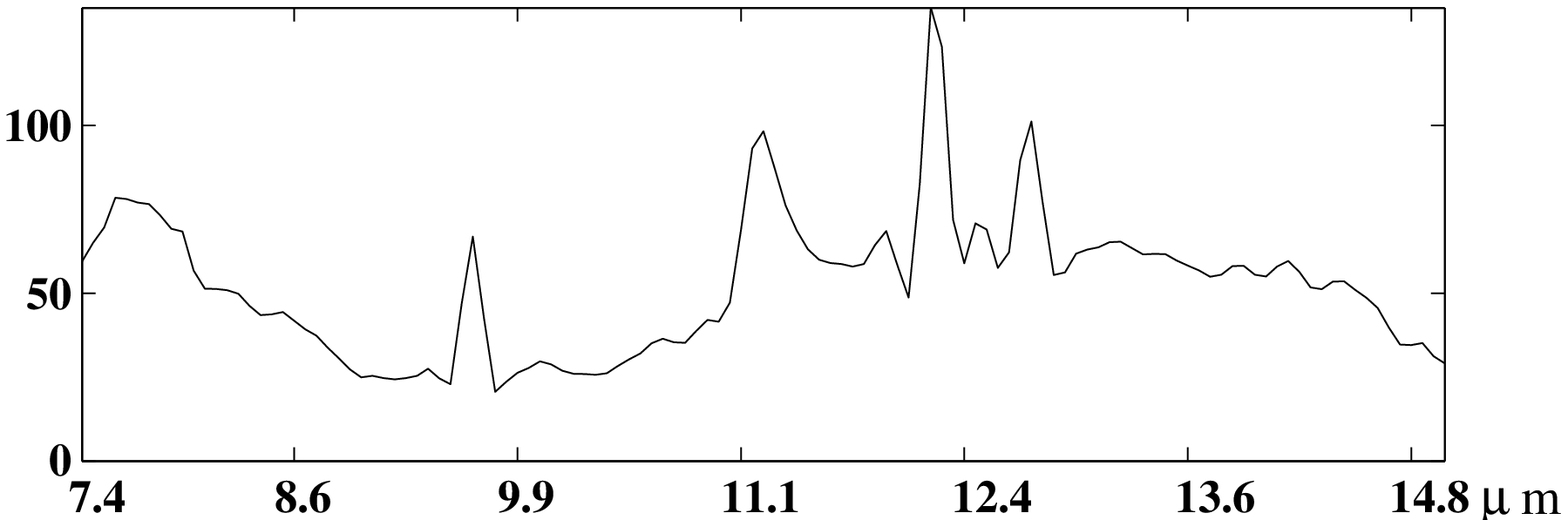}\label{fig:spec2}}
  \caption{Spectrum of the center point of the Fig. \ref{fig:imL2S}: (a) spectrum estimated 
with the conventional  method (b) spectrum estimated with 
   our method.}
  \label{fig:spectre}
\end{figure}

Our results (Fig. \ref{fig:imL2S} and \ref{fig:spec2}) can be compared with those obtained with
(Fig. \ref{fig:imAlain} and \ref{fig:spec1}, from \cite{Compiegne07}). A comparison of Fig. \ref{fig:imAlain} and \ref{fig:imL2S} clearly shows
that our approach provides more resolved images that bring out more structures
than the conventional approach. Note, in particular, the separation of the
two filaments on the left part of the Fig. \ref{fig:imL2S} obtained with 
our method, which remains
invisible after conventional processing. For comparison, Fig. \ref{fig:irac} shows the same object observed 
with the Infrared Array Camera (IRAC) of the Spitzer Space Telescope which has a  better native resolution since it observes at a shorter wavelength (4.5\;$\mu \mathrm{m}$). Here the same structures are observed, providing a strong argument in favour of the reality of the results provided by our method.  

A more precise analysis is done in section \ref{sec:etude-du-pouvoir}. 
It provides a quantitative evaluation of the resolution. 

Finally, the spectra reconstructed by 
our method (Fig.~\ref{fig:spec2}) have a resolution
slightly better
than the one reconstructed
by the conventional method (Fig.~\ref{fig:spec1}). The peaks characterizing
 the observed matter (gas and dust) are well positioned, narrower and
with a greater amplitude. However, ringing effects appear at the bases of 
the strongest peaks (Fig.~\ref{fig:spec2}). 
They could be explained by an over-evaluation
of the width $\sigma_{s}$ of the response of the grating, or by 
the gaussian approximation.

\subsection{Study of resolving power of our approach}\label{sec:etude-du-pouvoir}

This part is devoted to numerical quantification 
of the gain in angular resolution provided with our method, 
using the Rayleigh criterion, which is frequently 
used by astrophysicists: for the smaller resolvable detail, the first minimum of
 the image of one point source coincides with the maximum of another. 
In practice, two point sources with the same intensity  
and a flat strectrum are considered to be separated if the minimal flux between 
the two peaks is lower than 0.9 times the flux at the peak positions. 
The resolution is
studied in the $\beta$ direction only as this is the direction in which the
subslit scan is performed. 

\begin{figure}[htb]
  \centering
  \begin{tabular}{ccc}
    \includegraphics[width=0.12\textwidth]{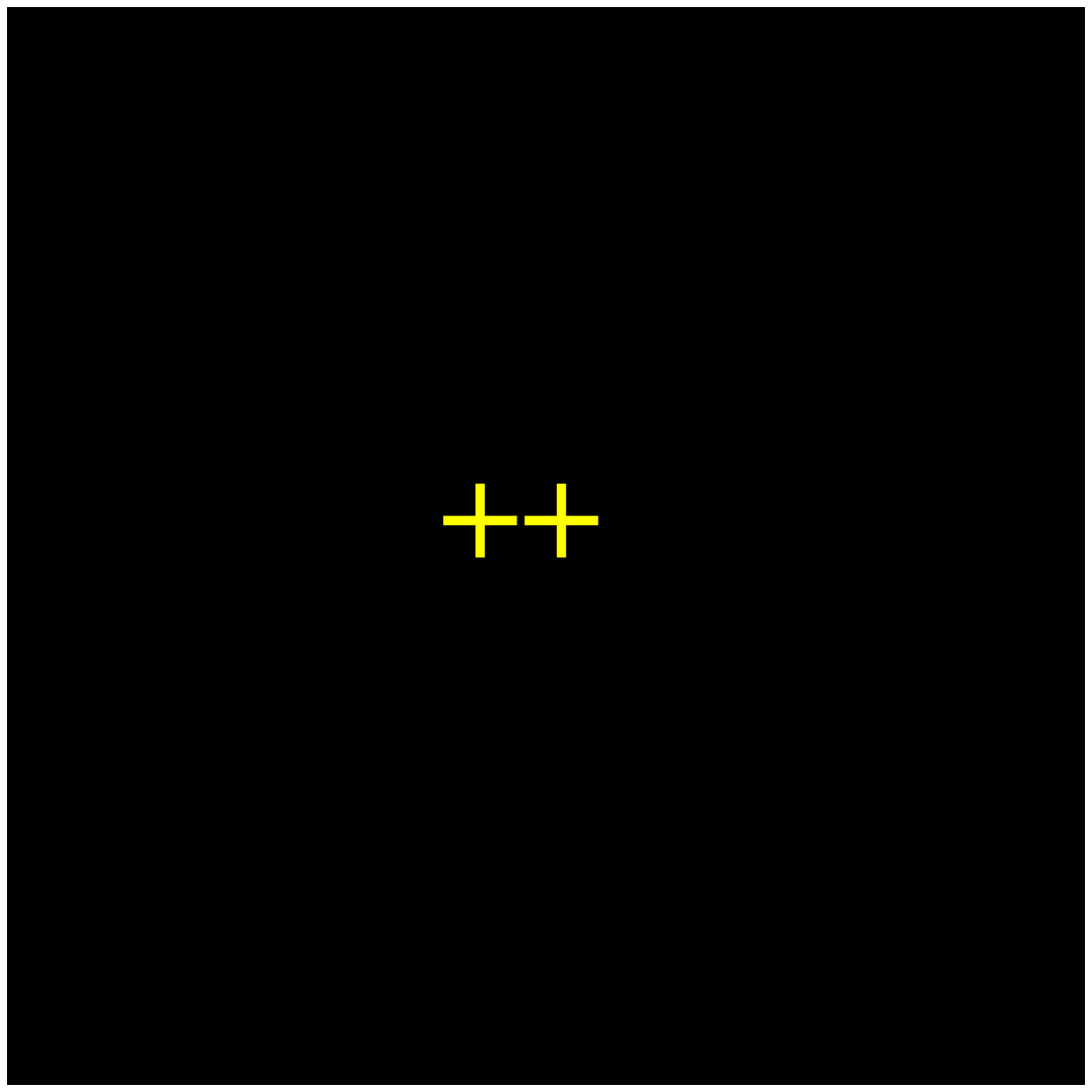}&
    \includegraphics[width=0.12\textwidth]{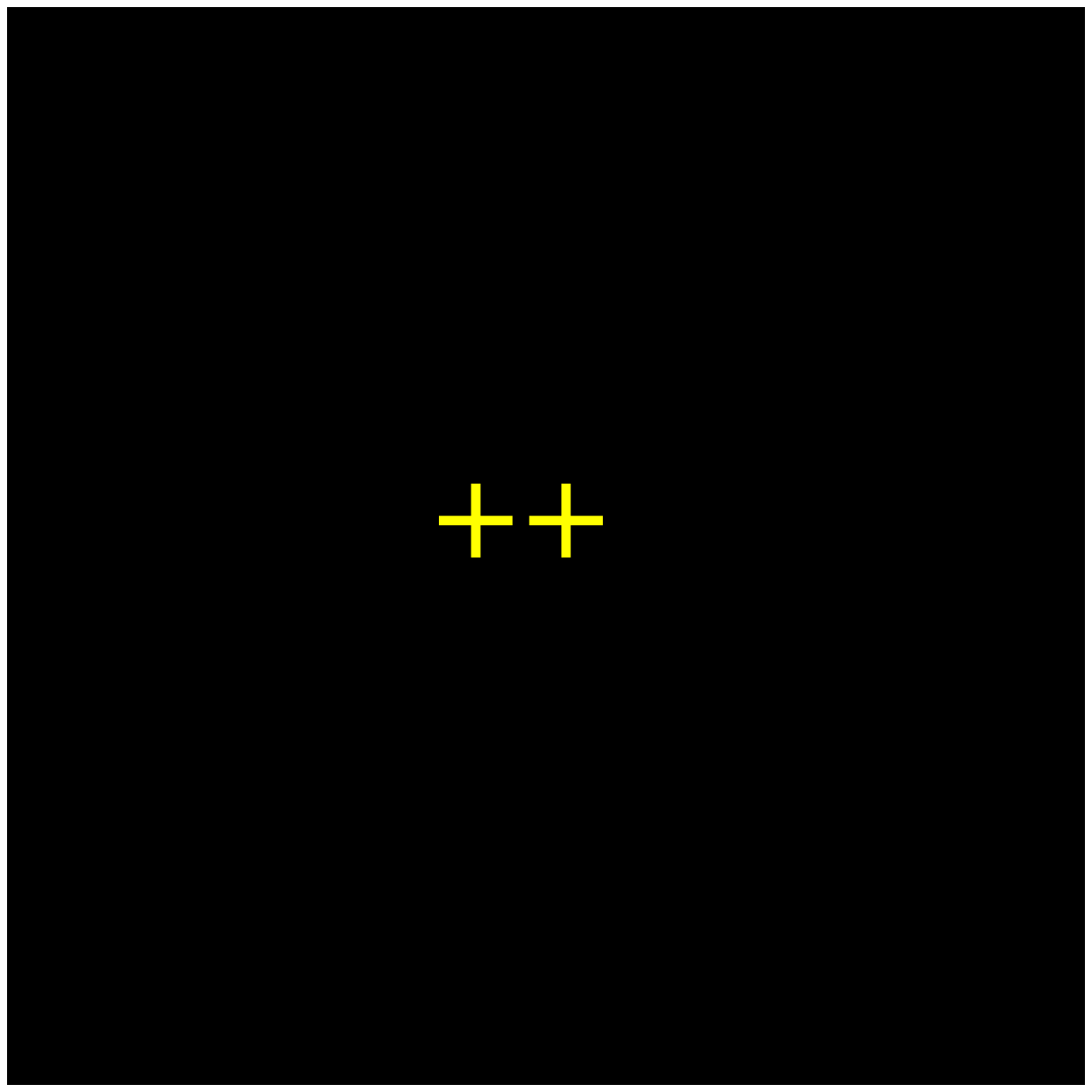}&
    \includegraphics[width=0.12\textwidth]{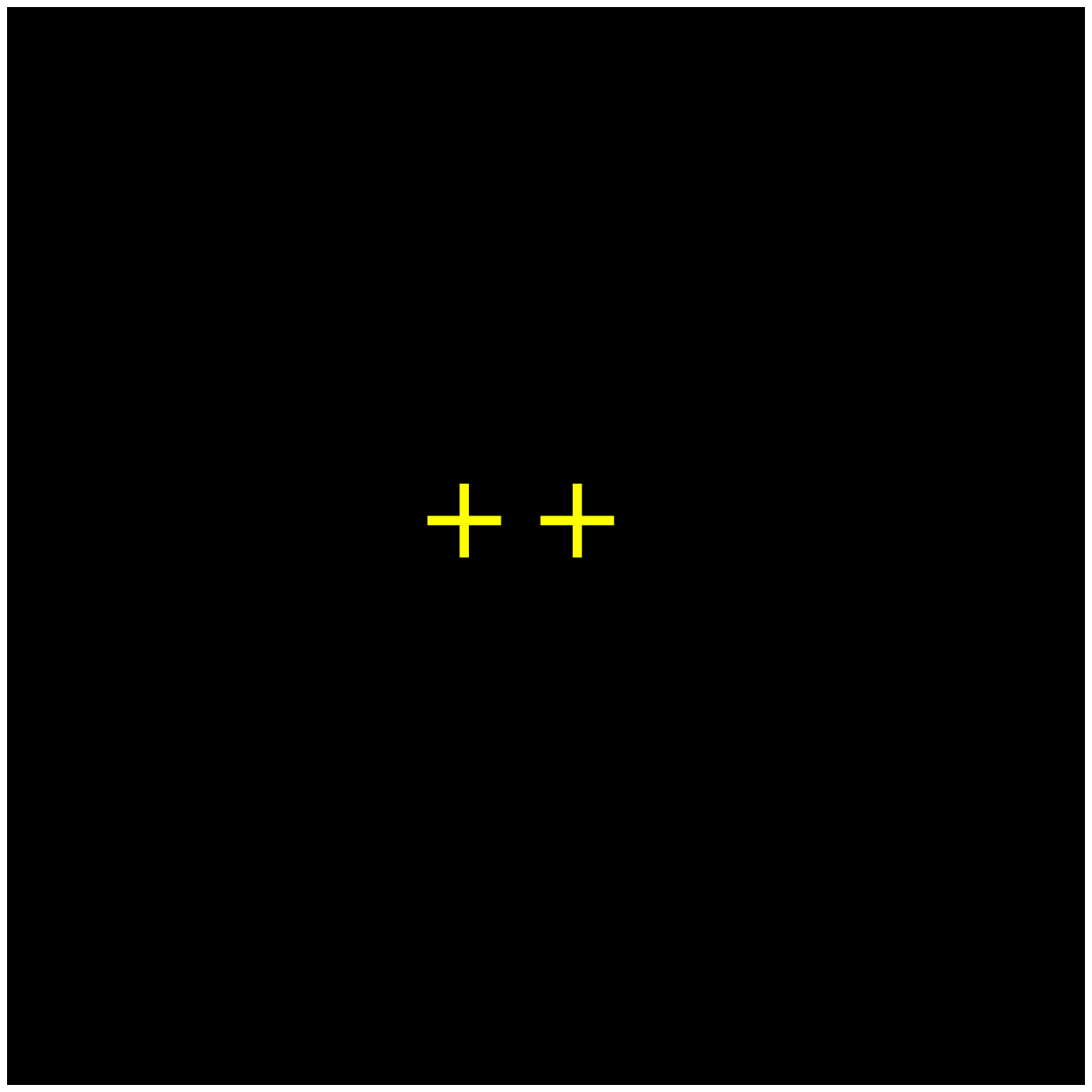}\\
    (a)&(b)&(c)\\
    \includegraphics[width=0.146\textwidth]{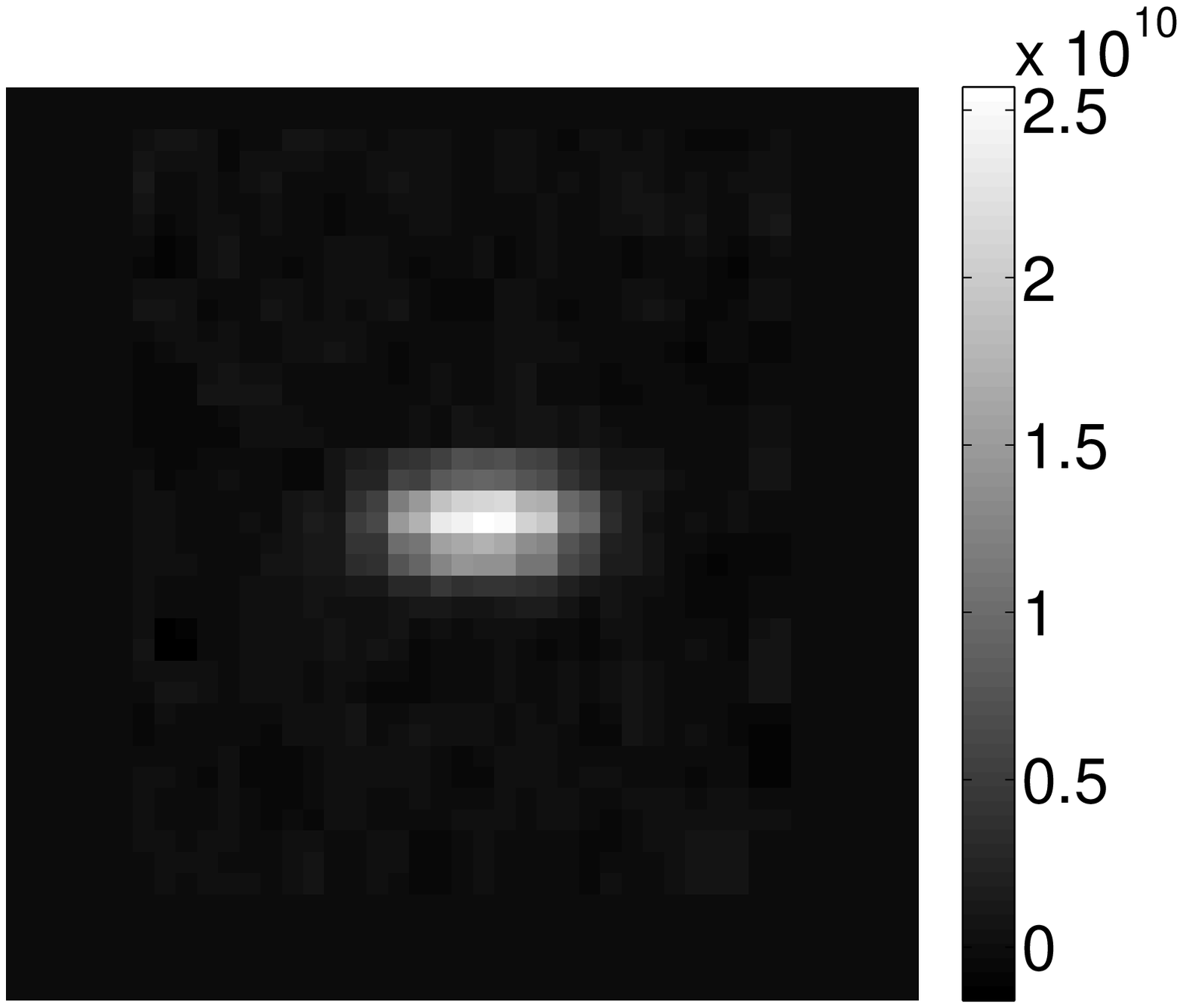}&
    \includegraphics[width=0.146\textwidth]{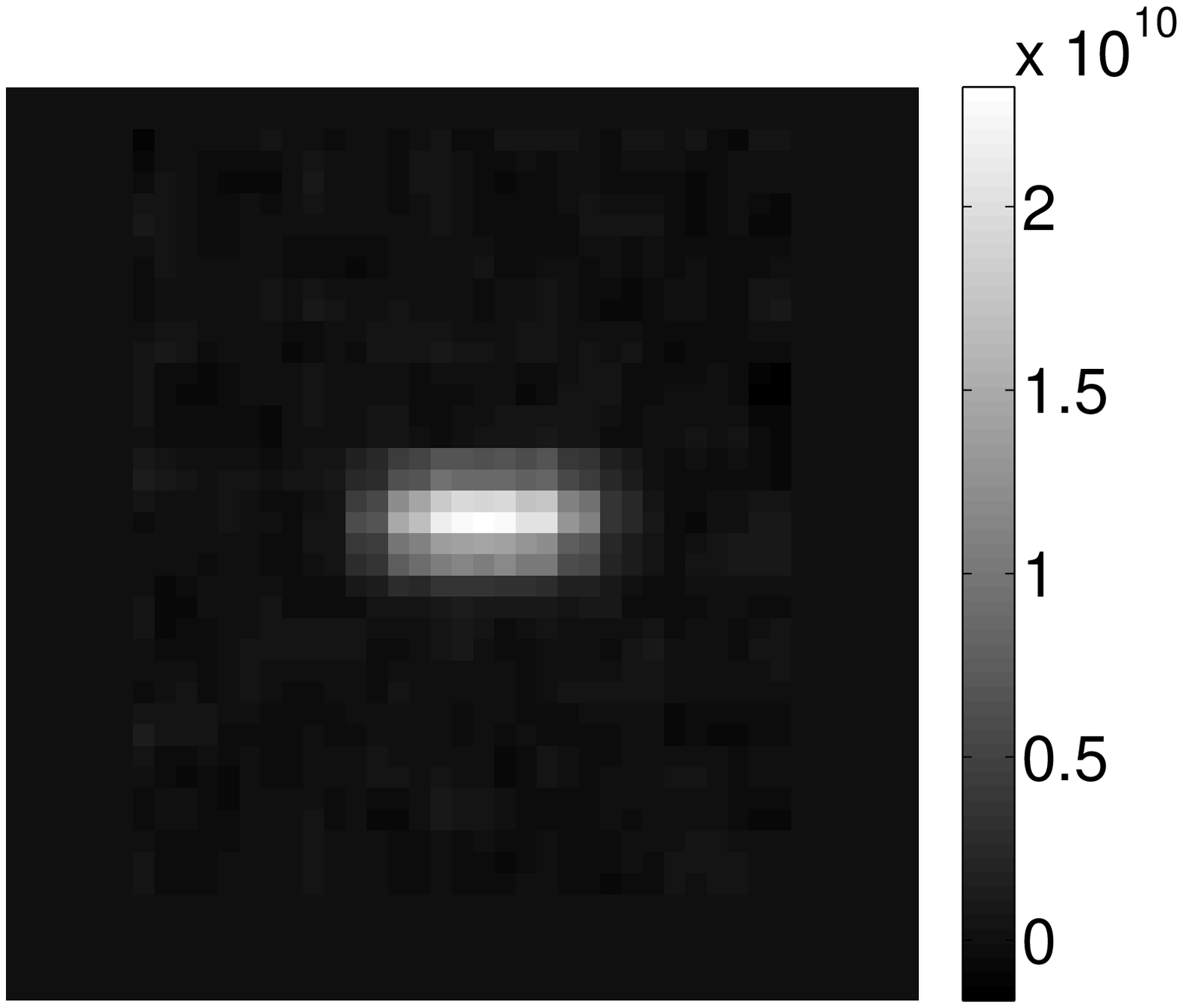}&
    \includegraphics[width=0.146\textwidth]{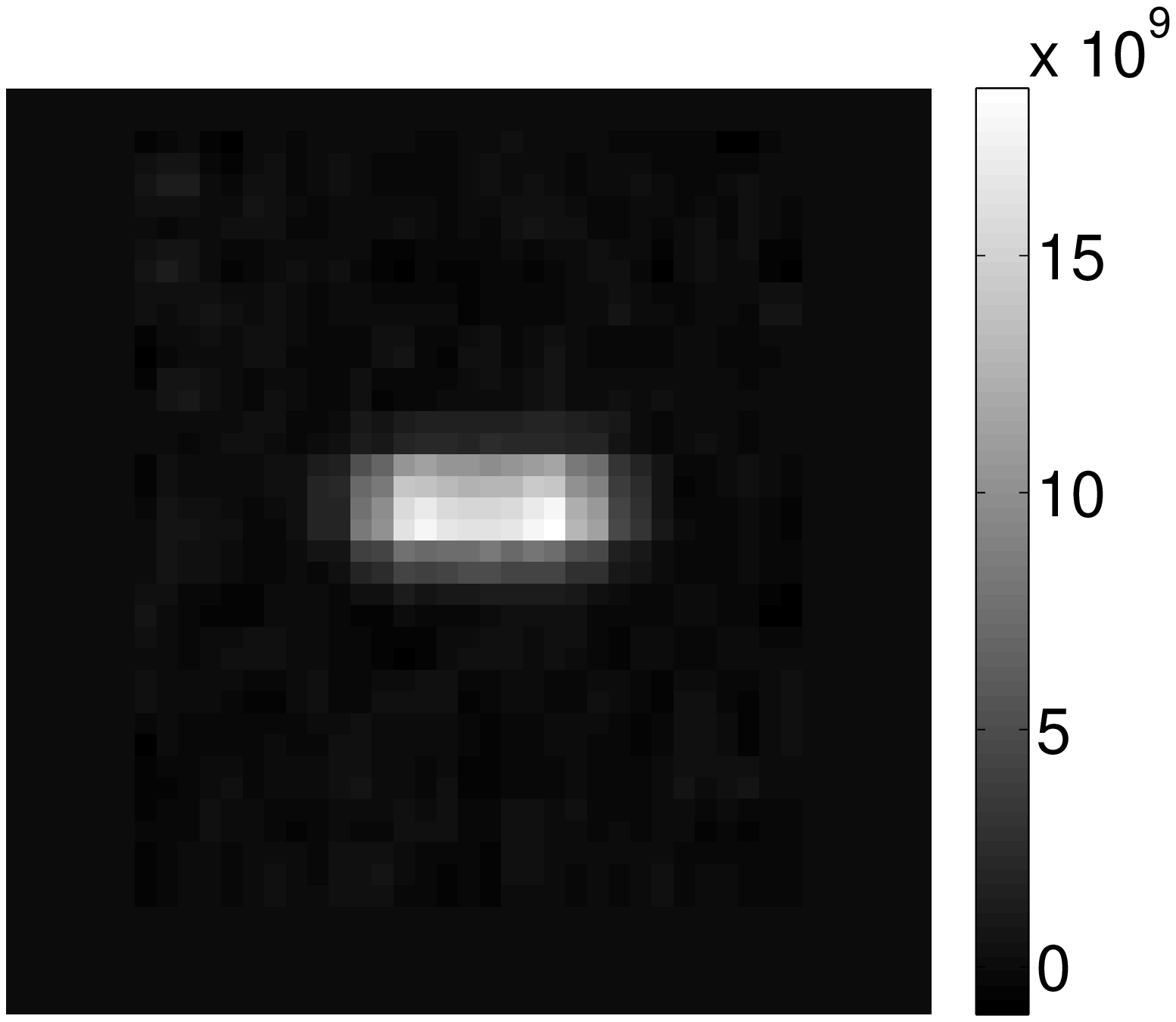}\\
    (d)&(e)&(f)\\
    \includegraphics[width=0.146\textwidth]{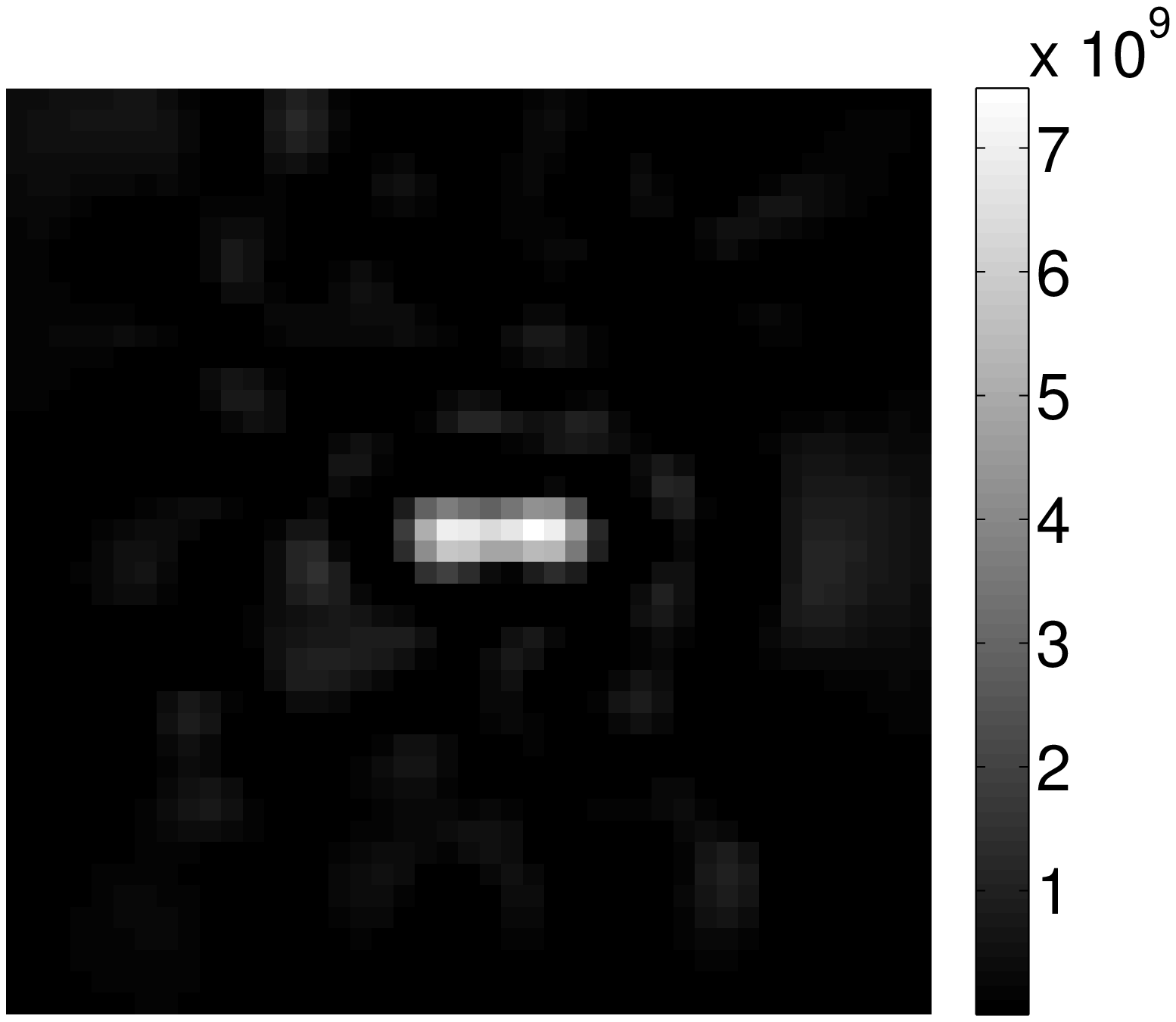}&
    \includegraphics[width=0.146\textwidth]{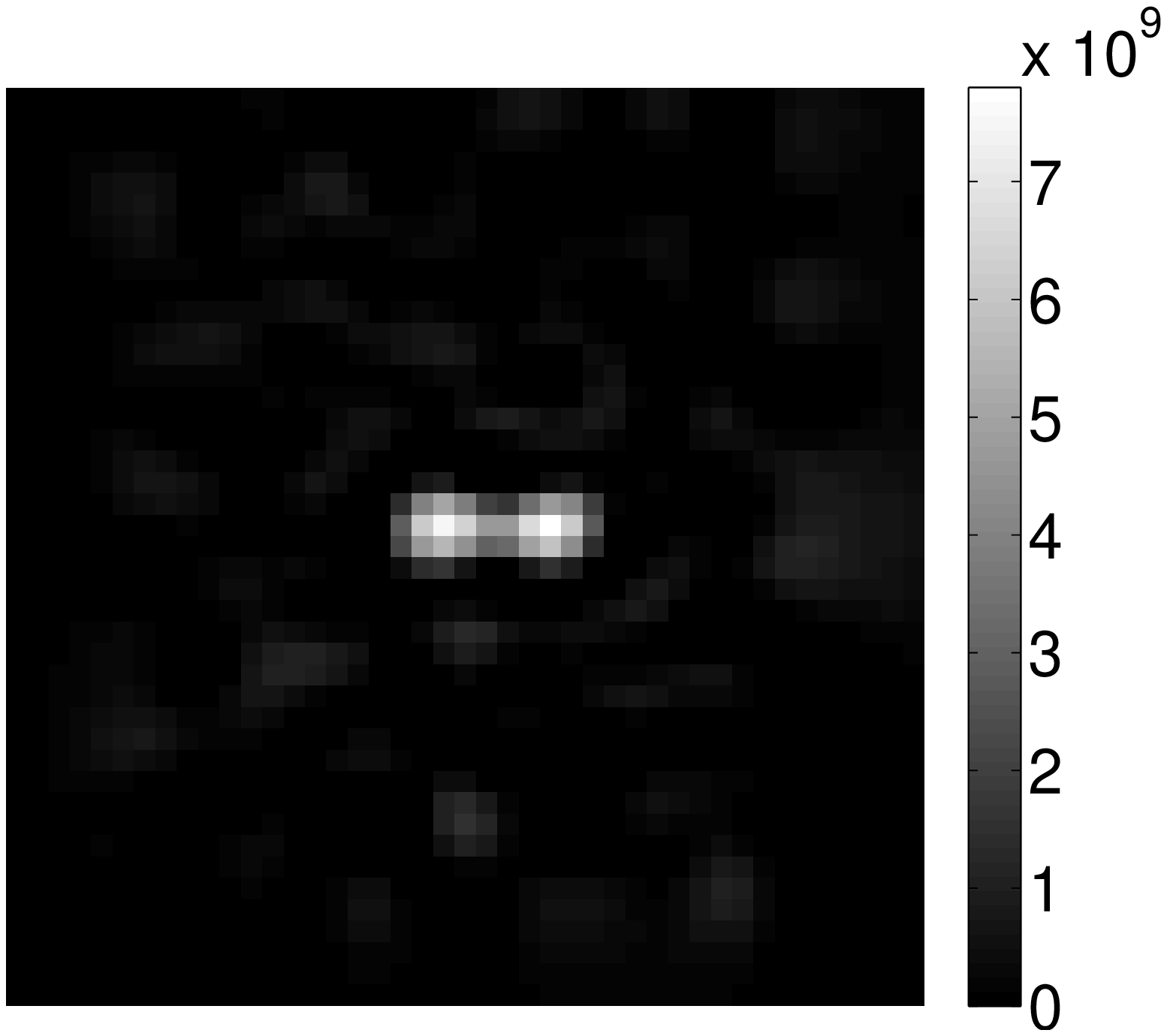}&
    \includegraphics[width=0.146\textwidth]{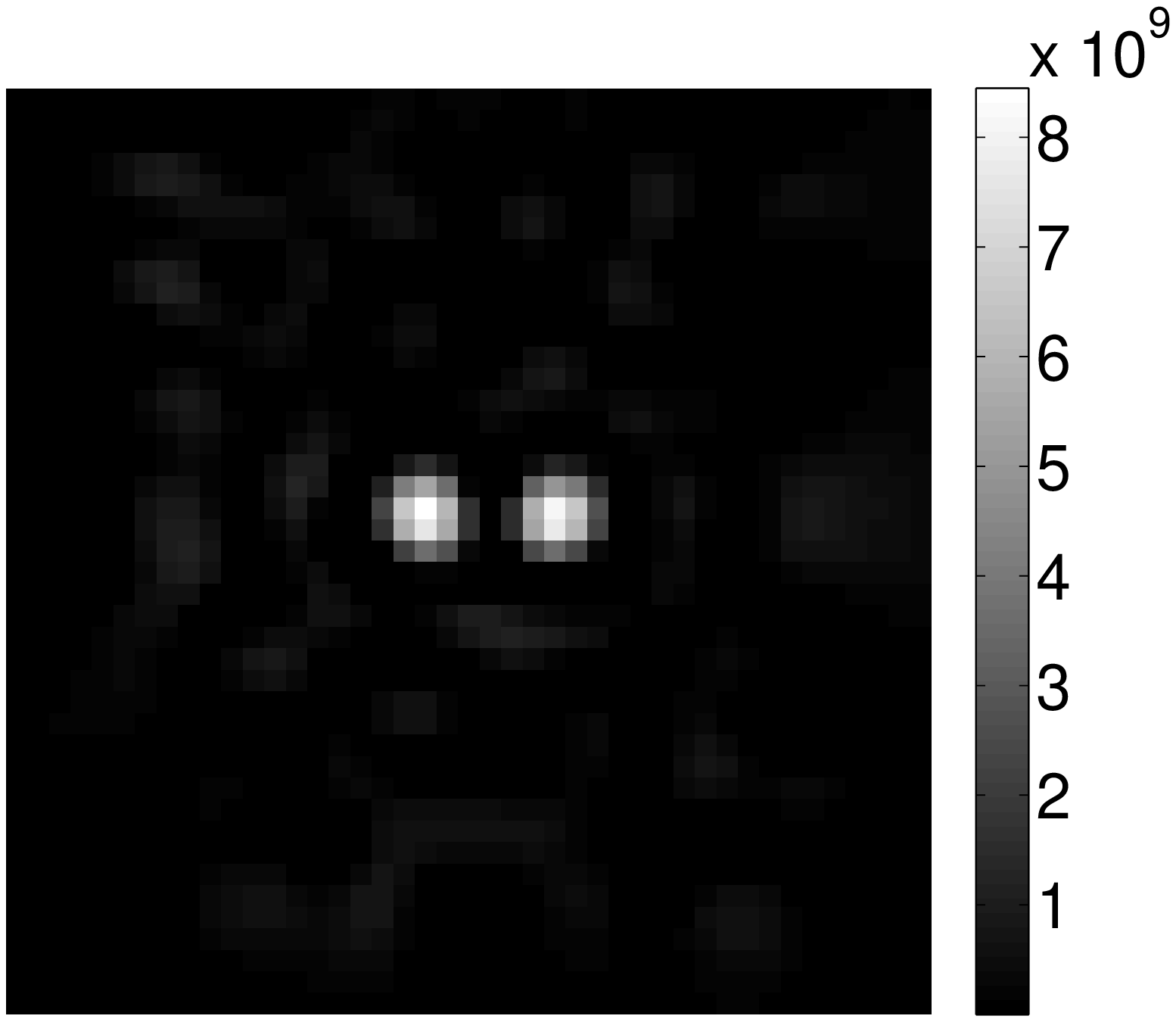}\\
    (g)&(h)&(i)
  \end{tabular}  
  \caption{Two peaks reconstruction for different $\delta$ : 
 (a-c) Visualisation of the peaks position   
for $\delta=3.4$, 4 and 5.4 arcsecond resp., (d-f) reconstruction with the conventional 
method for $\delta=3.4$, 4 and 5.4 arcsecond resp., (g-i) reconstruction with our over-resolution method for $\delta=3.4$,  4 and 5.4 arcsecond respectively. }
  \label{fig:ImagesRes}
\end{figure}

Two point sources are injected, at positions $\beta_1$ and $\beta_2$, 
respectively
(see Fig.~\ref{fig:ImagesRes}, top).
The corresponding data are simulated, and the reconstruction $\hat{\phi}(\beta)$ 
is performed.  As explained above, the two point sources are considered to be 
separated if  $\hat{\phi}(\cro{\beta_1+\beta_2}/2) \le 0.9 \times 
\hat{\phi}(\beta_1)$.
 The resolution is defined as the difference $\delta=\beta_2-\beta_1$ at which 
the two point sources start to be separated.

\begin{figure}[htb]
  \centering
  \begin{tabular}{c}
\includegraphics[width=0.37\textwidth]{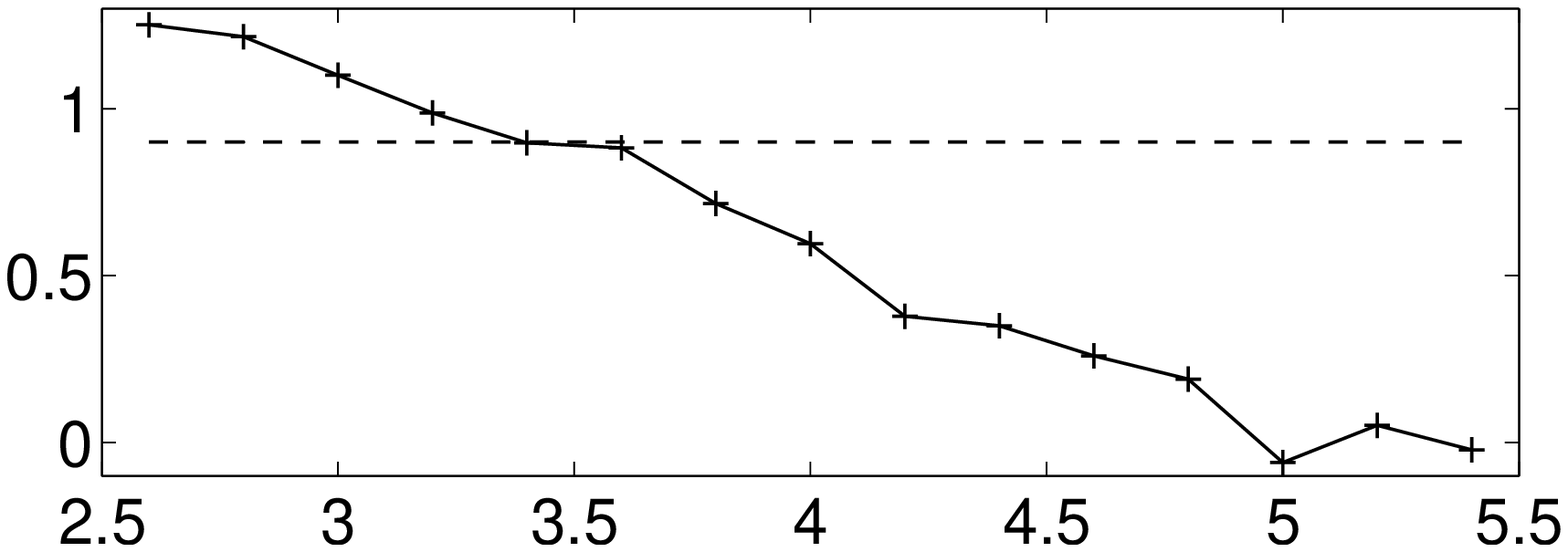}\\
(a)\\
  \includegraphics[width=0.37\textwidth]{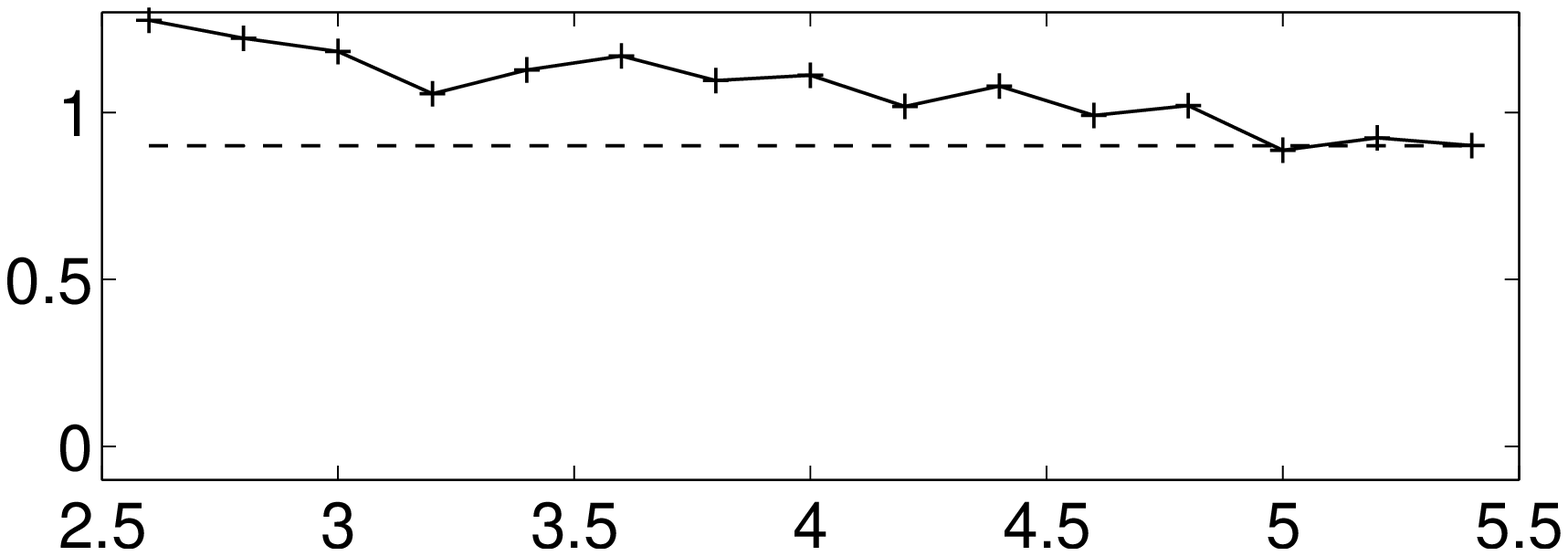}\\
(b)\\
  \end{tabular}
  \caption{Resolution of our method: the curve represents the ratio of the
  intensity at one peak to the intensity between the two peaks as a function
  of the distance between the peaks in arcseconds. The resolution is read
  at crossing of this curve and the dotted line (the ratio is 0.9). (a) Results
  obtained with our method. (b) Results obtained with the conventional  method} \label{fig:ResolF2}.
\end{figure}

Point sources are simulated for a set of differences $\delta$ between 2.4 and 
5.4 arcseconds and simulations are performed 
in the configuration of the real data 
(signal to noise ratio, energy of the data). Moreover, we use the regularization 
parameters 
 $\mu_{\alpha\beta}$ and $\mu_{\lambda}$ determined in section 
\ref{sec:donnees-simulees-1}. 
A number of reconstructions has been obtained. The ratio between the values of 
the reconstructed
function at $\beta_1$ and $(\beta_1+\beta_2)/2$ is calculated as a function of 
the difference $\delta$ between the two peaks. Results are shown in 
Fig. \ref{fig:ResolF2}. 

The computed resolution is 3.4  arcseconds (see Fig. \ref{fig:ResolF2}(a))
and 5 arcseconds (see Fig. \ref{fig:ResolF2}(b)) for our method 
and the conventional method, respectively.
Fig. \ref{fig:ImagesRes} illustrates this gain in angular resolution.
In the left column on Fig. \ref{fig:ImagesRes} ($\delta=3.4$) 
corresponds to  
the limit of resolution of our method. In this case, 
the peak is not separated with the conventional method 
(Fig. \ref{fig:ImagesRes} (d)). In the middle column on 
Fig. \ref{fig:ImagesRes}, our algorithm clearly separates 
the peak (Fig. \ref{fig:ImagesRes}(h)) and not the conventional method
(Fig.~\ref{fig:ImagesRes}(e)). In the right column, we observe 
a stain with our method is smaller than the conventional method.
Our method increases the resolution by a factor 1.5. 



\section{Conclusions}\label{sec:conclusions}

We have developed 
an original method for reconstructing the over-resolved 3D sky from
data provided by the IRS instrument. This method is based on:

\begin{enumerate}

\item a continuous variable model of the instrument based on
a precise integral physical description,

\item a decomposition of the continuous variable object over a family of 
Gaussian functions, which results in a linear, semi-parametric relationship,

\item an inversion in the framework of deterministic regularization based on
a quadratic criterion minimized by a gradient algorithm. 

\end{enumerate}

The first results on real data 
show that we are able to evidence spatial structures not detectable using conventional methods. 
The spatial resolution is improved by a factor 1.5.
This factor should increase using data with a motion between two acquisitions smaller than  
the half a slit width.

In the future, we plan to design highly efficient processing tools 
using our approach 
in particular
for the systematic processing of the data which will be taken 
with the next generation of infrared to milimeter space observatory
 (\textsc{Herschel}, \textsc{Planck}, ...).

%
\appendices

%
%

%
\bibliographystyle{IEEEtran}


\begin{IEEEbiography}
[{\includegraphics[width=1in,height=1.25in,clip,keepaspectratio]{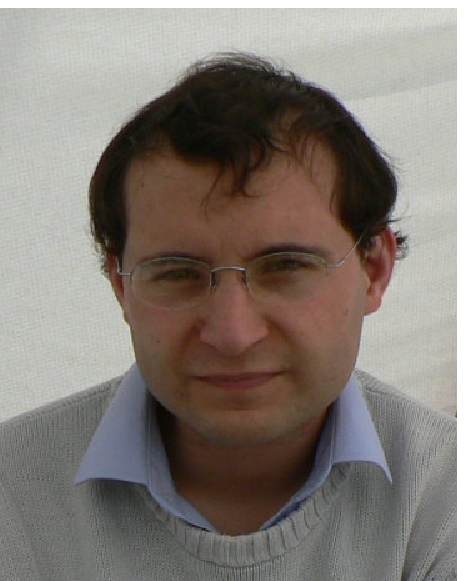}}]
{Thomas Rodet} was born in Lyon, France, in 1976. 
He received the Doctorat degree at Institut National Polytechnique de Grenoble, France, in 2002. \\
\indent
He is presently assistant professor in the D\'epartement de Physique at
 Universit\'e Paris-Sud 11 and researcher with the Laboratoire des Signaux et 
Syst\`emes (CNRS - Sup\'elec - UPS). He is interested in tomography methods 
and  Bayesian methods for inverse problems in astrophisical problems
(inversion of data taken from space observatory: Spitzer, Herschel, SoHO, STEREO).
\end{IEEEbiography} 

\begin{IEEEbiography}
  [{\includegraphics[width=1in,height=1.25in,clip,keepaspectratio]{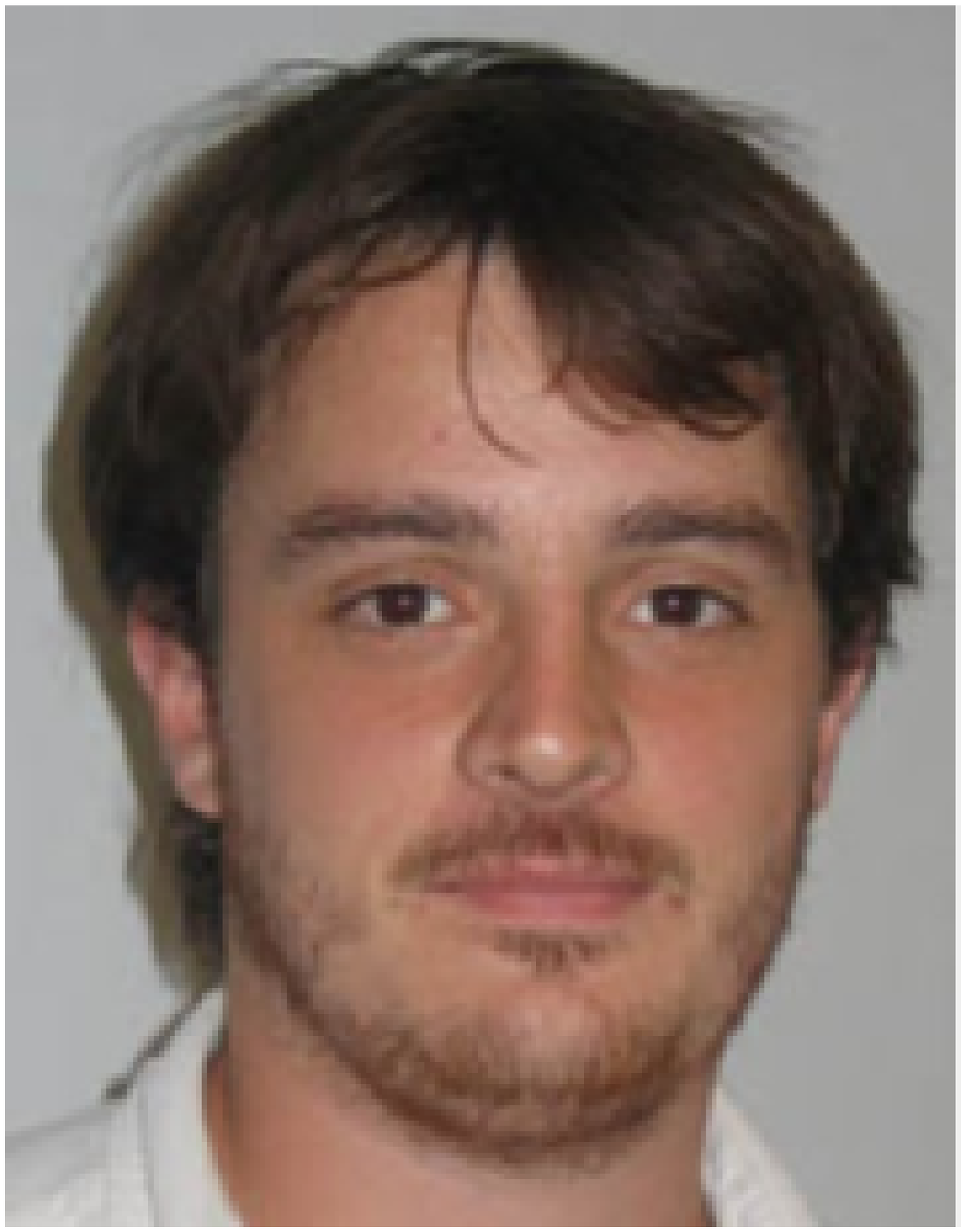}}]
  {Fran\c{c}ois Orieux} was born in Angers, France, in 1981. He
  graduated from the \'Ecole Sup\'erieure d'\'Electronique de l'Ouest
  in 2006 and obtain the M. S. degree from the University of Rennes 1.
  He currently pursuing the Ph. D. degree at the Universit\'e
  Paris-Sud 11.  His research interests are statistical image
  processing and Bayesian methods for inverse problems.
\end{IEEEbiography}

\begin{IEEEbiography}
[{\includegraphics[width=1in,height=1.25in,clip,keepaspectratio]{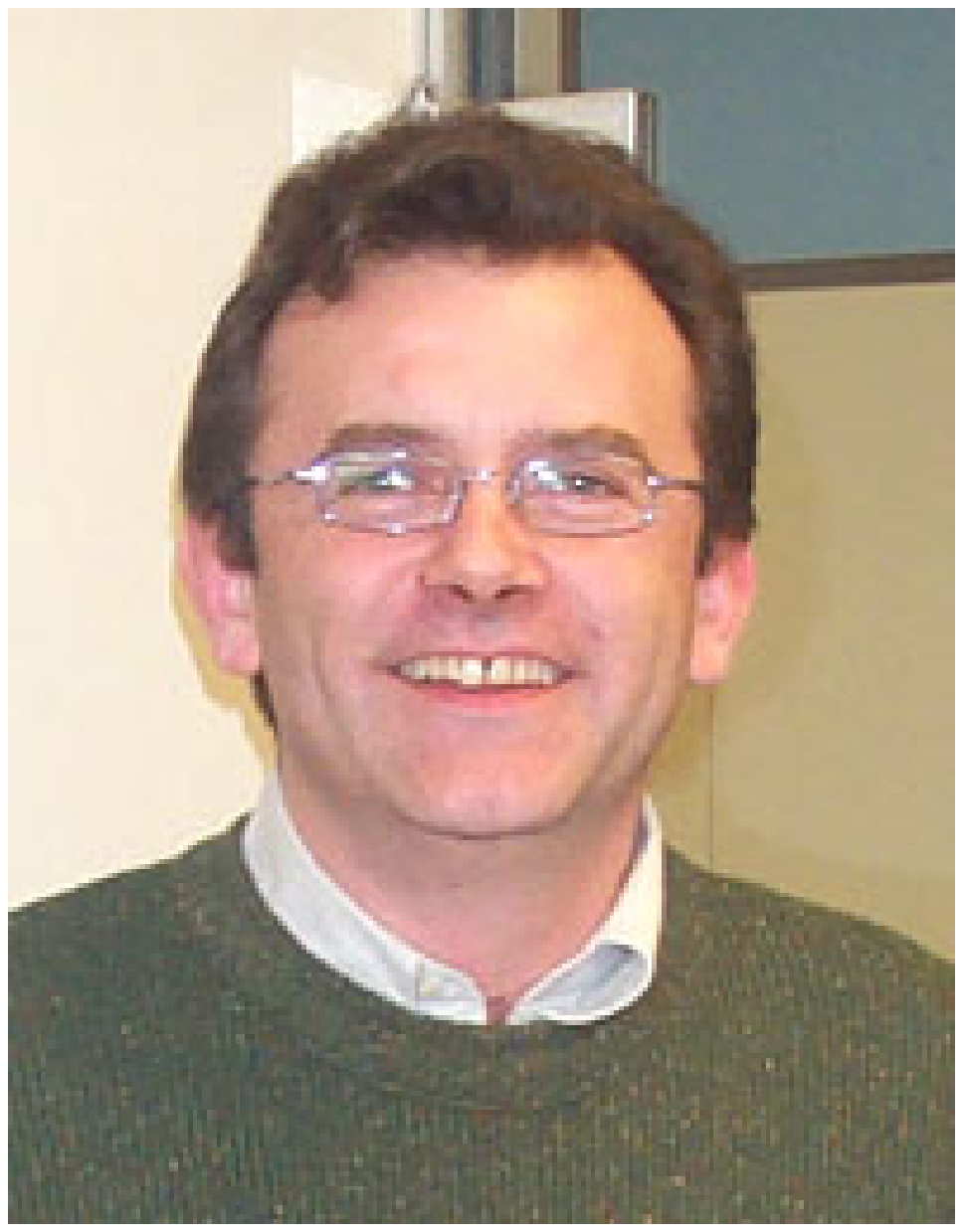}}]
{Jean-Fran\c{c}ois Giovannelli} was born in B\'eziers, France, in 1966. He graduated from the \'Ecole Nationale Sup\'erieure de l'\'Electronique et de ses Applications in 1990. He received the Doctorat degree in physics at Universit\'e Paris-Sud, Orsay, France, in 1995. \\
\indent
He is presently assistant professor in the D\'epartement de Physique at Universit\'e Paris-Sud and researcher with the Laboratoire des Signaux et Syst\`emes (CNRS - Sup\'elec - UPS). He is interested in regularization and Bayesian methods for inverse problems in signal and image processing. Application fields essentially concern astronomical, medical and geophysical imaging.
\end{IEEEbiography}

\begin{IEEEbiography}
[{\includegraphics[width=1in,height=1.25in,clip,keepaspectratio]{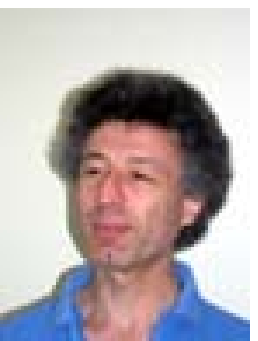}}]
{Alain Abergel} was born in Paris, France, in 1959. He received the Doctorat degree in astrophysics 
at the Paris\,6 University, France, in 1987, and is presently Professor in Physics and Astronomy at 
the Paris-Sud University.
He is interested in the interstellar medium of galaxies, and uses data taken from long wavelength 
space observatories (IRAS, COBE, ISO, Spitzer, ... ). He is Co-investigator for the Spectral 
and Photometric Imaging Receiver (SPIRE) instrument on-board the European Space Agency's Herschel Space Observatory. 
\end{IEEEbiography}

\end{document}